\documentclass[cernpreprint,atlasdraft=true,texlive=2016,UKenglish,texmf]{atlasdoc}
 
\usepackage{atlaspackage}
 
\usepackage{atlasbiblatex}
 
\usepackage{atlasphysics}
 
\graphicspath{{logos/}{figures/}}
 
\usepackage{ANA-HIGG-2018-28-PAPER-defs}
\usepackage{adjustbox}
\usepackage{multirow}
\usepackage[inline]{enumitem}
\usepackage{harpoon}
\usepackage{braket}
\usepackage[utf8]{inputenc}
 
\DeclareUnicodeCharacter{2212}{-}
 
\AtlasTitle{Higgs boson production cross-section measurements and their EFT interpretation in the $4\ell$ decay channel at $\sqrt{s}=$13~TeV with the ATLAS detector}

\AtlasAbstract{
Higgs boson properties are studied in the four-lepton decay channel (where lepton = $e$, $\mu$) using 139~fb$^{-1}$ of proton--proton collision data recorded
at $\sqrt{s}=$13~TeV by the ATLAS experiment at the Large Hadron Collider.  The inclusive cross-section times branching ratio for $H\to ZZ^*$ decay is measured to be $1.34 \pm 0.12$~pb for a Higgs boson with absolute rapidity below 2.5,  in  good agreement with the Standard Model prediction of $1.33 \pm 0.08$~pb.  Cross-sections times branching ratio are measured for the main Higgs boson production modes in several exclusive phase-space regions. The measurements are interpreted in terms of coupling modifiers and of the tensor structure of Higgs boson interactions using an effective field theory approach. Exclusion limits are set on the CP-even and CP-odd `beyond the Standard Model' couplings of the Higgs boson to vector bosons, gluons and top quarks.}
 
\author{The ATLAS Collaboration}
 
\AtlasRefCode{HIGG-2018-28}
 
\PreprintIdNumber{CERN-EP-2020-034}
 
\AtlasJournalRef{Eur. Phys. J. C 80 (2020) 957}
\AtlasDOI{10.1140/epjc/s10052-020-8227-9}
\AtlasJournal{EPJC}

\hypersetup{pdftitle={ATLAS document},pdfauthor={The ATLAS Collaboration}}
 
\begin{document}
 
\maketitle
 
\tableofcontents

\section{Introduction}
\label{sec:intro}
 
The observation of the Higgs boson by the ATLAS and CMS experiments~\cite{HIGG-2012-27, CMS-HIG-12-028} with the  Large Hadron Collider (LHC)  \runa~data set at centre-of-mass energies of $\sqrt{s}=$~7~\TeV\ and 8~\TeV~was a major step towards an understanding of the electroweak (EW) symmetry breaking mechanism~\cite{Englert:1964et,Higgs:1964pj,Guralnik:1964eu}. Tests of its spin and CP quantum numbers strongly indicate that the observed particle is of scalar nature and that the dominant   coupling structure is CP-even, consistent with the Standard Model (SM) expectation~\cite{CMS-HIG-12-041, HIGG-2013-01, HIGG-2013-17}. The measurements of the Higgs boson production and differential cross-sections, branching ratios, and the derived constraints on coupling-strength modifiers, assuming the SM coupling structure, have also shown no significant deviation from the predictions for the SM Higgs boson with a mass of 125~\GeV~\cite{diffXS, HIGG-2015-07, CMS-HIG-17-031,  HIGG-2018-57}. Furthermore, constraints have been set on various coupling parameters beyond the SM (BSM) that modify the tensor structure of the Higgs boson couplings to SM particles~\cite{HIGG-2013-17, CMS-HIG-14-018, CMS-HIG-14-035, HIGG-2015-06, CMS-HIG-17-011, HIGG-2016-22, HIGG-2016-21, CMS-HIG-18-002, CMS-HIG-17-034}.

Motivated by a clear Higgs boson signature and a high signal-to-background ratio in the $H \to ZZ^{*} \to 4\ell$ decay channel (where $\ell = e\text{ or }\mu$), the updated measurements of the Higgs boson coupling properties in this channel are presented using the entire Run~2 data set with \LumExact~of proton--proton ($pp$) collision data collected at $\sqrt{s}=13$~\TeV\  by the ATLAS detector between 2015 and 2018. Three types of results are presented in this paper: (i)  measurements of the Higgs boson production cross-sections times branching ratio, hereafter referred to as cross-sections, for the main production modes in several exclusive phase-space bins
in dedicated fiducial regions; (ii) interpretation of the measurements in terms of constraints on the Higgs boson coupling-strength modifiers within the $\kappa$-framework~\cite{Heinemeyer:2013tqa}; and (iii) interpretation of the measurements in terms of modifications to the tensor structure of Higgs boson couplings using an effective field theory (EFT) approach.
 
In addition to  a nearly four times higher integrated luminosity, there are several other important differences compared  to the previous results in this analysis channel~\cite{HIGG-2016-22}:
\begin{itemize}
\item an improved lepton isolation to mitigate the impact of additional $pp$ interactions in the same or neighbouring bunch crossings (pile-up),
\item an improved jet reconstruction using a particle flow algorithm~\cite{PERF-2015-09},
\item additional event categories for the classification of Higgs boson candidates,
\item new discriminants to enhance the sensitivity to distinguish the various production modes of the SM Higgs boson,
\item the use of data sidebands to constrain the dominant $ZZ^*$ background process,
\item a dedicated control region  to constrain the background in the reconstructed event categories probing \ttH\ production,
\item improved estimates of \zjets, \ttprod, and $WZ$ backgrounds, and
\item an EFT interpretation, based on a parameterisation of the cross-sections rather than a direct parameterisation of the reconstructed event yields.
\end{itemize}

\subsection{Simplified template cross-sections}
\label{subsec:intro_stxs}
 
In the framework of \textit{Simplified Template Cross Sections} (STXS)~\cite{bendavid2018les,YR4,Berger:2019wnu}, exclusive regions of phase space are defined for each Higgs boson production mechanism.
These phase-space regions, referred to as production bins, are defined to reduce the dependence on theoretical uncertainties
that directly fold into the measurements and at the same time maximise
the experimental sensitivity to measure the bins, enhance the contribution from possible BSM effects, and allow measurements from different Higgs boson decay modes to be combined. The number of production bins is limited to avoid loss of measurement sensitivity for a given amount of integrated luminosity.
 
\begin{figure}[htb!]
\hskip2.6cm\includegraphics[width=1.14\linewidth, angle=90]{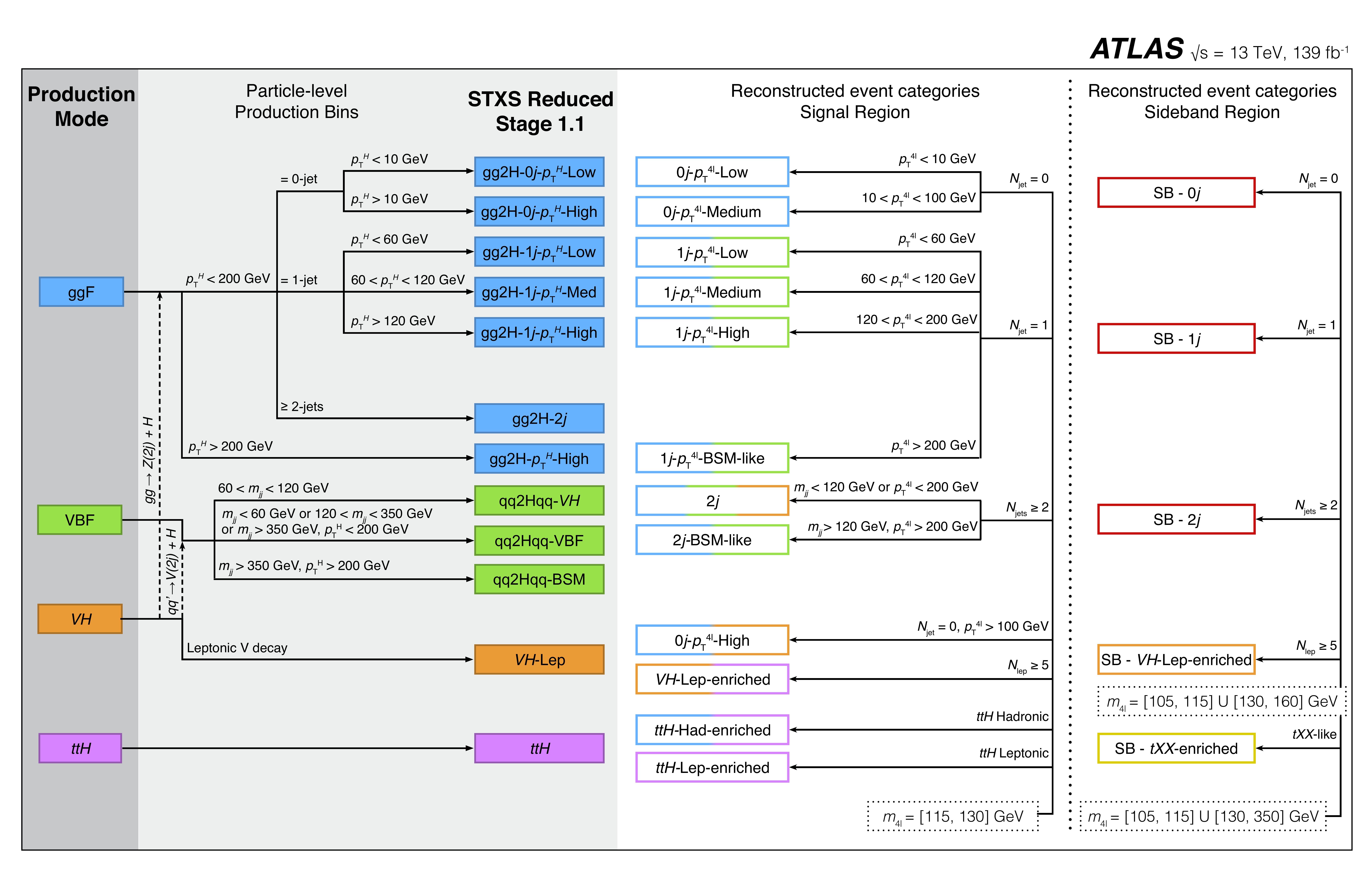}
\caption{Two sets (Production Mode Stage and Reduced Stage 1.1) of exclusive phase-space regions (production bins) defined at particle-level for the measurement of the Higgs boson production cross-sections (left and middle-left shaded panels), and the corresponding reconstructed event categories for signal (middle-right panel) and sidebands (right panel). The description of the production bins is given in Section~\ref{subsec:intro_stxs}, while the reconstructed
signal region and sideband event categories are described in Section~\ref{sec:categorization} and Section~\ref{sec:background}, respectively. The \bbH ($tH$) contribution is included in the \ggF (\STXSttH) production bins. The colours of each reconstructed event category box indicates the contributions from the relevant production processes.
}
\label{fig:stxs_bins}
\end{figure}

The definitions of the production bins used for this measurement are shown in the left panel of \figref{fig:stxs_bins}~(shaded area). All production bins are defined for Higgs bosons with rapidity $|y_H|<$~2.5 and no requirement is placed on the particle-level leptons.
Two sets of production bins with different granularity are considered, as a trade-off between statistical and theoretical uncertainties.
 
The first set of production bins (Production Mode Stage)~\cite{YR4} is defined according to the Higgs boson production modes: gluon--gluon fusion (\ggF), vector-boson fusion (\STXSVBF) and associated production with vector bosons (\STXSVH, where $V = W$ or $Z$) or top quark pairs (\STXSttH). Since $b$-jets from \bbH associated production are emitted at small angles relative to the beam axis and usually outside of the detector acceptance, the \bbH and \ggF Higgs boson production modes have similar signatures and acceptances. Their contributions are considered together with their relative ratio fixed to the SM prediction. In the following, the sum of their contributions is referred to as ggF. Similarly, single top production (\STXStH) is considered together with \ttH, with their relative ratio fixed to the SM prediction. In contrast to the Stage-0 production bins described in Ref.~\cite{YR4}, the \VH events with hadronic decays of the vector boson $V$ are included in the \VH production bin rather than in the \ggF or \STXSVBF bins. In this way, each of the four main Higgs boson production modes can be measured separately.
 
The second set of production bins (Reduced Stage 1.1) is more exclusive than the first one. Starting from the production bins of a more granular Stage~1.1 set~\cite{Berger:2019wnu}, several production bins are merged as the full set of bins cannot be measured separately in the $H\to ZZ^* \to 4\ell$ channel with the current data sample. The definitions of the bins are based on the multiplicity of particle-level jets, the Higgs boson transverse momentum $\pt^H$ and the invariant mass \Mjj\ of the two jets with the highest transverse momentum. Particle-level jets are built from all stable particles (particles with lifetime c$\mathrm{\tau} >$ 10~mm) including neutrinos, photons, and leptons from hadron decays or those produced in the parton shower.  The anti-$k_t$ jet reconstruction algorithm~\cite{Cacciari:2008gp,Cacciari:2011ma} with a radius parameter $R=0.4$ is used. All Higgs boson decay products, as well as the leptons and neutrinos from the decays of the associated $V$ bosons are excluded from the jet building, while the decay products from hadronically decaying associated $V$ bosons, are included. The jets are required to have $\pt>$~30~\GeV, with no restrictions on rapidity.
 
Events from \ggF production and $gg\rightarrow ZH$ production with a hadronically decaying $Z$ boson are split into seven common production bins. Six bins have a Higgs boson transverse momentum below 200~\GeV, while the seventh bin with Higgs boson transverse momentum above 200~\GeV\ (\STXSggToHHigh) is sensitive to contributions from BSM physics. For $\pt^H$ below 200~\GeV, further splits are made according to the jet multiplicity and $\pt^H$. Events with no jets are split into two bins with $\pt^H$ below and above 10~\GeV. Events with one jet are split into three bins with $\pt^H$  below 60~\GeV, between 60~\GeV\ and 120~\GeV, and above 120~\GeV. Finally, Higgs boson events with two or more jets are combined into one bin. The bins are respectively denoted by \STXSggToHZeroJL, \STXSggToHZeroJH, \STXSggToHOneJL, \STXSggToHOneJM, \STXSggToHOneJH and \STXSggToHTwoJ.
 
As described in Ref.~\cite{Berger:2019wnu}, \STXSVBF and \STXSVH production with hadronically decaying associated $V$ bosons represent the $t$-channel and $s$-channel contributions to the same electroweak $qqH$ production process and are therefore considered together for further splitting. Three bins are defined: one bin, sensitive to BSM contributions (\STXSqqtoHqqBSM), with $\pt^H$ above 200~\GeV\ and $\Mjj$ above 350~\GeV; one bin (\STXSqqtoHqqVHLike) with $\Mjj$ between 60~\GeV\ and 120~\GeV\ to target the \STXSVH production mode; and one bin (\STXSqqtoHqqRest) with the Higgs boson not satisfying these criteria to ensure sensitivity to the \STXSVBF process. $qqH$ events in which one or both jets have transverse momenta below the 30~\GeV\ threshold are treated as a part of the \STXSqqtoHqqRest bin.
 
The \STXSVH process with the associated $V$ boson decaying leptonically is considered separately (\STXSVHLep). The leptonic decay includes the decays into $\tau$-leptons and neutrino pairs. The \STXSttH production bin remains the same as in the Production Mode Stage.
 
The middle-right and right panels of \figref{fig:stxs_bins} summarise the corresponding categories of reconstructed events in which the cross-section measurements and background estimations are performed. These are described in detail in Section~\ref{sec:categorization}.

\subsection{Higgs boson couplings in the $\kappa$-framework}
\label{subsec:intro_kappa}
 
To probe physics beyond the SM, the measured production cross-sections are interpreted within a leading-order-motivated $\kappa$-framework~\cite{Heinemeyer:2013tqa}, in which a set of coupling modifiers $\vec{\kappa}$ is introduced to parameterise deviations from the SM predictions of the Higgs boson couplings to SM bosons and fermions. The framework assumes that the data originate from a single CP-even Higgs boson state with a mass of $125$~\GeV\ and the tensor coupling structure of the SM for its interactions. Only the coupling strengths are allowed to be modified by the BSM processes. The Higgs boson width is assumed to be small enough such that the narrow-width approximation is valid, allowing the Higgs boson production and decay to be factorised:
\begin{equation*}
\sigma\cdot \mathcal B \ (i\to H \to f) = \sigma_{i}(\vec{\kappa})\cdot\frac{\Gamma_{f}(\vec{\kappa})}{ \Gamma_{H}(\vec{\kappa})},
\end{equation*}
where $\sigma_{i}$ is the production cross-section via the initial state $i$, $\mathcal B$ and $\Gamma_{f}$ are the branching ratio and partial decay width for the decay into the final state $f$, respectively, and $\Gamma_{H}$ is the total width of the Higgs boson. For a Higgs boson production and decay process via couplings $i$ and $f$, respectively, coupling-strength modifiers are defined as
\begin{equation*}
\kappa^2_i = \frac{\sigma_i}{\sigma^{\textrm{SM}}_i} \ \ \ \textrm{and} \ \ \ \kappa^2_f = \frac{\Gamma_f}{\Gamma^{\textrm{SM}}_f},
\end{equation*}
so that
\begin{equation*}
\sigma\cdot \mathcal B \ (i\to H \to f) =
\kappa_i^2 \cdot \kappa_f^2 \cdot \sigma_i^{\textrm{SM}} \cdot \frac{\Gamma_f^{\textrm{SM}}}{\Gamma_H(\kappa_i^2,\kappa_f^2)}.
\end{equation*}

\subsection{Tensor structure of Higgs boson couplings in the effective field theory approach}
\label{subsec:intro_eft}
 
The $\kappa$-framework assumes that the tensor structure of the Higgs boson couplings is the same as in the SM\@. In order to probe for possible non-SM contributions to the tensor structure of the Higgs boson couplings, the measured simplified template cross-sections are interpreted using an EFT approach. In this approach, which exploits exclusive kinematical regions of the Higgs boson production and decay phase space, the BSM interactions are introduced via additional higher-dimensional operators $\mathcal{O}_i^{(d)}$ of dimension $d$, supplementing the SM Lagrangian $\cal{L}_{\textrm{SM}}$,
\begin{equation*}
\mathcal L_{\textrm{EFT}} = \mathcal L_{\textrm{SM}} +  \sum_{i}^{} \frac{C_i^{(d)}}{\Lambda^{(d-4)}} \mathcal O_i^{(d)} \ \ \ \text{for} \ d>4.
\end{equation*}
The parameters $C^{(d)}_i$ specify the strength of new interactions and are known as the \textit{Wilson coefficients}, and $\Lambda$ is the scale of new physics. Only dimension-six operators are considered for this paper, since the dimension-five and dimension-seven operators violate lepton and baryon number conservation and the impact of higher-dimensional operators is expected to be suppressed by more powers of the cutoff scale~$\Lambda$~\cite{Hays:180800442}. For energies less than the scale of new physics, only the ratio $c_i = C_i^{(d=6)}/\Lambda^2$ can be constrained by the data.
 
Constraints are set on the Wilson coefficients defined within the Standard Model Effective Field Theory (SMEFT) formalism~\cite{SMEFT} in the Warsaw basis ~\cite{wbasis}. The measurements in the $H\to ZZ^* \to 4\ell$ channel do not provide sensitivity for  simultaneous constraints on the full set of these coefficients. To reduce the number of relevant parameters, a minimal flavour-violating scenario is assumed and only operators affecting the Higgs boson cross-section at tree level are considered. Operators affecting only double Higgs boson production and those affecting the Higgs boson couplings to down-type quarks and leptons are neglected due to limited sensitivity. The impact of these operators on the total Higgs boson decay width is also neglected.
\begin{table}[!htbp]
\renewcommand*{\arraystretch}{1.0}
\caption{Summary of EFT operators in the SMEFT formalism that are probed in the $H\to ZZ^* \to 4\ell$ channel. The corresponding tensor structure in terms of the SM fields from Ref.~\cite{SMEFT} is shown together with the associated Wilson coefficients, the affected production vertices and the impact on the $H\to ZZ^*$ decay vertex. The Higgs doublet field and its complex conjugate are denoted as $H$ and $\widetilde{H}$, respectively. The left-handed quark doublets of flavour $p$ (the right-handed up-type quarks) are denoted $q_p$ ($u_r$). $V_{\mu\nu}$  ($\widetilde{V}_{\mu\nu}=\epsilon^{\mu\nu\rho\sigma}V_{\rho\sigma}$) is the (dual) field strength tensor for a given gauge field $V = G, W, B$.  The bosonic operators  with (without) a dual field strength tensor are CP-odd (CP-even). For the remaining operator with fermions ($\mathcal{O}_{uH}$), the CP-odd contribution is introduced through the non-vanishing imaginary part of the corresponding Wilson coefficient, denoted as $c_{\widetilde{u}H}$.}
\begin{center}
\begin{tabular}{lll | lll | c c}
\hline\hline
\multicolumn{3}{c|}{CP-even}&
\multicolumn{3}{c|}{CP-odd}&
\multicolumn{2}{c}{Impact on}\\
 
Operator&
Structure&
Coeff.&
Operator&
Structure&
Coeff.&
production&
decay\\
 
\hline
$\mathcal O_{uH}$ &
$H H^{\dagger}  \bar q_p u_r \tilde H $ &
$c_{uH}$ &
$\mathcal O_{uH}$ &
\textcolor{black}{$H H^{\dagger}  \bar q_p u_r \tilde H $} &
$c_{\widetilde{u}H}$ &
\STXSttH &
- \\

$\mathcal O_{HG}$ &
$H H^{\dagger} G^{A}_{\mu\nu} G^{\mu\nu A}$ &
$c_{HG}$ &
$\mathcal O_{H\widetilde{G}}$ &
$H H^{\dagger} \widetilde{G}^{A}_{\mu\nu} G^{\mu\nu A}$ &
$c_{H\widetilde{G}}$&
\ggF &
Yes \\
 
$\mathcal O_{HW}$ &
$H H^{\dagger} W^{l}_{\mu\nu} W^{\mu\nu l}$ &
$c_{HW}$ &
$\mathcal O_{H\widetilde{W}}$ &
$H H^{\dagger} \widetilde{W}^{l}_{\mu\nu} W^{\mu\nu l}$ &
$c_{H\widetilde{W}}$ &
\STXSVBF, \STXSVH  &
Yes \\
 
$\mathcal O_{HB}$ &
$H H^{\dagger} B_{\mu\nu} B^{\mu\nu }$ &
$c_{HB}$ &
$\mathcal O_{H\widetilde{B}}$ &
$H H^{\dagger} \widetilde{B}_{\mu\nu} B^{\mu\nu }$ &
$c_{H\widetilde{B}}$ &
\STXSVBF, \STXSVH  &
Yes \\
 
$\mathcal O_{HWB}$ &
$H H^{\dagger} \tau^l W^{l}_{\mu\nu} B^{\mu\nu}$ &
$c_{HWB}$ &
$\mathcal O_{H\widetilde{W}B}$ &
$H H^{\dagger} \tau^l \widetilde{W}^{l}_{\mu\nu} B^{\mu\nu}$ &
$c_{H\widetilde{W}B}$ &
\STXSVBF, \STXSVH  &
Yes \\
 
\hline\hline
 
\end{tabular}
\end{center}
\label{tab:OpSMEFT}
\end{table}
The remaining ten operators (see Table~\ref{tab:OpSMEFT}) comprise five CP-even and five CP-odd ones. The CP-even operators describing interactions between the Higgs boson and gluons and the top-Yukawa interactions are associated with the Wilson coefficients $c_{HG}$ and $c_{uH}$ from Ref.~\cite{SMEFT}, respectively. Similarly, the CP-even Higgs boson interactions with vector bosons are related to $c_{HW}$, $c_{HB}$, and $c_{HWB}$ that impact the \STXSVBF and \STXSVH production and the Higgs boson decay into $Z$ bosons. The Wilson coefficients for the corresponding CP-odd operators are $c_{\widetilde{u}H}$, $c_{H\widetilde{G}}$, $c_{H\widetilde{W}}$, $c_{H\widetilde{B}}$ and $c_{H\widetilde{W}B}$.
 
The constraints on the Wilson coefficients can be derived by comparing the expected with the measured simplified template cross-sections. For that purpose, the corresponding expected signal production cross-sections, the branching ratio and the signal acceptances are parameterised in terms of the Wilson coefficients. The dependence of signal production cross-sections on the EFT parameters can be obtained from its separation into three components:
\begin{equation*}
\sigma \propto
\left | \mathcal M_{\text{SMEFT}}  \right|^2=
\left |  \mathcal M_{\text{SM}}  +
\sum_i \frac{C_i}{\Lambda^2} \mathcal M_{i} \right |^2 =
\left | \mathcal{M}_\textrm{SM} \right |^2 +
\sum_i  2Re \left( \mathcal{M}_\textrm{SM}^* \mathcal{M}_i \right ) \frac{C_i}{\Lambda^2}  +
\sum_{ij} 2Re \left( \mathcal{M}_i^* \mathcal{M}_j \right ) \frac{C_i C_j}{\Lambda^4},
\end{equation*}
where the first term on the right-hand side is the squared matrix element for the SM, the second term represents the interference between the SM and dimension-six EFT amplitudes and the third term comprises the pure BSM contribution from dimension-six EFT operators alone. Following this expression, the dependence of the Higgs boson cross-section $\sigma^p (\vec{c})$ in a given production bin $p$ on a set of Wilson coefficients $\vec{c}$ is parameterised relative to the SM prediction $\sigma^p_{\textrm{SM}}$ as
\begin{equation}
\frac{\sigma^{p} (\vec{c})}{\sigma^{p}_\text{SM}} = 1 + \sum_i A^{p}_i c_i + \sum_{ij} B^{p}_{ij} c_i c_j,
\label{eq:sigma_parametrisation}
\end{equation}\\
where the coefficients $A^p_i$ and $B^p_{ij}$ are independent of $\vec{c}$ and are determined from simulation.
A similar procedure is applied to obtain from simulation the EFT parameterisation of the branching ratio $\BR^{4\ell}$ for the $H\to ZZ^*\to 4\ell$ decay from the partial $(\Gamma^{4\ell})$ and total decay width $(\Gamma^{\text{tot}})$ parameterisations,
\begin{equation}
\mathcal \BR^{4\ell} (\vec{c}) =
\frac{\Gamma^{4\ell} (\vec{c})}{\Gamma^{\text{tot}} (\vec{c})} =
\mathcal \BR^{4\ell}_{\text{SM}}  \cdot
\frac{1 + \sum_i A^{4\ell}_i c_i + \sum_{ij} B^{4\ell}_{ij} c_i c_j}{1 + \sum_f \left ( \sum_i A^{f}_i c_i + \sum_{ij} B^{f}_{ij} c_i c_j \right )},
\label{eq:br_EFT}
\end{equation}
where the total decay width is the sum of all partial decay widths $\Gamma^f$ related to the decay mode $f$.
The procedure for the parameterisation of the cross-sections and the branching ratios is described in more detail in Ref.~\cite{Hays:2290628}. The criteria employed in the selection of four-lepton candidates introduce an additional dependence of the signal acceptance on the EFT parameters. This is taken into account in the interpretation, as discussed in Section~\ref{sec:eft}.
\FloatBarrier
\section{ATLAS detector}
\label{sec:detector}
 
\newcommand{\AtlasCoordFootnote}{
ATLAS uses a right-handed coordinate system with its origin at the nominal interaction point (IP)
in the centre of the detector and the $z$-axis along the beam pipe.
The $x$-axis points from the IP to the centre of the LHC ring,
and the $y$-axis points upwards.
Cylindrical coordinates $(r,\phi)$ are used in the transverse plane,
$\phi$ being the azimuthal angle around the $z$-axis.
The pseudorapidity is defined in terms of the polar angle $\theta$ as $\eta = -\ln \tan(\theta/2)$.
Angular distance is measured in units of $\Delta R \equiv \sqrt{(\Delta\eta)^{2} + (\Delta\phi)^{2}}$.}
 
The ATLAS detector~\cite{PERF-2007-01,ATLAS-TDR-2010-19,PIX-2018-001} at the LHC is a multipurpose particle detector
with a forward--backward symmetric cylindrical geometry\footnote{\AtlasCoordFootnote} and a nearly $4\pi$ coverage in solid angle. It consists of an inner tracking detector (ID) surrounded by a thin superconducting solenoid, which provides a \SI{2}{\tesla} axial magnetic field, electromagnetic (EM) and hadron calorimeters, and a muon spectrometer (MS).
The inner tracking detector covers the pseudorapidity range $|\eta| < 2.5$.
It consists of silicon pixel, silicon microstrip, and transition radiation tracking detectors.
A lead/liquid-argon (LAr) sampling calorimeter provides electromagnetic energy measurements in the pseudorapidity range $|\eta| < 3.2$ with high granularity.
A steel/scintillator-tile hadron calorimeter covers the central pseudorapidity range ($|\eta| < 1.7$).
The endcap and forward regions are instrumented up to $|\eta| = 4.9$ with LAr calorimeters for both the EM and hadronic energy measurements.
The calorimeters are surrounded by the MS and
three large air-core toroidal superconducting magnets with eight coils each.
The field integral of the toroid magnets ranges between \num{2.0} and \num{6.0}~Tm across most of the detector.
The MS includes a system of precision tracking chambers and fast detectors for triggering, covering the region $|\eta| < 2.7$.
Events are selected using a first-level trigger implemented in custom electronics, which reduces the event
rate to a maximum of \num{100}~kHz using a subset of detector information. Software algorithms with access
to the full detector information are then used in the high-level trigger to yield a recorded event rate of
about \num{1}~kHz~\cite{TRIG-2016-01}.
 
\FloatBarrier

\section{Data set and event simulation}
\label{sec:simulation}
 
The full ATLAS Run 2 data set, consisting of $pp$ collision data at $\sqrt{s}$ = 13~\TeV\ taken between 2015 and 2018, is used for this analysis. The total integrated luminosity after imposing data quality requirements~\cite{collaboration2019atlas} is \Lum.

The production of the SM Higgs boson via gluon--gluon fusion, via vector-boson fusion,  with an associated vector boson and with a top quark pair was modelled with the \POWHEGBOXV{v2} Monte Carlo (MC) event generator~\cite{Alioli:2010xd,Frixione:2007vw,Nason:2004rx}. For \ggF, the PDF4LHC next-to-next-to-leading-order (NNLO) set of parton distribution functions (PDF) was used, while for all other production modes, the PDF4LHC next-to-leading-order (NLO) set was used~\cite{Butterworth:2015oua}.
 
The simulation of \ggF Higgs boson production used the \POWHEG method for merging the NLO Higgs boson + jet cross-section with the parton shower and the multi-scale improved NLO (\progname{MINLO}) method~\cite{Hamilton:2012np,Campbell:2012am,Hamilton:2012rf,Bagnaschi_2012} to simultaneously achieve NLO accuracy for the inclusive Higgs boson production. In a second step, a reweighting procedure (NNLOPS)~\cite{Hamilton:2013fea,Hamilton:2015nsa}, exploiting the Higgs boson rapidity distribution, was applied using the \progname{HNNLO} program~\cite{Catani:2007vq,Grazzini:2008tf} to achieve NNLO accuracy in the strong coupling constant \alphas.
The transverse momentum spectrum of the Higgs boson obtained with this sample is compatible with the fixed-order calculation from \progname{HNNLO} and the resummed calculation at next-to-next-to-leading-logarithm accuracy matched to NNLO fixed-order with \progname{Hres2.3}~\cite{Bozzi:2005wk,deFlorian:2011xf}.

The matrix elements of the \VBF, \VHprod, and \ttH production mechanisms were calculated up to NLO in QCD\@.
For \VH production, the \progname{MINLO} method was used to merge 0-jet and 1-jet events~\cite{Nason:2009ai,Hamilton:2012rf,Hamilton:2012np,Luisoni:2013kna,Hartanto:2015uka,Cullen:2011ac}.
The $gg\rightarrow ZH$ contribution was modelled at leading order (LO) in QCD\@.
 
The production of a Higgs boson in association with a bottom quark pair (\bbH) was simulated at NLO with \MGMCatNLOV{v2.3.3}~\cite{Alwall:2014hca,Wiesemann:2014ioa}, using the CT10 NLO PDF ~\cite{Lai:2010vv}. The production in association with a single top quark ($tH$+$X$ where $X$ is either $jb$ or $W$, defined in the following as $tH$) \cite{Demartin:2015uha,Demartin:2016axk} was
simulated at NLO with \MGMCatNLOV{v2.6.0} using the  NNPDF3.0nlo PDF set~\cite{Ball:2014uwa}.
 
For all production mechanisms, the \PYTHIAV{8}~\cite{Sjostrand:2007gs} generator was used for the \htollllbrief{} decay with $\ell = (e, \mu)$ as well as for parton showering, hadronisation and the underlying event. The contribution of the $Z\to \tau\tau$ decays is shown to have a negligible impact on the final result. The event generator was interfaced to \progname{EvtGen}~v1.2.0~\cite{Lange:2001uf} for simulation of the bottom and charm hadron decays.  For the \ggF, \VBF and \VH processes, the AZNLO~\cite{STDM-2012-23} set of tuned parameters was used, while the A14~\cite{ATL-PHYS-PUB-2014-021} set was used for \ttH, \bbH and $tH$ processes. All signal samples were simulated for a Higgs boson mass $m_H=$~125~\GeV.
 
For additional cross-checks, the \ggF sample was also generated with \MGMCatNLO. This simulation is accurate at NLO QCD accuracy for zero, one and two additional partons merged with the \progname{FxFx} merging scheme~\cite{Alwall:2014hca,Frederix:2012ps}. The events were showered using the \PYTHIAV{8} generator with the A14 set of tuned parameters.
 
The Higgs boson production cross-sections and decay branching ratios, as well as their uncertainties, are taken from Refs.~\cite{Dittmaier:2011ti,Dittmaier:2012vm,Heinemeyer:2013tqa,YR4,Djouadi:1997yw,Djouadi:2006bz,Dulat:2015mca,Harland-Lang:2014zoa,Ball:2014uwa}.
The \ggF production is calculated with next-to-next-to-next-to-leading order (N$^{3}$LO) accuracy in QCD and has NLO electroweak (EW) corrections applied~\cite{Aglietti:2004nj,Bonetti:2018ukf,Actis:2008ts,Actis:2008ug,Pak:2009dg,Harlander:2009my,Harlander:2009bw,Harlander:2009mq,Dulat:2018rbf,Anastasiou:2015ema,Anastasiou:2016cez}. For \VBF production, full NLO QCD and EW calculations are used with approximate NNLO QCD corrections~\cite{Ciccolini:2007jr,Ciccolini:2007ec,Bolzoni:2010xr}.
The $qq$- and $qg$-initiated \VH production is calculated at NNLO in QCD and NLO EW corrections are applied~\cite{Brein:2012ne,Harlander:2018yio, Brein:2003wg,Brein:2011vx,Denner:2011id,Altenkamp:2012sx,Harlander:2014wda,Denner:2014cla,Ciccolini:2003jy}, while $gg$-initiated \VH production is calculated at NLO in QCD\@. The \ttH~\cite{Beenakker:2002nc,Dawson:2003zu, Yu:2014cka,Frixione:2015zaa}, \bbH~\cite{Dawson:2003kb,Dittmaier:2003ej,Harlander:2011aa} and \STXStH~\cite{Demartin:2015uha,Demartin:2016axk} processes are calculated to NLO accuracy in QCD\@. The total branching ratio is calculated in the SM for the \htollllbrief{} decay with $m_H=$~125~\GeV\ and $\ell$ = ($e$, $\mu$) using \progname{PROPHECY4F}~\cite{Bredenstein:2006rh,Bredenstein:2006nk}, which includes the complete NLO EW corrections, and the interference effects between identical final-state fermions. Due to the latter, the expected branching ratios of the 4$e$ and 4$\mu$ final states are about 10\% higher than the branching ratios to $2e2\mu$ and $2\mu2e$ final states.
Table~\ref{tab:MCSignal} summarises the predicted SM production cross-sections and branching ratios for the
\htollllbrief{} decay for $m_H=125$~\GeV.

\begin{table*}[htbp]
\centering
\caption{The predicted SM Higgs boson production cross-sections ($\sigma$) for \ggF,
\VBF and five associated production modes
in $pp$ collisions for $m_H=125$~\gev\ at    $\sqrt{\mathrm{s}}=13~\tev$~\cite{Dittmaier:2011ti,Dittmaier:2012vm,Heinemeyer:2013tqa,YR4,Djouadi:1997yw,Djouadi:2006bz,Dulat:2015mca,Harland-Lang:2014zoa,Ball:2014uwa,Aglietti:2004nj,Bonetti:2018ukf,Actis:2008ts,Actis:2008ug,Pak:2009dg,Harlander:2009my,Harlander:2009bw,Harlander:2009mq,Dulat:2018rbf,Anastasiou:2015ema,Anastasiou:2016cez,Ciccolini:2007jr,Ciccolini:2007ec,Bolzoni:2010xr,Brein:2012ne,Harlander:2018yio, Brein:2003wg,Brein:2011vx,Denner:2011id,Altenkamp:2012sx,Harlander:2014wda,Denner:2014cla,Ciccolini:2003jy,Beenakker:2002nc,Dawson:2003zu, Yu:2014cka,Frixione:2015zaa,Dawson:2003kb,Dittmaier:2003ej,Harlander:2011aa,Demartin:2015uha,Demartin:2016axk,Djouadi:1997yw,Boselli:2015aha, Bredenstein:2006rh,Bredenstein:2006ha,Bredenstein:2006nk}.
The quoted uncertainties correspond to the total theoretical systematic uncertainties calculated by adding in quadrature the uncertainties due to missing higher-order corrections and PDF$+\alphas$.  The decay branching ratios (\BR) with the associated
uncertainty for \htoZZ\ and \htollllbrief, with $\ell = e, \mu$, are also given.
\label{tab:MCSignal}}
\vspace{0.1cm}
{\renewcommand{\arraystretch}{1.1}
 
\begin{tabular}{ll|r}
\hline \hline
 
\multicolumn{2}{l|}{Production process}&  \multicolumn{1}{c}{$\sigma$~[pb]}\\
\hline

\ggF& $\left(gg\to H \right)$&
$48.6 \pm 2.4\phantom{00}$    \\
 
\VBF& $\left(qq'\to Hqq' \right)$&
$3.78  \pm 0.08\phantom{0}$    \\
 
\WH&  $\left( \WHprod \right)$  &
$1.373 \pm 0.028$ \\
 
\ZH&  $\left( \ZHprod \right)$ &
$0.88 \pm 0.04\phantom{0}$    \\
 
\ttH&  $\left(\ttHprod \right)$ &
$0.51 \pm 0.05\phantom{0}$ \\
 
\bbH&  $\left(\bbHprod \right)$ &
$0.49 \pm 0.12\phantom{0}$ \\
 
$tH$ &  $\left(\tHprod \right)$ &
$0.09 \pm 0.01\phantom{0}$ \\
 
\hline
 
\multicolumn{2}{l|}{Decay process}&  \multicolumn{1}{c}{\BR~[$\cdot$ 10$^{-4}$]}\\
\hline
\multicolumn{2}{l|}{\htoZZ}       &   $262  \pm 6\phantom{.000}$ \\
\multicolumn{2}{l|}{\htollllbrief}&  $\phantom{00}1.240 \pm 0.027$ \\

\hline\hline
\end{tabular}
}
\end{table*}
 
For the study of the tensor structure of Higgs boson couplings within an effective field theory approach, several samples with different values of EFT parameters were simulated at LO in QCD separately for the $\ggF+\bbH$, $\VBF + V(\to qq)H$, $qq\to Z(\to \ell\ell)H$, $qq\to W(\to \ell\nu)H$, \ttH, $tHW$ and $tHjb$ production modes using \MGMCatNLO and the NNPDF23lo PDF.
The BSM signal is defined by the flavour symmetric \progname{SMEFTsim\_A\_U35\_MwScheme\_UFO\_v2.1} model~\cite{SMEFT,Aebischer_2018}, which incorporates the SMEFT dimension-six operators in the standard Universal FeynRules Output  format created using the FeynRules framework~\cite{Feynrules1,Feynrules2}. The light quarks ($u$, $d$, $s$ and $c$) and leptons are assumed to be massless in the model. The generated events were showered with \PYTHIAV{8}, using the CKKW-L matching scheme to match matrix element and parton shower computations with different jet multiplicities~\cite{Sjostrand:2007gs}. The A14 set of tuned parameters was used. All processes were simulated in the four-flavour scheme, apart from the $tHW$~production, for which the five-flavour scheme  was used~\cite{Alwall:2014hca}.

The \zzstar\ continuum background from quark--antiquark annihilation was modelled using \SHERPAV{v2.2.2}~\cite{Gleisberg:2008ta,Gleisberg:2008fv,Cascioli:2011va,Bothmann:2019yzt},
which provides a matrix element calculation accurate to NLO in $\alphas$  for 0-jet and 1-jet final states and LO accuracy for 2-jets and 3-jets final states.
The merging with the \progname{Sherpa} parton shower~\cite{Schumann:2007mg} was performed using the ME+PS@NLO prescription~\cite{Hoeche:2012yf}.
The NLO EW corrections were applied as a function of the invariant mass $m_{\zzstar}$ of the $ZZ^*$ system~\cite{Biedermann:2016yvs,Biedermann:2016lvg}.
 
The gluon-induced \zzstar\ production was modelled by \SHERPAV{v2.2.2} \cite{Gleisberg:2008ta,Gleisberg:2008fv,Cascioli:2011va} at LO in QCD  for 0-jet and 1-jet final states.
The higher-order QCD effects for the $gg\rightarrow ZZ^{*}$ continuum production cross-section were calculated
for massless quark loops~\cite{Caola:2015psa,Caola:2015rqy,Campbell:2016ivq} in the heavy top-quark
approximation~\cite{Melnikov:2015laa}, including the interference with $gg\rightarrow H^{*} \rightarrow ZZ$ processes~\cite{Bonvini:2013jha,Li:2015jva}. The $gg \to ZZ$ simulation was scaled by a $K$-factor of 1.7~$\pm$~1.0, which is defined as the ratio of the higher-order to the leading-order cross-section predictions.
 
Production of \zzstar\ via vector-boson scattering was simulated with the \SHERPAV{v2.2.2}~\cite{Bothmann:2019yzt} generator. The LO-accurate matrix elements were matched to a parton shower using the MEPS@LO prescription.
 
For all \zzstar\ processes modelled using \SHERPA, the NNPDF3.0nnlo PDF set~\cite{Ball:2014uwa} was used,
along with a dedicated set of tuned parton-shower parameters.
 
For additional checks, the $q\bar{q}$-initiated \zzstar\ continuum background was also modelled using \POWHEGBOXV{v2} and \MGMCatNLO, using the CT10~\cite{Lai:2010vv} and the PDF4LHC NLO PDF set, respectively. For the former, the matrix element was generated at NLO accuracy in QCD and effects of singly resonant
amplitudes and interference effects due to $Z/\gamma^*$ were included. For the latter, the simulations are accurate to NLO in QCD for zero and one additional parton merged with the \progname{FxFx} merging scheme. For both, the \PYTHIAV{8} generator was used for the modelling of parton showering, hadronisation, and the underlying event. The AZNLO and A14 sets of tuned parameters were used for the simulations performed with \POWHEGBOXV{v2} and \MGMCatNLO generators, respectively.
 
The \emph{WZ} background~\cite{Nason:2013ydw} was modelled at NLO accuracy in QCD using \POWHEGBOXV{v2} with the CT10 PDF set and was interfaced to \PYTHIAV{8}, using the AZNLO set of tuned parameters for modelling of parton showering, hadronisation, and the underlying event
and to \progname{EvtGen} v1.2.0 for the simulation of bottom and charm hadron decays. The
triboson backgrounds \emph{ZZZ}, \emph{WZZ}, and \emph{WWZ}  with four or more prompt leptons (\emph{VVV}) were modelled at NLO accuracy for the inclusive process and at LO for up to two additional parton emissions using \SHERPAV{v2.2.2}.
 
The simulation of \ttZ\ events with both top quarks decaying semileptonically and the $Z$ boson decaying leptonically was performed with \MGMCatNLO using the
NNPDF3.0nlo~\cite{Ball:2014uwa} PDF set
interfaced to \PYTHIAV{8} using the A14 set of tuned parameters,
and
the total cross-section was normalised to a prediction computed at NLO in the QCD and EW couplings~\cite{Frixione:2015zaa}. For modelling comparisons, \SHERPAV{v2.2.1} was used to simulate \ttZ\ events at LO\@. The $tWZ$, \ttWW, \ttWZ, \ttZg, \ttZZ, \ttt, \tttt\ and $tZ$ background processes were simulated with \MGMCatNLO interfaced to \PYTHIAV{8}, using the A14 set of tuned parameters. These processes are collectively referred to as the \tXX\ process.
 
The modelling of events containing $Z$ bosons with associated jets ($Z+\,$jets)
was performed using the \SHERPAV{v2.2.1} generator. Matrix elements were calculated for up to two partons at NLO and four partons at LO using
\progname{Comix} \cite{Gleisberg:2008fv} and \progname{OpenLoops} \cite{Cascioli:2011va},
and merged with the \SHERPA parton shower~\cite{Schumann:2007mg} using
the \progname{ME+PS@NLO} prescription~\cite{Hoeche:2012yf}. The NNPDF3.0nnlo PDF set is used in conjunction
with dedicated set of tuned parton-shower parameters.
 
The \ttprod\ background was modelled using \POWHEGBOXV{v2} with the NNPDF3.0nlo PDF set. This simulation was interfaced to \PYTHIAV{8}, using the A14 set of tuned parameters,
for parton showering, hadronisation, and the underlying event, and to \progname{EvtGen} v1.2.0 for heavy-flavour hadron decays.
Simulated $Z+\,$jets and  \ttprod\  background samples were normalised to the data-driven estimates described in Section~\ref{sec:background}.

Generated events were processed through the ATLAS detector simulation~\cite{SOFT-2010-01} within the $\GEANT4$ framework~\cite{GEANT4} and reconstructed in the same way as collision data.
Additional $pp$ interactions in the same and nearby
bunch crossings were included in the simulation.  Pile-up events were generated using \PYTHIAV{8} with the A2 set of tuned parameters~\cite{ATL-PHYS-PUB-2012-003} and the MSTW2008LO PDF set~\cite{Martin:2009iq}. The simulation samples were weighted to
reproduce the distribution of the number of interactions per bunch crossing observed in data.
\FloatBarrier

\section{Event selection}
\label{sec:selection}
\label{sub:selection}
 
\subsection{Event reconstruction}
\label{subsec:selection_reco}

The selection and categorisation of the Higgs boson candidate events rely on the
reconstruction and identification of electrons, muons, and jets, closely following the analyses reported in Refs.~\cite{HIGG-2016-25, HIGG-2016-22}.
 
Proton--proton collision vertices are constructed from reconstructed trajectories of charged particles in the ID with transverse momentum $\pt>$~500~\MeV. Events are required to have at least one collision vertex with at least two associated tracks. The vertex with the highest $\sum{\pt^2}$ of reconstructed tracks is selected as the primary vertex of the hard interaction. The data are subjected to quality requirements to reject events in which detector components were not operating correctly.
 
Electron candidates are reconstructed from energy clusters in the electromagnetic calorimeter that are matched to ID tracks~\cite{EGAM-2018-01}. A Gaussian-sum filter algorithm~\cite{ATLAS-CONF-2012-047} is used to compensate for radiative energy losses in the ID for the track reconstruction, while a dynamical, topological cell-based approach for cluster building is used to improve the  energy resolution relative to the previous measurements in Refs.~\cite{HIGG-2016-25, HIGG-2016-22}, in particular for the case of bremsstrahlung photons. Electron identification is based on a likelihood discriminant combining the measured track properties, transition radiation response, electromagnetic shower shapes and the quality of the track--cluster matching.
The `loose' likelihood criteria, applied in combination with track hit requirements, provide an electron reconstruction and identification efficiency of
at least 90\% for isolated electrons with $\pt > 30$~\GeV\ and 85\%--90\% below~\cite{EGAM-2018-01}. Electrons are required to have $\et>$~7~\GeV\ and pseudorapidity $|\eta|<$~2.47, with their energy calibrated as described in Ref.~\cite{EGAM-2018-01}.
 
Muon candidate reconstruction~\cite{PERF-2015-10} within the range $|\eta|<2.5$ is primarily performed by a global fit to fully reconstructed tracks in the ID and the MS, with a `loose'~\cite{PERF-2015-10} identification criterion applied. This criterion has an efficiency of
at least 98\% for isolated muons with $\pt = 5$~\GeV\ and rises to 99.5\% at higher \pt.
At the centre of the detector ($|\eta|<$~0.1),
which has a reduced MS geometrical coverage, muons are also identified by matching a fully reconstructed ID track to either an MS track segment or a calorimeter energy deposit consistent with a minimum-ionising particle (calorimeter-tagged muons). For these two cases, the muon momentum is measured from the ID track alone. In the forward MS region (2.5~$<|\eta|<~$~2.7), outside the full ID coverage,  MS tracks with hits in
the three MS layers are accepted and combined with forward ID tracklets, if they exist (stand-alone muons). Calorimeter-tagged muons are required to have $\pt>$~15~\GeV. For all other muon candidates, the transverse momentum is required to be greater than 5~\GeV. The muon momentum is calibrated using the procedure described in Ref.~\cite{PERF-2015-10}.
Muons with transverse impact parameter greater than 1~mm are rejected.\footnote{The transverse impact parameter $d_0$ of a charged-particle track is defined in the transverse plane as the distance from the primary vertex to the track's point of closest approach. The longitudinal impact parameter $z_0$ is the distance in the $z$ direction between this track point and the primary vertex.} Additionally, muons and electrons are required to have a longitudinal impact parameter ($|z_0\sin\theta|$) less than 0.5~mm.
 
Jets are reconstructed using a particle flow algorithm~\cite{PERF-2015-09} from noise-suppressed positive-energy topological clusters~\cite{PERF-2014-07} in the calorimeter using the anti-$k_t$ algorithm~\cite{Cacciari:2008gp,Cacciari:2011ma} with a radius parameter $R$~=~0.4. Energy deposited in the calorimeter by charged particles is subtracted and replaced by the momenta of tracks that are matched to those topological clusters.
Compared to only using topological clusters, jets reconstructed with the particle flow algorithm with $\pt>30~\GeV$ have approximately 10\% better  transverse momentum resolution. The two different algorithms have similar resolution for \pt above 100~\GeV.
The jet four-momentum is corrected for the calorimeter's non-compensating response, signal losses due to noise threshold effects, energy lost in non-instrumented regions, and contributions from pile-up~\cite{PERF-2016-04,PERF-2015-09,ATLAS-CONF-2016-035}. Jets are required to have $\pt>$~30~\GeV\ and $|\eta|<$~4.5.
Jets from pile-up with $|\eta|<$~2.5 are suppressed using a jet-vertex-tagger multivariate discriminant ~\cite{ATLAS-CONF-2014-018,PERF-2014-03}.
Jets with $|\eta|<$~2.5 containing $b$-hadrons are identified using the MV2c10 $b$-tagging algorithm~\cite{ATL-PHYS-PUB-2017-013, PERF-2016-05}, and its  60\%, 70\%, 77\% and 85\% efficiency working points are combined into a pseudo-continuous $b$-tagging weight~\cite{FTAG-2018-01} that is assigned to each jet.

Ambiguities are resolved
if electron, muon, or 
jet candidates overlap in geometry or share the same detector information.
If the two calorimeter energy clusters from the two electron candidates overlap, the electron with the higher \et is retained. If a reconstructed electron and muon share the same ID track, the muon is rejected if it is calorimeter-tagged; otherwise the electron is rejected. Reconstructed jets geometrically overlapping in a cone of radial size $\Delta R$ = 0.1 (0.2) with a muon (an electron) are also removed.
 
The missing transverse momentum vector, \METVector, is defined as the negative vector sum of the transverse momenta of all the identified and calibrated leptons, photons and jets and the remaining unclustered energy, where the latter is estimated from low-$\pt$ tracks associated with the primary vertex but not assigned to any lepton, photon, hadronically decaying $\tau$-lepton or jet candidate~\cite{PERF-2016-07, ATLAS-CONF-2018-023}. The missing transverse momentum (\MET) is defined as the magnitude of \METVector.
 
\subsection{Selection of the Higgs boson candidates}
\label{subsec:selection_sel}
 
A summary of the event selection criteria is given in Table\,\ref{tab:higgscuts}.
\begin{table}[!htbp]
\begin{center}
\caption{Summary of the criteria applied to the selected Higgs boson candidate in each event. The mass threshold $m_{\textrm{min}}$ is defined in Section~\ref{subsec:selection_reco}. \label{tab:higgscuts}}
\vspace{0.2cm}
\resizebox{\textwidth}{!}{
\renewcommand{\arraystretch}{1.1}
\begin{tabular}{lccc}
\hline
\hline\\[-10pt]
\multicolumn{4}{c}{\textsc{Trigger}}\\[3pt]
\hline
 
\multicolumn{4}{l}{Combination of single-lepton, dilepton and trilepton triggers} \\
 
\hline\\[-10pt]
\multicolumn{4}{c}{\textsc{Leptons and Jets}}\\[3pt]
\hline
\textsc{Electrons}&
\multicolumn{3}{l}{$\et\ > 7\ \GeV$ and $|\eta| < 2.47$} \\
\multicolumn{1}{l}{\textsc{Muons}} &
\multicolumn{3}{l}{$\pt\ > 5\ \GeV$ and $|\eta| < 2.7$, calorimeter-tagged: $\pt > 15$~\GeV} \\
\multicolumn{1}{l}{\textsc{Jets}}&
\multicolumn{3}{l}{$\pt\ > 30\ \GeV$ and $|\eta| < 4.5$} \\
 
\hline\\[-10pt]
\multicolumn{4}{c}{\textsc{Quadruplets}}\\[3pt]
\hline
 
\multicolumn{4}{l}{All combinations of two same-flavour and opposite-charge lepton pairs} \\
\multicolumn{4}{l}{- Leading lepton pair: lepton pair with invariant mass $m_{12}$ closest to the $Z$ boson mass $m_Z$}\\
\multicolumn{4}{l}{- Subleading lepton pair: lepton pair with invariant mass $m_{34}$ second closest to the $Z$ boson mass $m_Z$}\\
\multicolumn{4}{l}{Classification according to the decay final state: 4$\mu$, 2$e$2$\mu$, 2$\mu$2$e$, 4$e$}\\
 
\hline\\[-10pt]
\multicolumn{4}{c}{\textsc{Requirements on each quadruplet}}\\[3pt]
\hline
 
\multicolumn{1}{l}{\textsc{Lepton}}&
\multicolumn{3}{l}{- Three highest-$\pt$ leptons must have $\pt$ greater than $20, 15\text{ and } 10$~\GeV} \\			
\multicolumn{1}{l}{\textsc{reconstruction}}&
\multicolumn{3}{l}{- At most one calorimeter-tagged or stand-alone muon} \\
 
\hline
\multicolumn{1}{l}{\textsc{Lepton pairs}}&
\multicolumn{3}{l}{- Leading lepton pair: $50 < m_{12} < 106$~\GeV} \\
\multicolumn{1}{l}{\textsc{}}&
\multicolumn{3}{l}{- Subleading lepton pair: $m_{\textrm{min}}< m_{34} < 115$~\GeV} \\
&
\multicolumn{3}{l}{- Alternative same-flavour opposite-charge lepton pair:  $m_{\ell\ell} > 5$~\GeV} \\
&
\multicolumn{3}{l}{- $\Delta R(\ell,\ell')>0.10$ for all lepton pairs} \\
 
\hline
\multicolumn{1}{l}{\textsc{Lepton isolation}}&
\multicolumn{3}{l}{- The amount of isolation $\et$ after summing the track-based and 40\% of the } \\
\multicolumn{1}{l}{\textsc{}}&
\multicolumn{3}{l}{~ calorimeter-based contribution must be smaller than 16\% of the lepton $\pt$} \\
 
\hline
\textsc{Impact parameter}&
\multicolumn{3}{l}{- Electrons: $|d_0|/\sigma(d_0)<5$} \\
\textsc{significance}   &
\multicolumn{3}{l}{- Muons: $|d_0|/\sigma(d_0)<3$} \\

\hline
\textsc{Common vertex}         &
\multicolumn{3}{l}{- $\chi^{2}$-requirement on the fit of the four lepton tracks to their common vertex}\\

\hline\\[-10pt]
\multicolumn{4}{c}{\textsc{Selection of the best quadruplet}}\\[3pt]
\hline
\multicolumn{4}{l}{- Select  quadruplet with $m_{12}$ closest to $m_Z$ from one decay final state } \\
\multicolumn{4}{l}{~ in decreasing order of priority: 4$\mu$, 2$e$2$\mu$, 2$\mu$2$e$ and 4$e$} \\
 
\multicolumn{4}{l}{- If at least one additional (fifth) lepton with $\pt>12$~\GeV\ meets the isolation, impact parameter} \\
\multicolumn{4}{l}{~ and angular separation criteria, select the quadruplet with the highest matrix-element value} \\
 
\hline\\[-10pt]
\multicolumn{4}{c}{\textsc{Higgs boson mass window}}\\[3pt]
\hline
\multicolumn{4}{l}{- Correction of the four-lepton invariant mass due to the FSR photons in $Z$ boson decays} \\
\multicolumn{4}{l}{- Four-lepton invariant mass window in the signal region: $115 <m_{4\ell} < 130$~\GeV\ } \\
\multicolumn{4}{l}{- Four-lepton invariant mass window in the sideband region: } \\
\multicolumn{4}{l}{~ $105 <m_{4\ell} < 115$~\GeV\ or $130 <m_{4\ell} < 160\ (350)$~\GeV} \\

\hline\hline
\end{tabular}
}
\end{center}
\end{table}
Events were triggered by a combination of single-lepton, dilepton and trilepton triggers with different transverse momentum thresholds. Single-lepton triggers with the lowest thresholds had strict identification and isolation requirements. Both the high-threshold single-lepton triggers and the multilepton triggers had looser selection criteria.
Due to an increasing peak luminosity, these thresholds increased slightly during the data-taking periods~\cite{ATL-DAQ-PUB-2016-001,ATL-DAQ-PUB-2017-001}. For single-muon triggers, the \pt threshold ranged from between 20 and 26~\GeV, while for single-electron triggers, the \pt threshold ranged from 24 to 26~\GeV. The global trigger efficiency for signal events passing the final selection is about 98\%.
 
In the analysis, at least two same-flavour and opposite-charge lepton pairs (hereafter referred to as lepton pairs) are required in the final state,
resulting in one or more possible lepton quadruplets in each event.
The three highest-$\pt$ leptons in each quadruplet are required to have transverse momenta above 20~\GeV, 15~\GeV\ and 10~\GeV, respectively. To minimise the background contribution from non-prompt muons,
at most one calorimeter-tagged or stand-alone muon is allowed per quadruplet.
 
The lepton pair with the invariant mass $m_{12}$ ($m_{34}$) closest (second closest) to the $Z$ boson mass~\cite{PhysRevD.98.030001} in each quadruplet is referred to as the leading (subleading) lepton pair.
Based on the lepton flavour, each quadruplet is classified into one of the following decay final states: 4$\mu$, 2$e$2$\mu$, 2$\mu$2$e$ and 4$e$, with the first two leptons always representing  the leading lepton pair.
In each of these final states, the quadruplet with $m_{12}$ closest to the $Z$ boson mass has priority to be considered for the selection of the final Higgs boson candidate.
In case additional prompt leptons are present in the event, the priority may change due to the matrix-element based pairing as described later on.
All quadruplets are therefore required to pass the following selection criteria.
 
To ensure that the leading lepton pair from the signal originates from a $Z$ boson decay, the leading lepton pair is required to satisfy 50~\GeV~$<m_{12}<$~106~\GeV. The subleading lepton pair is required to have a mass $m_{\mathrm{min}}<m_{34}<$~115~\GeV, where $m_{\mathrm{min}}$ is 12~\GeV\ for the four-lepton invariant mass $m_{4\ell}$ below 140~\GeV, rising linearly to 50~\GeV\ at $m_{4\ell}=$~190~\GeV\ and then remaining at 50~\GeV\ for all higher $m_{4\ell}$ values. This criterion suppresses the contributions from processes in which an on-shell $Z$ boson is produced in association with a leptonically decaying meson or virtual photon. In the 4$e$ and 4$\mu$ final states, the two alternative opposite-charge lepton pairings within a quadruplet are required to have a dilepton mass above 5~\GeV\ to suppress the $J/\psi$ background.
All leptons in the quadruplet are required to have an angular separation of $\Delta R>$~0.1.

Each electron (muon) track is required to have a transverse impact parameter significance $|d_0/\sigma(d_0)|<5$~(3), to suppress the background from heavy-flavour hadrons. Reducible background from the $Z$+jets and \ttprod\ processes is further suppressed by imposing track-based and calorimeter-based isolation criteria on each lepton~\cite{ATL-PHYS-PUB-2017-022,PERF-2015-10}. A scalar $\pt$ sum (track isolation) is made from the tracks with $\pt > 500$~\MeV\ which either originate from the primary vertex or have $|z_0\sin\theta | < 3$~mm if not associated with any vertex and lie within a cone of $\Delta R =$~0.3 around the muon or electron. Above a lepton $\pt$ of 33~\GeV, this cone size falls linearly with $\pt$ to a minimum cone size of 0.2 at 50~\GeV. Similarly, the scalar  \et sum (calorimeter isolation) is calculated from the positive-energy topological clusters that are not associated with a lepton track in a cone of $\Delta R =$~0.2 around the muon or electron. The sum of the track isolation and 40\% of the calorimeter isolation is required to be less than
16\% of the lepton $\pt$. The calorimeter isolation is corrected for electron shower leakage, pile-up and underlying-event contributions. Both isolations are corrected for track and topological cluster contributions from the remaining three leptons. The pile-up dependence of this isolation selection is improved compared with that of the previous measurements~\cite{HIGG-2016-25,HIGG-2016-22,ATLAS-CONF-2018-018} by optimising the criteria used for exclusion of tracks associated with a vertex other than the primary vertex and by the removal of topological clusters associated with tracks. The signal efficiency of the isolation criteria is greater than 80\%, improving the efficiency by about 5\% compared with the previous analysis for the same  background rejection.

The four quadruplet leptons are required to originate from a common vertex point. A requirement corresponding to a signal efficiency of better than 99.5\% is imposed on the $\chi^2$ value from the fit of the four lepton tracks to their common vertex.

If there is more than one decay final state per event with the priority quadruplet ($m_{12}$ closest to $m_Z$) satisfying the selection criteria, the quadruplet from the final state with highest selection efficiency, i.e.~ordered 4$\mu$, 2$e$2$\mu$, 2$\mu$2$e$ and 4$e$, is chosen as the Higgs boson candidate.
 
\begin{figure}[!htbp]
\centering
\subfloat[]{
\includegraphics[width=0.45\linewidth]{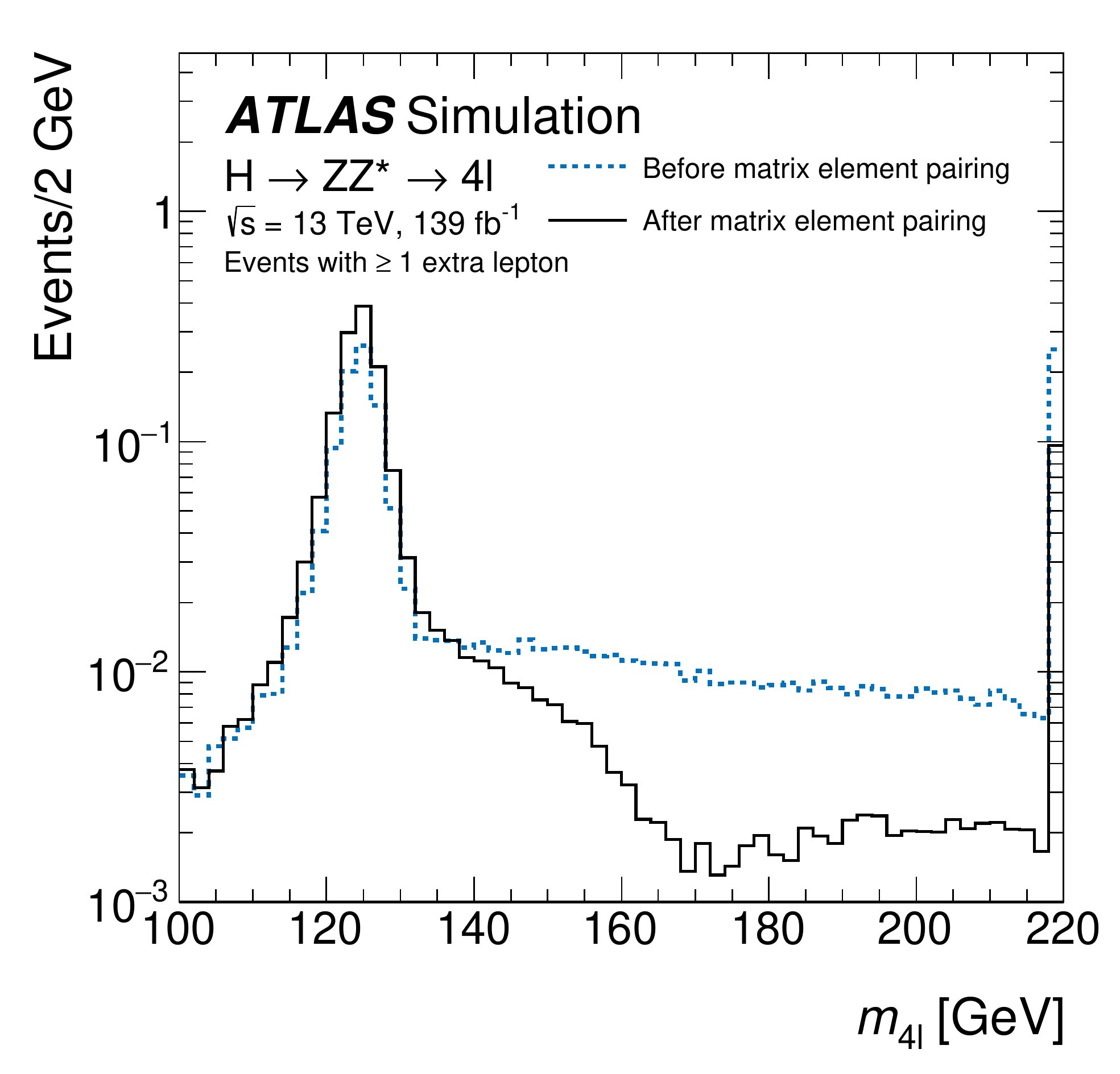}\label{fig:MEPairEffect}}
\subfloat[]{
\includegraphics[width=0.45\linewidth]{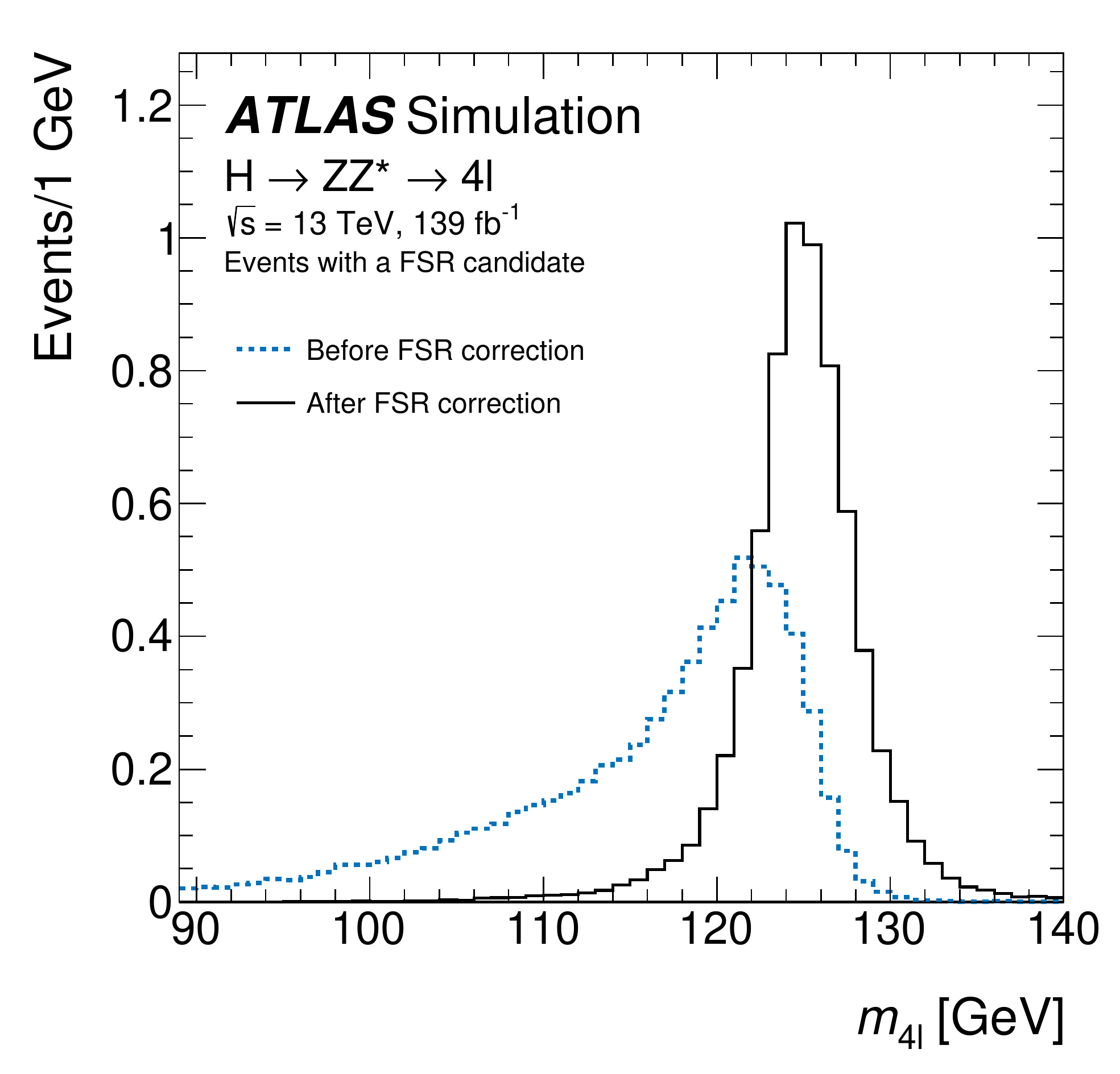}\label{fig:FSRCorrEffect}}
\caption{Impact on the expected invariant mass distribution of the selected Higgs boson candidates due to (a) matrix-element-based pairing for candidates with at least one extra lepton and (b) accounting for final-state radiation for candidates with an FSR candidate. For (a), the overflow events are included in the last bin.
}
\label{fig:FSR_ME_impact}
\end{figure}

In the case of \VH or \ttH production, there may be additional prompt leptons present in the event, together with the selected quadruplet.  Therefore, there is a possibility that one or more of the leptons selected in the quadruplet do not originate from a Higgs boson decay, but rather from the $V$ boson leptonic decay or the top quark semileptonic decay. To improve the lepton pairing in such cases, a matrix-element-based pairing method assuming the SM tensor structure is used for all events containing at least one additional lepton with $\pt>$12~\GeV\ and satisfying the same identification, isolation and angular separation criteria as the four quadruplet leptons~\cite{HIGG-2016-25, HIGG-2016-22}. For all possible quadruplet combinations that satisfy the selection, a matrix element for the Higgs boson decay is computed at LO using the \MGMCatNLO~\cite{Alwall:2014hca} generator, with the reconstructed lepton momentum vectors as inputs to the calculation. The quadruplet with the largest matrix-element value is selected as the Higgs boson candidate. This method leads to a 50\% improvement in correctly identifying the leptons in the quadruplet as those originating from a Higgs boson decay if an extra lepton is identified. The impact of the matrix element on the expected invariant mass distribution is shown in \figref{fig:MEPairEffect}.

To improve the four-lepton invariant mass reconstruction, the reconstructed final-state radiation (FSR) photons in $Z$ boson decays are accounted for using the same strategy as the previous publications~\cite{HIGG-2016-25, HIGG-2016-22}. Collinear FSR candidates are defined as candidates with $\Delta R <0.15$ to the nearest lepton in the quadruplet. Collinear FSR candidates are considered only for muons from the leading lepton pair, while non-collinear FSR candidates are considered for both muons and electrons from leading and subleading $Z$ bosons.
 
Collinear FSR candidates are selected from reconstructed photon candidates and from electron candidates that share an ID track with the muon. Further criteria are applied to each candidate, based on the following discriminants: the fraction, $f_1$, of cluster energy in the front segment of the EM calorimeter divided by the total cluster energy to reduce backgrounds from muon ionisation; the angular distance, $\Delta R_{\textrm{cluster},\mu}$, between the candidate EM cluster and the muon; and the candidate $\pt$, which must be at least 1~\GeV. For all selected electron candidates and for photon candidates with $\pt < 3.5~\GeV$, a requirement of $f_1>0.2$ and $\Delta R_{\textrm{cluster},\mu}<0.08$ is imposed. The collinear photon candidates with $\pt > 3.5~\GeV$ are selected if  $f_1>0.1$ and $\Delta R_{\textrm{cluster},\mu}<0.15$. Non-collinear FSR candidates are selected only from reconstructed isolated photons meeting the `tight'  criteria~\cite{EGAM-2018-01,PERF-2017-02} and satisfying $\pt > 10~\GeV$ and $\Delta R_{\textrm{cluster},\ell}>0.15$.
 
Only one FSR candidate is included in the quadruplet, with preference given to collinear FSR and to the candidate with the highest $\pt$. An FSR candidate is added to the lepton pair if the invariant mass of the lepton pair is between 66~\GeV\ and 89~\GeV\ and if the invariant mass of the lepton pair and the photon is below 100~\GeV. Approximately 3\% of reconstructed Higgs boson candidates have an FSR candidate and its impact on the expected invariant mass distribution is shown in \figref{fig:FSRCorrEffect}.

The Higgs boson candidates within a mass window of 115~\GeV~$<m_{4\ell}<$~130~\GeV\ are selected as the signal region.  Events failing this requirement but that are within a mass window of 105~\GeV~$<m_{4\ell}<$~115~\GeV\ or 130~\GeV~$<m_{4\ell}<$~160~(350)~\GeV\ are assigned to the sideband regions used to estimate the leading backgrounds as described in Section~\ref{sec:background}.

The selection efficiencies of the simulated signal in the fiducial region $|y_H|<2.5$, where $y_H$ is the Higgs boson rapidity, are about 33\%, 25\%, 19\% and 16\%, in the 4$\mu$, 2$e$2$\mu$, 2$\mu$2$e$ and 4$e$ final states, respectively.
 
\FloatBarrier

\section{Event categorisation and production mode discrimination}
\label{sec:categorization}
\label{sec:cat}
In order to be sensitive to different production bins in the framework of simplified template cross-sections, the selected Higgs boson candidates in the mass window $115~\GeV < m_{4\ell} < 130~\GeV$ are classified into several dedicated reconstructed event categories. In addition, the events in the mass sidebands are also categorised for purposes of  background estimation described in Section~\ref{sec:background}.
In general, more than one production mode contributes to each reconstructed event category, as well as various background processes. For this reason, multivariate discriminants are  introduced in most of the mutually exclusive reconstructed event categories to distinguish between these contributions.

\subsection{Event categorisation}
\label{subsec:cat_cat}
 
For signal events, the classification is performed in the order shown in the middle-right panel of \figref{fig:stxs_bins} (from bottom to top) and as described below. First, those events classified as enriched in the \ttH process are split according to the decay mode of the two $W$ bosons from the top quark decays. For semileptonic and dileptonic decays (\CatttHlep), at least one additional lepton with $\pt>$~12~\GeV\footnote{
The additional lepton is a lepton candidate as defined in Section~\ref{subsec:selection_reco}. It is also required to satisfy the same isolation, impact parameter and angular separation requirements as the leptons in the quadruplet.} together with
at least two \textit{b}-tagged jets (with 85\% $b$-tagging efficiency), or
at least five jets among which at least one \textit{b}-tagged jet (with 85\% $b$-tagging efficiency) or
at least two  jets among which at least one \textit{b}-tagged jet (with 60\% $b$-tagging efficiency) is required.
For the fully hadronic decay (\CatttHhad), there must be either
at least five jets among which at least two \textit{b}-tagged jets (with 85\% $b$-tagging efficiency) or
at least four jets among which at least one \textit{b}-tagged jet (with 60\% $b$-tagging efficiency).
Events with additional leptons but not satisfying the jet requirements define the next category enriched in \VH production events with leptonic vector-boson decay (\CatVHLep).
 
The remaining events are classified according to their reconstructed jet multiplicity into events with no jets, exactly one jet or at least two jets. Events with at least two reconstructed jets are divided into two categories: one is a `BSM-like' category (\CatTwoJBSM) and the other (\CatTwoJ) contains the bulk of events with significant contributions from the \VBF and \VH production modes in addition to \ggF. The \CatTwoJBSM category requires the invariant mass $m_{jj}$ of the two leading jets to be larger than 120~\GeV\ and the  four-lepton transverse momentum, \ptH, to be larger than 200~\GeV; the remaining events are placed in the \CatTwoJ category.
 
Events with zero or one jet in the final state are expected to be mostly from the \ggF process. Following the particle-level definition of production bins in Section~\ref{subsec:intro_stxs}, the 1-jet~category is further split into four categories with \ptH\ smaller than 60~\GeV\ (\CatOneJL), between 60 and 120~\GeV\ (\CatOneJM),  between 120 and 200~\GeV\ (\CatOneJH), and larger than 200~\GeV\ (\CatOneJBSM).
 
The largest number of \ggF events and the highest \ggF purity are expected in the zero-jet category. The zero-jet category is split into three categories with \ptH\ smaller than 10~\GeV~(\CatZeroJL), between 10 and 100~\GeV~(\CatZeroJM) and above 100~\GeV~(\CatZeroJH). The first two categories follow the production bin splitting, and the last category improves the discrimination between \VH ($V \to \ell \nu / \nu\nu$) and \ggF.
 
As illustrated in \figref{fig:stxs_bins}, there is a dedicated reconstructed event category for each production bin except for  \STXSggToHTwoJ, \STXSqqtoHqqVHLike and \STXSqqtoHqqRest. These production bins are largely measured from the 2-jet reconstruction category, and to a lesser extent from the 1-jet categories, using multivariate discriminants (see Section~\ref{subsec:cat_nn}). The \STXSggToHHigh production bin is measured simultaneously in all reconstructed event categories with high transverse momentum of the four-lepton system, independent of the reconstructed jet multiplicity.

The rightmost panel of \figref{fig:stxs_bins} shows the background event classification. For estimating the \tXX~process from the mass sideband, a \tXX-enriched sideband category (\CatSBtXX) is defined, which includes events with at least two jets including at least one tagged as a $b$-jet with 60\% efficiency and \MET $>$ 100~\GeV~in the $m_{4\ell}$ mass range 105--115~\GeV~or 130--350~\GeV. This region is dominated by \ttZ\ (87\%) and has small contributions from \ttprod, \tttt,  $tWZ$, \ttW, \ttWW, \ttWZ, \ttZg, \ttZZ and $tZ$. The \tXX~process is expected to give the largest contribution in `\ttH-like' categories. The large mass range for this category, larger than for the non-resonant $ZZ$ as discussed next, allows better statistical precision for the estimate of this background.
 
For the estimation of non-resonant \zzstar\ production, events not meeting the criteria for the \CatSBtXX category and in the $m_{4\ell}$ mass range 105--115~\GeV\ or 130--160~\GeV\ are split according to the number of reconstructed jets: exactly zero jets (\CatSBZeroJ), exactly one jet (\CatSBOneJ) or at least two jets (\CatSBTwoJ). This mass range limits the contribution from the single-resonance process, $Z \to 4\ell$, and from the on-shell $ZZ$~process. Similarly, events in the same mass range with an extra reconstructed lepton separately form the \CatSBVHL category, which is enriched with signal events containing leptons from the associated $V$ leptonic decay or the top quark semileptonic decay. This category is mainly designed to improve the expected sensitivity for \STXSVHLep by about 5\%, having a $VH$ purity of about 19\%.

\begin{figure}
\begin{center}\includegraphics[width=0.99\columnwidth]{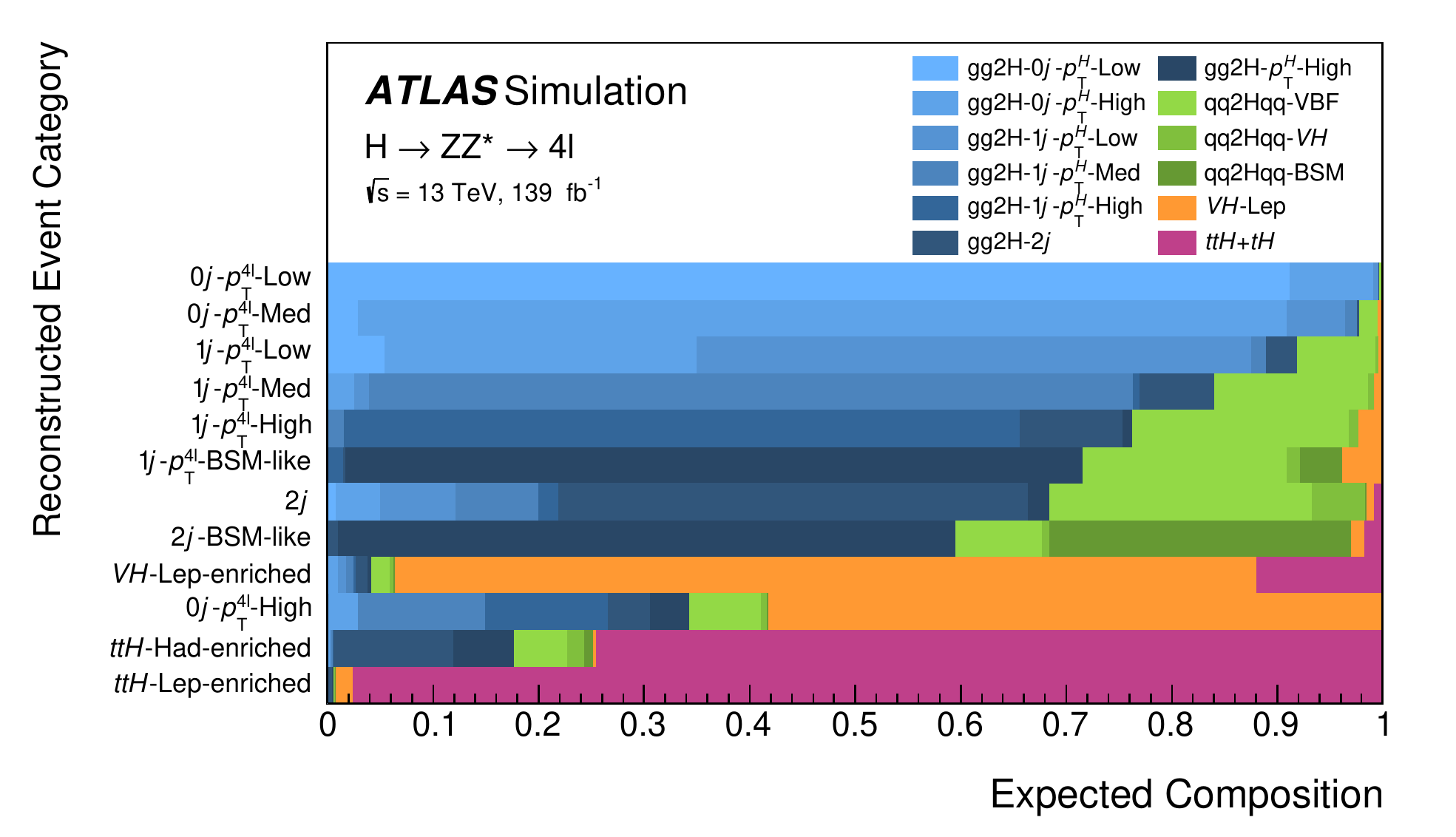}
\caption{Standard Model signal composition in terms of the Reduced Stage-1.1 production bins in each reconstructed event category. The \bbH contributions are included in the \ggF production bins.}
\label{fig:sig_compositionSTXS1p1}
\end{center}
\end{figure}

The expected number of signal events is shown in Table~\ref{tab:event_yields_signal} for each reconstructed event
category separately for each production mode. The \ggF\ and \bbH\ contributions are shown separately to compare their relative contributions, but both belong in the same (\ggF)
production bin. The highest \bbH\ event yield is expected in the \CatZeroJ categories since the jets tend to be more forward than in the \ttH\ process, thus escaping
the acceptance of the \ttH\ selection criteria. The sources of uncertainty in these expectations are detailed in Section~\ref{sec:systematics}. The signal composition in terms
of the Reduced Stage-1.1 production bins is shown in \figref{fig:sig_compositionSTXS1p1}.
 
The separation of the contributions from different production bins, such as
the \STXSggToHTwoJ, \STXSqqtoHqqVHLike and \STXSqqtoHqqRest components contributing in categories with two or more jets, is improved by means of discriminants obtained using multivariate data analysis, as described in the following section.
 
\begin{table*}[htbb!]
\centering
\footnotesize
\caption{The expected number of SM Higgs boson events with $m_H=$~125~\GeV\ for an integrated luminosity of \Lum\ at $\sqrt{s}=$~13~\TeV\ in each reconstructed event signal ($115<m_{4\ell}<130$~\GeV) and sideband ($m_{4\ell}$ in 105--115~\GeV\ or 130--160~\GeV\ for \zzstar, 130--350~\GeV\ for \tXX) category, shown separately for each production bin of the Production Mode Stage. The \ggF and \bbH yields are shown separately but both contribute to the same (\ggF) production bin, and $ZH$ and $WH$ are reported separately but are merged together for the final result. Statistical and systematic uncertainties, including those for total SM cross-section predictions, are added in quadrature. Contributions that are below 0.2\% of the total signal in each reconstructed event category are not shown and are replaced by `$-$'.}
\label{tab:event_yields_signal}
\vspace{0.1cm}
\setlength{\tabcolsep}{1cm}
{\renewcommand{\arraystretch}{1.3}
\resizebox{\textwidth}{!}{
\begin{tabular}{* {1} {@{\hspace{2pt}}l@{\hspace{2pt}}}* {7} {@{\hspace{2pt}}c@{\hspace{2pt}}}}
\hline\hline
\noalign{\vspace{0.05cm}}
Reconstructed & \multicolumn{6}{c}{SM Higgs boson production mode}\\
event category  & \ggF & \VBF  &   \WH   & \ZH& \ttH   +  \tH& \bbH\\ \hline
 
Signal &  \multicolumn{6}{c}{$115<m_{4\ell}<130$ \GeV}\\  \hline
 
\CatZeroJL  & $   23.9 \pm    3.5~~ $ & $  0.073 \pm  0.006 $ & $ 0.0173 \pm 0.0031 $ & $ 0.0131 \pm 0.0023 $ & $               - $ & $   0.17 \pm   0.09 $ & \\
\CatZeroJM  & $     74 \pm      8~~ $ & $   1.03 \pm   0.15 $ & $   0.37 \pm   0.05 $ & $   0.40 \pm   0.05 $ & $               - $ & $    0.8 \pm    0.4 $ & \\
\CatZeroJH  & $  0.109 \pm  0.026 $ & $ 0.0157 \pm 0.0024 $ & $  0.056 \pm  0.005 $ & $  0.173 \pm  0.016 $ & $ 0.00065 \pm 0.00023 $ & $               - $ & \\
\CatOneJL  & $     31 \pm      4~~ $ & $   1.99 \pm   0.11 $ & $   0.52 \pm   0.05 $ & $   0.35 \pm   0.04 $ & $               - $ & $   0.41 \pm   0.21 $ & \\
\CatOneJM  & $   17.3 \pm    2.8~~ $ & $   2.50 \pm   0.18 $ & $   0.52 \pm   0.06 $ & $   0.40 \pm   0.04 $ & $ 0.0078 \pm 0.0013 $ & $   0.09 \pm   0.04 $ & \\
\CatOneJH  & $    3.6 \pm    0.8 $ & $   0.84 \pm   0.07 $ & $  0.158 \pm  0.015 $ & $  0.166 \pm  0.016 $ & $ 0.0044 \pm 0.0006 $ & $  0.011 \pm  0.006 $ & \\
\CatOneJBSM  & $   0.87 \pm   0.23 $ & $  0.246 \pm  0.020 $ & $  0.060 \pm  0.007 $ & $  0.054 \pm  0.006 $ & $ 0.00156 \pm 0.00032 $ & $ 0.0009 \pm 0.0005 $ & \\
\CatTwoJ  & $     25 \pm      5~~ $ & $    8.5 \pm    0.6 $ & $   1.94 \pm   0.15 $ & $   1.69 \pm   0.13 $ & $   0.46 \pm   0.04 $ & $   0.30 \pm   0.15 $ & \\
\CatTwoJBSM  & $    1.9 \pm    0.6 $ & $   1.08 \pm   0.05 $ & $  0.120 \pm  0.016 $ & $  0.122 \pm  0.016 $ & $  0.075 \pm  0.007 $ & $ 0.0021 \pm 0.0010 $ & \\
\CatVHLep  & $  0.050 \pm  0.011 $ & $  0.019 \pm  0.004 $ & $   0.80 \pm   0.07 $ & $  0.245 \pm  0.021 $ & $  0.166 \pm  0.013 $ & $ 0.0027 \pm 0.0014 $ & \\
\CatttHhad  & $   0.15 \pm   0.16 $ & $  0.021 \pm  0.004 $ & $  0.020 \pm  0.005 $ & $  0.055 \pm  0.013 $ & $   0.75 \pm   0.07 $ & $  0.020 \pm  0.011 $ & \\
\CatttHlep  & $ 0.0019 \pm 0.0022 $ & $ 0.00019 \pm 0.00008 $ & $ 0.0046 \pm 0.0026 $ & $ 0.0032 \pm 0.0018 $ & $   0.41 \pm   0.04 $ & $               - $ & \\

\hline
Sideband & \multicolumn{5}{c}{$105<m_{4\ell}<115$ \GeV~or $130<m_{4\ell}<160$ \GeV}\\
\hline
 
\CatSBZeroJ  & $    4.2 \pm    0.5 $ & $  0.050 \pm  0.010 $ & $  0.096 \pm  0.011 $ & $  0.042 \pm  0.005 $ & $               - $ & $  0.044 \pm  0.022 $ & \\
\CatSBOneJ  & $   2.37 \pm   0.29 $ & $  0.241 \pm  0.024 $ & $  0.100 \pm  0.013 $ & $  0.063 \pm  0.008 $ & $ 0.0049 \pm 0.0009 $ & $  0.023 \pm  0.012 $ & \\
\CatSBTwoJ  & $   1.25 \pm   0.26 $ & $   0.43 \pm   0.05 $ & $  0.119 \pm  0.014 $ & $  0.103 \pm  0.012 $ & $  0.109 \pm  0.010 $ & $  0.016 \pm  0.008 $ & \\
\CatSBVHL  & $  0.015 \pm  0.005 $ & $ 0.0029 \pm 0.0011 $ & $  0.084 \pm  0.008 $ & $  0.104 \pm  0.010 $ & $  0.065 \pm  0.006 $ & $ 0.0013 \pm 0.0007 $ & \\

\hline
& \multicolumn{5}{c}{$105<m_{4\ell}<115$ \GeV~or $130<m_{4\ell}<350$ \GeV}\\
\hline
 
\CatSBttV  & $  0.001 \pm  0.010 $ & $ 0.00012 \pm 0.00009 $ & $ 0.0006 \pm 0.0004 $ & $ 0.0008 \pm 0.0004 $ & $  0.068 \pm  0.008 $ & $               - $ & \\
 
\hline
Total  & $    186 \pm     14~~ $ & $   17.0 \pm    0.8~~ $ & $    5.0 \pm    0.4 $ & $   3.97 \pm   0.29 $ & $   2.13 \pm   0.18 $ & $    1.9 \pm    1.0 $ & \\

\hline\hline
\end{tabular}}
}
\end{table*}

\subsection{Multivariate production mode discriminants}
\label{subsec:cat_nn}
 
To further increase the sensitivity of the cross-section measurements in the production bins (Section~\ref{subsec:intro_stxs}), multivariate discriminants using neural networks (NNs) \cite{Guest:2018}
are introduced in many of the reconstructed signal event categories as observables used in the statistical fit, described in Section~\ref{subsec:stxs_stxs}. The NN architecture and training procedure are defined using Keras with TensorFlow~\cite{chollet2015keras,tensorflow2015-whitepaper}.
These networks are trained using several discriminating observables, as defined in Table~\ref{tab:BDT_NN_trainingVars}, on simulated SM Higgs boson signals with $m_{H} = 125~\GeV$ or non-Higgs-boson background.  Due to the low number of  signal events expected in the \CatZeroJH, \CatOneJBSM and \CatttHlep categories, only the observed yield is used as the discriminant in these categories.

Two types of NNs are used: feed-forward multilayer perceptron (MLP) and recurrent (RNN)~\cite{RNN1,Goodfellow-et-al-2016,Guest:2018,tensorflow2015-whitepaper,chollet2015keras}.
Each NN discriminant combines two RNNs, one for the \pt-ordered variables related to the four leptons in the quadruplet and one for variables related to jets, and an MLP with additional variables related to the full event. The jet RNN accepts inputs from up to three jets.
The outputs of the MLP and the two RNNs are chained into another MLP to complete an NN discriminant, which is trained to approximate the posterior probability for an event to originate from a   given process. This is used in each reconstructed event category to discriminate between two or three processes, e.g.\ \ggF, \VBF and $ZZ$ background in the \CatOneJL category. The variables used to train the MLP and RNNs for each category along with the processes being separated are summarised in Table~\ref{tab:BDT_NN_trainingVars}.
\begin{table}[h!]
\setlength{\tabcolsep}{20pt}
\renewcommand{\arraystretch}{1.4}
\caption{The input variables used to train the MLP, and the two RNNs for the four leptons and the jets (up to three). For each category, the processes which are classified by an NN, their corresponding input variables and the observable used are shown. For example, there are eight input variables for the Lepton RNN being trained if \ptl\ and \etal\ are listed.  Leptons and jets are denoted by `$\ell$' and `$j$'. See the text for the definitions of the variables.
}
\label{tab:BDT_NN_trainingVars}
\begin{adjustbox}{width=1.0\textwidth,center}
\begin{tabular}{l | c | c c c | c}
\hline
\noalign{\vspace{0.05cm}}
Category & Processes & MLP & Lepton RNN & Jet RNN & Discriminant \\
\noalign{\vspace{0.05cm}}
\hline
\hline
\noalign{\vspace{0.05cm}}
 
\CatZeroJL &
\multirow{2}{*}{\ggF, \zzstar} &
\ptH, \dZZ, \monetwo, \mthreefour,  &
\multirow{2}{*}{\ptl, \etal}   &
\multirow{2}{*}{-}   &
\multirow{2}{*}{\NNggF}   \\
 
\CatZeroJM &
&
\costhetastar, \costhetaOne, \phiZZ
&
&
&\\ \hline
 
\multirow{2}{*}{\CatOneJL} &
\multirow{2}{*}{\ggF, \VBF, \zzstar} &
\ptH, \ptj, \etaj,  &
\multirow{2}{*}{\ptl, \etal}   &
\multirow{2}{*}{-}  &
\NNVBF~ for $\NNZZ < 0.25$\\
 
&
&
\dRj, \dZZ
&
&
& \NNZZ~ for $\NNZZ > 0.25$\\ \hline

\multirow{2}{*}{\CatOneJM} &
\multirow{2}{*}{\ggF, \VBF, \zzstar} &
\ptH, \ptj, \etaj, \MET,  &
\multirow{2}{*}{\ptl, \etal}   &
\multirow{2}{*}{-}  &
\NNVBF~ for $\NNZZ < 0.25$\\
 
&
&
\dRj, \dZZ, \etaH
&
&
&
\NNZZ~ for $\NNZZ > 0.25$\\ \hline
 
\multirow{2}{*}{\CatOneJH} &
\multirow{2}{*}{\ggF, \VBF} &
\ptH, \ptj, \etaj,   &
\multirow{2}{*}{\ptl, \etal}   &
\multirow{2}{*}{-}  &
\multirow{2}{*}{\NNVBF}\\
 
&
&
\MET, \dRj, \etaH
&
&
&\\ \hline

\multirow{2}{*}{\CatTwoJ} &
\multirow{2}{*}{\ggF, \VBF, \VH} &
\multirow{2}{*}{\Mjj, \pTHjj} &
\multirow{2}{*}{\ptl, \etal}   &
\multirow{2}{*}{\ptj, \etaj} &
\NNVBF~ for $\NNVH < 0.2$\\
 
&
&
&
&
&\NNVH~ for $\NNVH > 0.2$\\ \hline
 
\CatTwoJBSM &
\ggF, \VBF &
\ZeppetaZZ, \pTHjj   &
\ptl, \etal    &
\ptj, \etaj &
\NNVBF \\ \hline
 
\multirow{2}{*}{\CatVHLep} &
\multirow{2}{*}{\VH, \ttH} &
\njets, \njetsb,  &
\multirow{2}{*}{\ptl}   &
\multirow{2}{*}{-}  &
\multirow{2}{*}{\NNttH}\\
 
&
&
\MET, \HT
&
&
& \\ \hline
 
\multirow{2}{*}{\CatttHhad} &   \multirow{2}{*}{\ggF, \ttH, \tXX}
&
\ptH, \Mjj,   &  \
\multirow{2}{*}{\ptl, \etal}   &
\multirow{2}{*}{\ptj, \etaj}  &
\NNttH~ for $\NNtXX < 0.4$
\\
 
&
&
\dRj, \njetsb,
&
&
&
\NNtXX~ for $\NNtXX > 0.4$\\
 
\noalign{\vspace{0.05cm}}
\hline
 
\hline
\hline
\end{tabular}
\end{adjustbox}
\end{table}

The NN training variables not previously defined are listed as follows. The kinematic discriminant \dZZ~\cite{HIGG-2013-21}, defined as the difference between the logarithms of the squared matrix elements for the signal decay (same as in Section~\ref{sec:selection}) and squared matrix elements for the background process, is used to distinguish \ggF from the non-resonant $ZZ$ background. Three angles~\cite{HIGG-2013-01} are used to further distinguish these processes:
the cosine of the leading $Z$ boson's production angle $\theta^*$ in the four-lepton rest frame;
the cosine of $\theta_1$ defined as the angle between the negatively charged lepton of the leading $Z$ in the leading $Z$ rest frame and the direction of flight of the leading $Z$ in the four-lepton rest frame; and the angle $\phi_{ZZ}$,
between the two $Z$ decay planes in the four-lepton rest frame.
The angular separation of the leading jet from the $4\ell$ system, \dRj, is used to distinguish \VBF or \ttH from \ggF.
For categories with two or more jets, kinematic variables that also include the information from the two leading jets are used: the invariant mass, \mjj; the transverse momentum of the $4\ell$ and the 2-jet system, \pTHjj; and the Zeppenfeld variable, $\eta_{ZZ}^{\text{Zepp}}= \left|\eta_{4\ell} -  \frac{\eta_{j1}+\eta_{j2}}{2}\right |$~\cite{Rainwater_1996}.
The number of reconstructed jets, \njets, the number of $b$-tagged jets at 70\% tagging efficiency, \njetsb, and the scalar sum of the \pt of all reconstructed jets, \HT, are used to identify the \ttH process.
 
Depending on the category and the number of processes being targeted, the NN has two or three output nodes. The value computed at each node represents the probability, with an integral of one, for the event to originate from the given process. For example, for the 0-jet category, two probabilities are evaluated, $\mathrm{NN_{ggF}}$ and $\mathrm{NN}_{ZZ}$. As these two values are a linear transformation of each other, only one output, $\mathrm{NN_{ggF}}$, is used as a discriminant in the fit model.
In categories with three targeted processes, only two of the three corresponding output probabilities are independent. In a given category, a selection is applied on one of the three output probabilities to split the events in two subcategories. This output probability is then used as the discriminant for the subcategory of events passing the selection,  while for the other subcategory one of the two remaining output probabilities is used. The selection criterion is chosen so as to provide the largest purity of the targeted process for events passing the selection.
For example, in the $1$-jet category, $\mathrm{NN_{VBF}}$ and $\mathrm{NN}_{ZZ}$ are used. The subcategory of events with $\mathrm{NN}_{ZZ}$ larger than 0.25 uses $\mathrm{NN}_{ZZ}$ as the discriminant in the fit model, while $\mathrm{NN_{VBF}}$ is used in the remaining subcategory. The subcategory definitions and observables used in all reconstructed event categories are summarised in Table~\ref{tab:BDT_NN_trainingVars}.

\FloatBarrier

\section{Background contributions}
\label{sec:background}
\subsection{Background processes with prompt leptons}
 
Non-resonant SM \zzstar\ production via $qq$ annihilation, gluon--gluon fusion and vector-boson scattering can result in four prompt leptons in the final state and constitutes the largest background for the analysis. While for the previous analyses~\cite{HIGG-2016-25, HIGG-2016-22}, simulation was exclusively used to estimate both the shape and normalisation, in this analysis the normalisation is constrained by a data-driven technique. This allows the systematic uncertainty to be reduced
by removing both the theoretical and luminosity uncertainties contributing to the normalisation uncertainty.
 
\begin{figure}[!htbp]
\centering
\subfloat[]{
\includegraphics[width=0.45\linewidth]{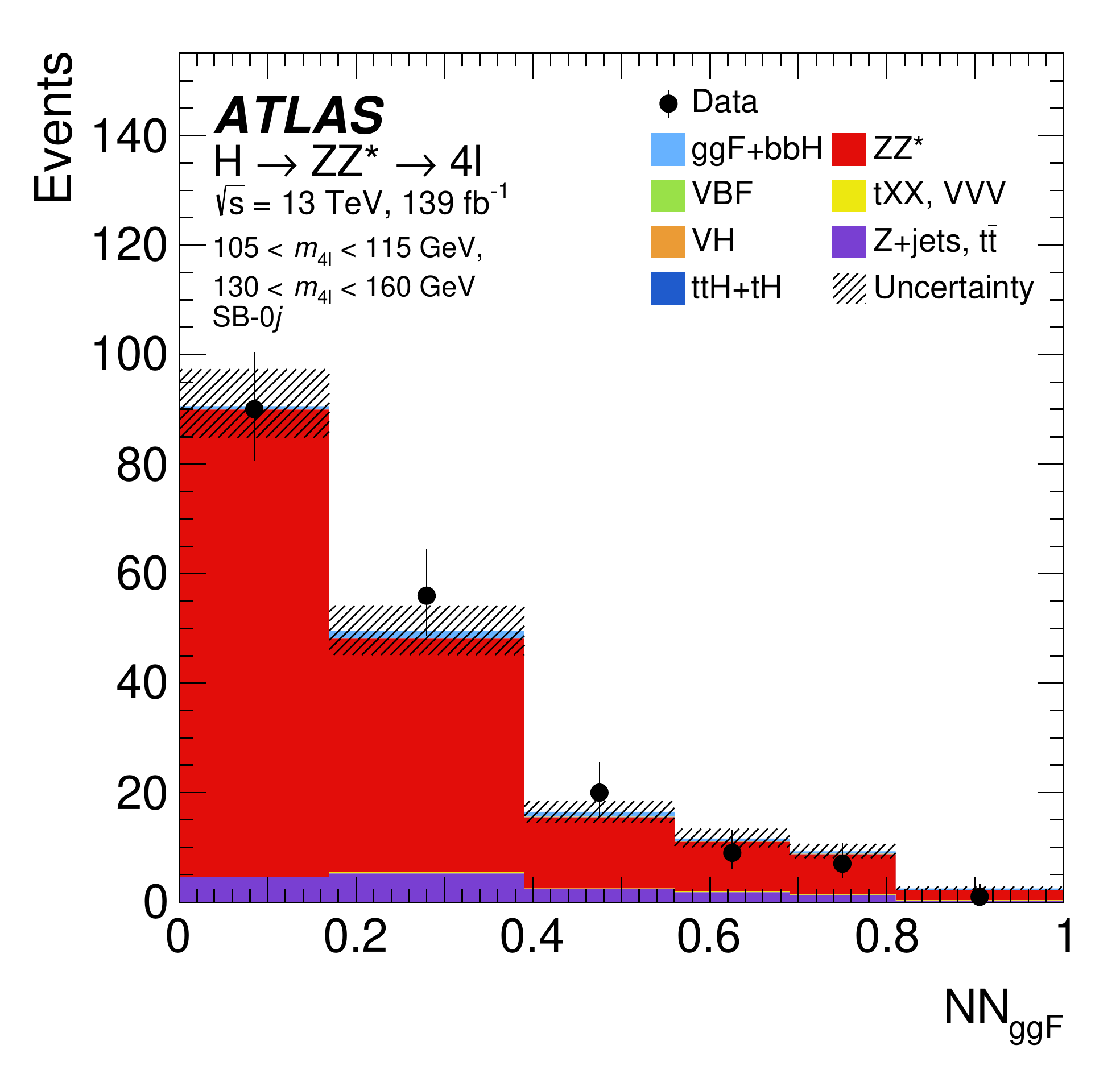}\label{fig:SB_NN}}
\subfloat[]{
\includegraphics[width=0.45\linewidth]{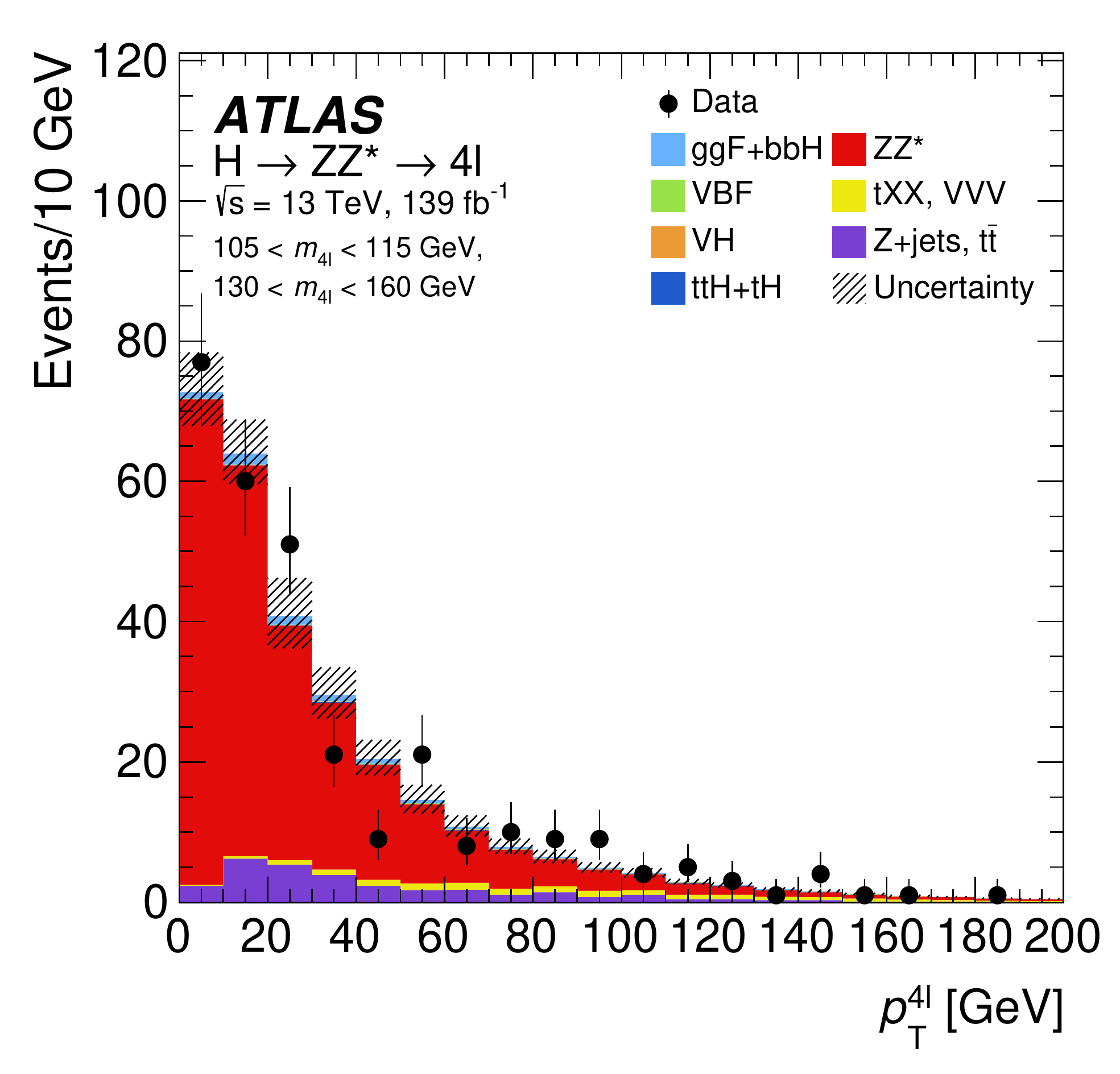}\label{fig:SB_Pt4l}}	\\
\subfloat[]{
\includegraphics[width=0.45\linewidth]{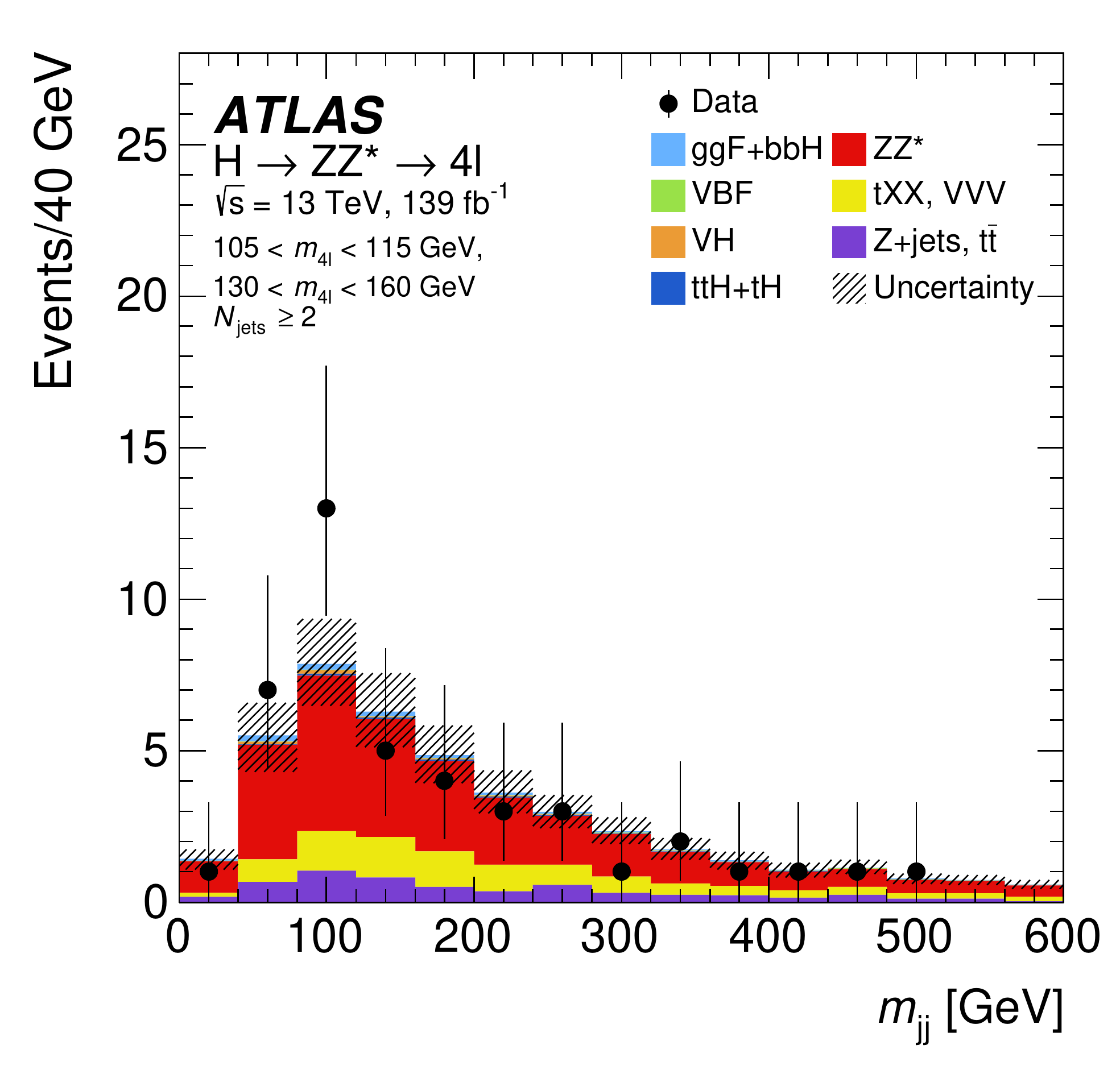}\label{fig:SB_Mjj}}
\subfloat[]{
\includegraphics[width=0.45\linewidth]{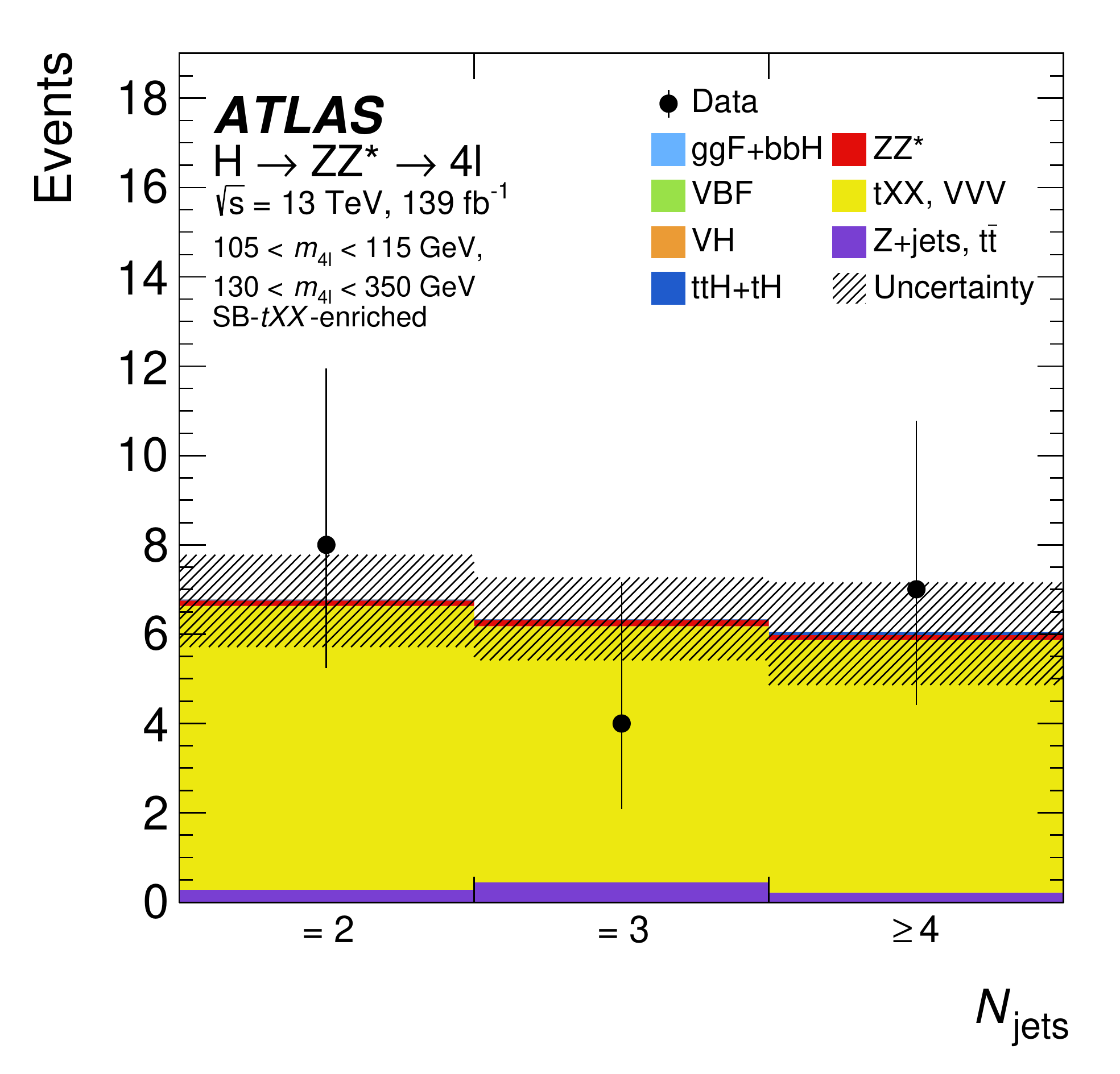}\label{fig:SB_ttX}}
\caption{The observed and expected  (post-fit) distributions for an integrated luminosity of \Lum\  at $\sqrt{\mathrm{s}}=$~13~\TeV\ in the different background enriched regions: (a) $\mathrm{NN_{ggF}}$ in the \CatSBZeroJ sideband region, (b) \ptfourl\  in the sideband region combining the \CatSBZeroJ, \CatSBOneJ and \CatSBTwoJ categories, (c) \mjj\ in the  \CatSBTwoJ category, and (d) \njets\ in the \CatSBtXX region. The SM Higgs boson signal is assumed to have a mass of $m_{H}$ = 125~\GeV. The uncertainty in the prediction is shown by the hatched band, calculated as described in Section~\ref{sec:systematics}.
\TheoryBlurb
}
\label{fig:ZZBkgPlot}
\end{figure}

As outlined in Section~\ref{subsec:cat_cat}, to estimate the normalisation, sideband categories in the $m_{4\ell}$ mass region 105--115~\GeV~and 130--160~\GeV\ are defined according to the jet multiplicity (\CatSBZeroJ, \CatSBOneJ, \CatSBTwoJ). The normalisation of the \zzstar~background is simultaneously fitted with a common normalisation factor for signal region and sideband categories with the same jet multiplicity. For example, the \zzstar\ background is scaled by a common factor for \CatTwoJ, \CatTwoJBSM and \CatSBTwoJ categories. The background shape templates for $\mathrm{NN}$ discriminants and the expected fraction of events in relevant reconstructed signal-region event categories are obtained from simulation. As shown in \figref{fig:SB_NN}, good agreement is found between data and simulation for the shape of the NN observable. All expected distributions are shown after the final fit to the data for the Production Mode measurement (see Section~\ref{sec:stxs}) and are referred to as post-fit distributions in the following. The simulated distributions of the observables \ptfourl~and \mjj\ employed for the
prediction of event fractions in each event category also agree with data, as seen in Figures~\ref{fig:SB_Pt4l} and \ref{fig:SB_Mjj} respectively. The estimation of the \zzstar~process in the jet multiplicity bins removes one of the leading theoretical uncertainties~\cite{ATL-PHYS-PUB-2017-005}.
Due to the limited sensitivity and the low expected yield, the normalisation of \zzstar\ in \ttH-like categories is estimated from simulation.
 
Similarly, backgrounds affecting the \ttH-like categories are estimated simultaneously from an enriched sample selected in a dedicated sideband region (\CatSBtXX), with the mass cut extended up to 350~\GeV\ to improve the statistical precision of the estimate. The normalisation of the $tXX$ process is simultaneously fitted across the \CatttHlep, \CatttHhad and \CatSBtXX categories.
The \njets\ observable distribution, which is used to predict the event fractions in each category, is shown in~\figref{fig:SB_ttX} and agrees with data.
In all other categories, the sensitivity of the $tXX$ measurement is limited due to a small number of expected $tXX$ events and its normalisation is estimated from simulation.
 
The contribution from \emph{VVV} processes is estimated 
for all categories using the simulated samples presented in Section~\ref{sec:simulation}.

\subsection{Background processes with non-prompt leptons}

Other processes, such as \zjets, \ttprod, and $WZ$, containing at least one jet, photon or lepton from a hadron decay that is misidentified as a prompt lepton, also contribute to the background. These `reducible' backgrounds are significantly smaller than the non-resonant $ZZ^*$ background and are estimated from data using different approaches for the \llmumu\ and \llee\ final states~\cite{HIGG-2016-22, HIGG-2016-25}.
 
In the \llmumu\ final states, the normalisation of the \zjets\ and \ttprod\ backgrounds are determined by performing fits to the invariant mass of the leading lepton pair in
dedicated independent control regions. The shape of the invariant mass distribution for each region is parameterised using simulated samples. In contrast to the previous analyses~\cite{HIGG-2016-25, HIGG-2016-22}, this fit is performed independently for each reconstructed event category, which removes the use of simulation to estimate the event fractions in these categories.
 
The control regions used to estimate this background are defined by closely following the requirements outlined in Section~\ref{subsec:selection_sel}. The definition and modified requirements for each of the four control regions are:
\begin{enumerate}
\item an enhanced heavy-flavour control region with inverted impact-parameter and relaxed isolation requirements on the subleading lepton pair and relaxed vertex $\chi^2$ requirements,
\item an enhanced \ttprod\ $e\mu+\mu \mu$ control region with an opposite-flavour leading lepton pair $e\mu$ and relaxed impact-parameter, isolation, and opposite-sign charge requirements on the subleading lepton pair $\mu \mu$, as well as relaxed vertex $\chi^2$ requirements,
\item an enhanced light-flavour control region with inverted isolation requirements for at least one lepton in the subleading lepton pair, and
\item a same-sign $\ell\ell+\mu^{\pm}\mu^{\pm}$  control region with relaxed impact-parameter and isolation requirements.
\end{enumerate}
The first two are the primary control regions used to estimate \zjets\ and \ttprod, and the latter two improve the estimate by reducing the statistical error of the fitted normalisation.

The background normalisations are obtained separately for the \zjets\ and \ttprod\ background processes using the simultaneous fit in the four control regions. The normalisation $n_i^{CR}$ in each control region CR for the background process $i$ is expressed as a fraction, $n_i^{CR} = t_i^{CR} \times N_i^{VR}$, of the normalisation $N_i^{VR}$ in a dedicated relaxed validation region (VR). $N_i^{VR}$ is used as the common parameter when fitting the normalisations in the different CRs. The transfer factor $t_i^{CR}$ is the ratio of the background contribution in the relaxed validation region and the given control regions. The relaxed validation region is defined by following the requirements outlined in Section~\ref{subsec:selection_sel} but by relaxing the impact-parameter and isolation requirements on the subleading lepton pair. This region contains a substantially larger number of events compared with the other four control regions, allowing a more reliable prediction of the shapes of the NN distributions. The shapes of the background NN distribution are then extrapolated together with the corresponding background normalisation from the relaxed validation to the signal region by means of additional transfer factors $T_i$. Transfer factors $t_i^{CR}$ and $T_i$ to extrapolate the background contributions from the control regions to the relaxed validation region and from there to the signal region are estimated from simulation and validated in several additional data control regions
 
The \llee\ control-region selection requires the electrons in the subleading lepton pair to have the same charge, and relaxes the identification, impact parameter and isolation
requirements on the electron candidate with the lowest transverse energy. This fake electron candidate, denoted by $X$, can be a light-flavour jet, an electron from photon
conversion or an electron from heavy-flavour hadron decay. The heavy-flavour background is determined from simulation. Good agreement is observed between simulation and data in a heavy-flavour enriched control region.
 
The remaining background is separated into light-flavour and photon conversion background components using the sPlot method~\cite{splot} which is performed on electron candidates $X$, separately for each reconstructed category in bins of the jet multiplicity and the transverse momentum of the electron candidate. The size of the two background components is obtained from a fit to the number of hits from the electron candidate $X$ in the innermost ID layer in the \llee\ data control region, where a hit indicates either a hadron track or an early conversion. A hit in the next-to-innermost pixel layer is used when the electron falls in a region that was either not instrumented with an innermost pixel layer module or where the module was not operating. The templates of the final discriminants for the mentioned fit of the light-flavour and photon conversion background components are obtained from simulated $Z+X$ events with an on-shell $Z$ boson decay candidate accompanied by an electron $X$ selected using the same criteria as in the  \llee\ control region. The simulated $Z+X$ events are also used to obtain the transfer factor for the $X$ candidate for the extrapolation of the light-flavour and photon conversion background contributions from the \llee\ control region to the signal region, after correcting the simulation to match the data in dedicated control samples of $Z+X$ events. The extrapolation to the signal region is also performed in bins of the electron transverse momentum and the jet multiplicity, separately for each reconstructed event category.
A method similar to that for the \llmumu\ final state is used to extract the NN shape, where the fractions of events from light-flavour jets and photon conversions are estimated from simulation and corrected transfer factors are used.
\FloatBarrier

\section{Systematic uncertainties}
\label{sec:systematics}
The systematic uncertainties are categorised into experimental and theoretical
uncertainties. The first category includes uncertainties in lepton and jet reconstruction,
identification, isolation and trigger efficiencies, energy resolution and scale, and uncertainty in the total integrated
luminosity. Uncertainties from the procedure used to derive the data-driven background estimates are also
included in this category. The second category includes uncertainties in theoretical modelling of the signal and
background processes.
 
The uncertainties can affect the signal acceptance, selection efficiency and discriminant distributions as well as the background
estimates.
The dominant sources of uncertainty and their effect are described in the following subsections. The impact of these uncertainties on the measurements is summarised in Table~\ref{tab:RankingSummarXS}.

\begin{table*}[!htbp]
\centering
\caption{The impact of the dominant systematic uncertainties (in percent) on the cross-sections in production bins of the Production Mode Stage and the Reduced Stage 1.1. Similar sources of systematic uncertainties are grouped together: luminosity (Lumi.),
electron/muon reconstruction and identification efficiencies and pile-up modelling ($e$, $\mu$, pile-up), jet energy scale/resolution and $b$-tagging efficiencies (Jets, flav.\ tag), uncertainties in reducible background (reducible bkg), theoretical uncertainties in \zzstar\ background and \tXX\ background, and theoretical uncertainties in the signal due to parton distribution function (PDF), QCD scale (QCD) and  parton showering algorithm (Shower). The uncertainties are rounded to the nearest 0.5\%, except for the luminosity uncertainty, which is measured to be 1.7\% and increases for the \VH signal processes
due to the simulation-based normalisation of the $VVV$ background.}
\label{tab:RankingSummarXS}
\vspace{0.1cm}
{\renewcommand{\arraystretch}{1.2}
 
\resizebox{\linewidth}{!}{
\begin{tabular}{c |cccc |ccccc}
\hline\hline
\multirow{3}{*}{Measurement} &
\multicolumn{4}{c|}{Experimental uncertainties [\%]}&
\multicolumn{5}{c}{Theory uncertainties [\%]}\\

& \multirow{2}{*}{Lumi.} &
\multicolumn{1}{c}{$e$, $\mu$,}&
\multicolumn{1}{c}{Jets,}&
\multicolumn{1}{c|}{Reducible} &
\multicolumn{2}{c}{Background} &
\multicolumn{3}{c}{Signal}\\
 
&
&
\multicolumn{1}{c}{pile-up} &
\multicolumn{1}{c}{flav. tag} &
\multicolumn{1}{c|}{bkg}&
\multicolumn{1}{c}{\zzstar} &
\multicolumn{1}{c}{\tXX} &
\multicolumn{1}{c}{PDF} &
\multicolumn{1}{c}{QCD} &
\multicolumn{1}{c}{Shower}\\

\hline
\multicolumn{10}{c}{Inclusive cross-section}\\
\hline
&  $ 1.7 $
&$ 2.5 $
&$ \phantom{0}0.5 $
&$ < 0.5 $
&$ \phantom{<} \ 1\phantom{.0} $
&$ < 0.5 $
&$ < 0.5 $
&$ \phantom{0}1\phantom{.0} $
&$ \phantom{0}2\phantom{.0} $ \\
 
\hline
\multicolumn{10}{c}{Production mode cross-sections}\\
\hline
\STXSggF
& $ 1.7 $
&$ 2.5 $
&$ \phantom{0}1\phantom{.0} $
&$ < 0.5 $
&$ \phantom{0}1.5 $
&$ < 0.5 $
&$ \phantom{<} \ 0.5 $
&$ \phantom{0}1\phantom{.0} $
&$ \phantom{0}2\phantom{.0} $ \\
\STXSVBF
& $ 1.7 $
&$ 2\phantom{.0} $
&$ \phantom{0}4\phantom{.0} $
&$ < 0.5 $
&$ \phantom{0}1.5 $
&$ < 0.5 $
&$ \phantom{<} \ 1\phantom{.0} $
&$ \phantom{0}5\phantom{.0} $
&$ \phantom{0}7\phantom{.0} $ \\
\STXSVH
& $ 1.9 $
&$ 2\phantom{.0} $
&$ \phantom{0}4\phantom{.0} $
&$ \phantom{<} \ 1\phantom{.0} $
&$ \phantom{0}6\phantom{.0} $
&$ < 0.5 $
&$ \phantom{<} \ 2\phantom{.0} $
&$ 13.5 $
&$ \phantom{0}7.5 $ \\
\STXSttH
& $ 1.7 $
&$ 2\phantom{.0} $
&$ \phantom{0}6\phantom{.0} $
&$ < 0.5 $
&$ \phantom{0}1\phantom{.0} $
&$ \phantom{<} \ 0.5 $
&$ \phantom{<} \ 0.5 $
&$ 12.5 $
&$ \phantom{0}4\phantom{.0} $ \\
 
\hline
\multicolumn{10}{c}{Reduced Stage-1.1 production bin cross-sections}\\
\hline
\STXSggToHZeroJL
& $ 1.7 $
&$ 3\phantom{.0}   $
&$ \phantom{0}1.5 $
&$ \phantom{<} \ 0.5  $
&$ \phantom{0}6.5 $
&$ < 0.5 $
&$ < 0.5 $
&$ \phantom{0}1\phantom{.0}   $
&$ \phantom{0}1.5 $ \\
\STXSggToHZeroJH
& $ 1.7 $
&$ 3\phantom{.0}   $
&$ \phantom{0}5\phantom{.0}   $
&$ < 0.5 $
&$ \phantom{0}3\phantom{.0}   $
&$ < 0.5 $
&$ < 0.5 $
&$ \phantom{0}0.5 $
&$ \phantom{0}5.5 $ \\
\STXSggToHOneJL
& $ 1.7 $
&$ 2.5 $
&$ 12\phantom{.0}  $
&$ \phantom{<} \ 0.5  $
&$ \phantom{0}7\phantom{.0}   $
&$ < 0.5 $
&$ < 0.5 $
&$ \phantom{0}1\phantom{.0}   $
&$ \phantom{0}6\phantom{.0}   $ \\
\STXSggToHOneJM
& $ 1.7 $
&$ 3\phantom{.0}   $
&$ \phantom{0}7.5 $
&$ < 0.5 $
&$ \phantom{0}1\phantom{.0}   $
&$ < 0.5 $
&$ < 0.5 $
&$ \phantom{0}1.5 $
&$ \phantom{0}5.5 $ \\
\STXSggToHOneJH
& $ 1.7 $
&$ 3\phantom{.0} $
&$ 11\phantom{.0}  $
&$ \phantom{<} \ 0.5  $
&$ \phantom{0}2\phantom{.0}   $
&$ < 0.5 $
&$ < 0.5 $
&$ \phantom{0}2\phantom{.0}   $
&$ \phantom{0}7.5 $ \\
\STXSggToHTwoJ
& $ 1.7 $
&$ 2.5 $
&$ 16.5$
&$ \phantom{<} \ 1\phantom{.0}    $
&$ 12.5$
&$  \phantom{<} \ 0.5   $
&$ < 0.5 $
&$ \phantom{0}2.5 $
&$ 10.5 $ \\
\STXSggToHHigh
& $ 1.7 $
&$ 1.5 $
&$ \phantom{0}3\phantom{.0}   $
&$ \phantom{<} \ 0.5  $
&$ \phantom{0}3.5 $
&$ < 0.5 $
&$ < 0.5 $
&$ \phantom{0}2\phantom{.0}   $
&$ \phantom{0}3.5 $ \\
\STXSqqtoHqqVHLike
& $ 1.8 $
&$ 4\phantom{.0}   $
&$ 17\phantom{.0}  $
&$ \phantom{<} \ 1\phantom{.0}    $
&$ \phantom{0}4\phantom{.0}   $
&$  \phantom{<} \ 1 \phantom{.0}      $
&$ \phantom{<} \ 0.5   $
&$ \phantom{0}5.5 $
&$ \phantom{0}8\phantom{.0} $ \\
\STXSqqtoHqqRest
& $ 1.7 $
&$ 2\phantom{.0}   $
&$ \phantom{0}3.5 $
&$ < 0.5 $
&$ \phantom{0}5\phantom{.0}   $
&$ < 0.5 $
&$ < 0.5 $
&$ \phantom{0}6\phantom{.0}   $
&$ 10.5 $ \\
\STXSqqtoHqqBSM
& $ 1.7 $
&$ 2\phantom{.0}   $
&$ \phantom{0}4\phantom{.0}   $
&$ < 0.5 $
&$ \phantom{0}2.5 $
&$ < 0.5 $
&$ < 0.5 $
&$ \phantom{0}3\phantom{.0}   $
&$ \phantom{0}8\phantom{.0} $ \\
\STXSVHLep
& $ 1.8 $
&$ 2.5 $
&$ \phantom{0}2\phantom{.0}   $
&$ \phantom{<} \ 1\phantom{.0}  $
&$ \phantom{0}2\phantom{.0}   $
&$  \phantom{<} \ 0.5   $
&$ < 0.5 $
&$ \phantom{0}1.5 $
&$ \phantom{0}3\phantom{.0} $ \\
\STXSttH
& $ 1.7 $
&$ 2.5 $
&$ \phantom{0}5\phantom{.0}   $
&$ \phantom{<} \ 0.5  $
&$ \phantom{0}1\phantom{.0}   $
&$  \phantom{<} \ 0.5   $
&$ < 0.5 $
&$ 11\phantom{.0}  $
&$ \phantom{0}3\phantom{.0} $ \\
 
\hline\hline
\end{tabular}
}}
\end{table*}

\subsection{Experimental uncertainties}
\label{subsec:systematics_exp}

The uncertainty in the combined 2015--2018 integrated luminosity is 1.7\%~\cite{ATLAS-CONF-2019-021}, obtained using the LUCID-2 detector~\cite{LUCID2} for the primary luminosity measurements. This uncertainty affects the signal and the normalisation of the simulated background estimates when not constrained by the data sidebands.
 
The uncertainty in the predicted yields due to pile-up modelling ranges between 1\% and 2\% and is derived by varying the
average number of pile-up events in the simulation to cover the
uncertainty in the ratio of the predicted to measured inelastic cross-sections~\cite{STDM-2015-05}.
 
The electron (muon) reconstruction, isolation and identification efficiencies, and the energy (momentum) scale and resolution are
derived from data using large samples of $J/\psi\rightarrow \ell\ell$ and $Z\rightarrow \ell\ell$ decays~\cite{EGAM-2018-01, PERF-2015-10}.
Typical uncertainties in the predicted yields for the relevant decay channels due to the identification and reconstruction efficiency uncertainties are below $1\%$ for muons and
1\%--2\% for electrons. The uncertainty in the expected yields due to the muon and electron isolation efficiency is also taken into account, with the typical size being 1\%. The uncertainties in the trigger efficiencies have a negligible impact.
The uncertainties in the electron and muon energy and momentum scale and resolution are small and also have a negligible impact on the measurements.
 
The uncertainties in the jet energy scale and resolution are in the range 1\%--3\%~\cite{ PERF-2016-04}. The impact of these uncertainties is more relevant for the \VH, \VBF and \ttH production mode cross-sections (3\%--5\%)
and for all the Reduced Stage-1.1 cross-section measurements, including the \ggF process split into the different \njets\ exclusive production bins (5\%--20\%).
 
The uncertainty in the calibration of the $b$-tagging algorithm, which is derived from dileptonic \ttprod\ events, amounts to a few percent over most of the jet \pt\ range~\cite{PERF-2016-05}.
This uncertainty is only relevant in the \CatttH category, with its expected impact being approximately 1\% in the \ttH cross-section measurement. The uncertainties associated with the \MET reconstruction have a negligible impact.

A shift in the simulated Higgs boson mass corresponding to the precision of the Higgs boson measurement, $m_{H} = 125.09 \pm 0.24$ \GeV\ \cite{HIGG-2014-14}, is shown to have a negligible impact on the signal acceptance.
A small dependency of the $\mathrm{NN_{ggF}}$ discriminant shape in the \CatZeroJL\ and \CatZeroJM\ categories on $m_{H}$ is observed for the signal (below 2\% in the highest NN score bins) and is included in the signal model. This uncertainty affects the measurement of \ggF production, as well as the measurements in other production bins with large \ggF contamination.

For the data-driven measurement of the reducible background, three sources of uncertainty are considered: statistical uncertainty, overall systematic uncertainty for each of \llmumu\ and \llee, and a shape systematic uncertainty that varies with the reconstructed event category. Since the yields are estimated by using a statistical fit to a control data region with large statistics, the inclusive background estimate has a relatively small (3\%) statistical uncertainty. The systematic uncertainty for \llmumu\ and the heavy-flavour component of \llee\ is estimated by comparing the lepton identification, isolation and impact parameter significance efficiency between data and simulated events in a separate region, enriched with on-shell $Z$ boson decays accompanied by an electron or a muon. For both the \llmumu\ and \llee\ estimates, the difference in efficiency is assigned as the uncertainty in the extrapolation of the yield estimate from the control region to the signal region. For the \llee\ light-flavour component, the efficiency is derived from an enriched control region with a systematic uncertainty estimated by varying the assumed light- and heavy-flavour components. These inclusive uncertainties (6\%) are treated as correlated across the reconstructed event categories. Finally, there are additional uncorrelated uncertainties (8\%--70\%) in the fraction of the reducible background in each event category due to the statistical precision of the simulated samples.
 
\subsection{Theoretical uncertainties}
\label{subsec:systematics_theory_couplings}
 
The theoretical modelling of the signal and background processes is affected by uncertainties due to missing higher-order corrections, modelling of parton showers and the underlying event, and PDF$+\alphas$ uncertainties.
 
The impact of the theory systematic uncertainties on the signal depends on the kind of measurement
that is performed. For signal-strength measurements, defined as the measured cross-section divided by the SM prediction, or interpretation of cross-section using the EFT approach, each
source of theory uncertainty affects both the acceptance and the predicted SM cross-section. For the cross-section measurements, only effects on the acceptance need to be considered.
 
The impact of the theory systematic uncertainties on the background depends on the method of estimating the normalisation. If simulation is used, the uncertainties in the acceptance and the predicted SM cross-section are included. If the normalisation is estimated from a data-driven method, only the impact on the relative event fractions between categories is considered.

One of the dominant sources of theoretical uncertainty is the prediction of the \ggF
process in the different \njets~ categories.
The \ggF process gives a large contribution in categories with at least two jets. To estimate the variations due to the impact of higher-order contributions not included in the calculations
and migration effects on the \njets~\ggF cross-sections, the approach described in Refs.~\cite{YR4,Stewart:2011cf}
is used, which exploits the latest predictions for the inclusive jet cross-sections. In particular, the uncertainty from the choice of factorisation and
renormalisation scales, the choice of resummation scales, and the migrations between the 0-jet and 1-jet phase-space bins or between the 1-jet and $\geq 2$-jet bins are considered~\cite{YR4,Stewart:2013faa, Liu:2013hba, Boughezal:2013oha}.
The impact of QCD scale variations on the Higgs boson \pT\ distribution is taken into account as an
additional uncertainty. The uncertainty in higher-order corrections to the Higgs boson \pT\ originating from the assumption of
infinite top quark mass in
the heavy-quark loop is also taken into account by comparing the \pT\ distribution predictions to finite-mass calculations.
An additional uncertainty in the acceptance of the \ggF process in \VBF topologies ~\cite{Gangal:2013nxa} due to missing higher orders in QCD
in the calculation
is estimated by variations of the renormalisation and factorisation scales using fixed-order calculations with MCFM~\cite{MCFM}.
An additional uncertainty in the Higgs boson \pT\ distribution, derived by varying the renormalisation, factorisation and NNLOPS scale in the simulation, in the 0-jet topology is considered. This is particularly relevant when measuring the inclusive \ggF cross-section using the \ptH\ categories for events with no jet activity. To account for higher-order corrections to $p_{\text{T}}^{Hjj}$, which is used as an NN input variable, the uncertainty is derived by comparing the predicted distribution obtained using \progname{Powheg NNLOPS} and \MGMCatNLO with the FxFx merging scheme.

For the \VBF production mode, the uncertainty due to missing higher orders in QCD is parameterised using the scheme outlined in Ref.~\cite{bendavid2018les}. The migration effects due to the selection criteria imposed on the number of jets, transverse momentum of the Higgs boson, transverse momentum of the Higgs boson and the leading dijet system and the invariant mass of the two leading jets, used to define the full Stage 1.1 STXS production bins, are computed by varying the renormalisation and factorisation scales by a factor of two. The uncertainties are cross-checked with fixed-order calculations. Similarly, for the \VH production mode with the associated $V$ decaying leptonically, the scale variations are parameterised as migration effects due to the selection criteria imposed on the number of jets and the transverse momentum of the associated boson~\cite{ATL-PHYS-PUB-2018-035}.
 
For the \VH production mode with the associated $V$ decaying hadronically and the \ttH production mode, the uncertainty due to missing higher orders in QCD is obtained by varying the  renormalisation and factorisation scales by a factor of two. The configuration with the largest impact, as quantified by the relative difference between the varied and the nominal configuration, is chosen to define the uncertainty in each experimental category. These uncertainties are treated as uncorrelated among the different production modes. Due to the limited accuracy of the simulated samples, the uncertainties evaluated using this method for the total cross-sections are larger than those described in Ref.~\cite{YR4}.
 
The uncertainties in the acceptance due to the modelling of parton showers and the underlying event are estimated with AZNLO tune eigenvector variations and by comparing the acceptance
using the parton showering algorithm from \PYTHIAV{8} with that from \HERWIGV{7}~\cite{Bellm:2015jjp} for all signal processes. The uncertainty due to each AZNLO tune variation is taken as correlated among the different production modes while the difference between the parton showering algorithms is treated as an uncorrelated
uncertainty. The uncertainties due to higher-order corrections to the Higgs boson decay are modelled using the \progname{PROPHECY4F}~\cite{Bredenstein:2006rh,Bredenstein:2006ha} and \progname{Hto4L}~\cite{Boselli:2015aha,Boselli:2017pef} generators. These corrections are below 2\% and have a negligible impact on the results.
A 100\% uncertainty is assigned to heavy-flavour quark production modelling for the \ggF\ contribution entering in the \CatttH category. This has a negligible impact on the results.

The impact of the PDF uncertainty is estimated with the
thirty eigenvector variations of the \progname{PDF4LHC\_nlo\_30} Hessian PDF set following the PDF4LHC recommendations~\cite{Butterworth:2015oua}. The modification of the predictions
originating from each eigenvector variation is added as a separate source of uncertainty in the model. The same procedure is applied for the \ggF, \VBF, \VH and \ttH processes, enabling correlations to be taken into account in the fit model.

The impacts of the theoretical uncertainties, as described above, on the shape of NN discriminants are also considered. For \ggF production, a further cross-check is performed by
comparing the NN
shapes in the corresponding
categories as predicted by \progname{Powheg NNLOPS} and \progname{MadGraph5\_aMC@NLO} with the FxFx
merging scheme. All the NN shapes from the two generators agree within the scale variations and, therefore, no additional shape uncertainty is included.

For signal-strength measurements, an additional uncertainty related to the $H\to ZZ^*$ branching ratio prediction~\cite{Bredenstein:2006rh,Bredenstein:2006ha} is included in the measurement.

Since the normalisation of the \zzstar\ process in most reconstructed event categories is constrained by performing a simultaneous fit to sideband regions enriched in this contribution
together with the signal regions,
most of the theoretical uncertainty in the normalisation for this background vanishes. Nevertheless,
uncertainties in the shapes of the discriminants for the \zzstar\ background and in the relative contribution of this background between the sidebands and the signal regions are taken into account.
The uncertainties due to missing higher-order effects in QCD are estimated by varying the
factorisation and renormalisation QCD scales by a factor of two; the impact of the PDF uncertainty
is estimated by using the MC replicas of the \progname{NNPDF3.0} PDF set. Uncertainties due to parton shower modelling for the \zzstar\ process are considered as well. The impact of these uncertainties is below 2\% for all production mode cross-sections measured. In addition, a comparison between \progname{Sherpa} and \progname{Powheg} is also taken as an additional source of systematic uncertainty. This model uncertainty is treated as uncorrelated among the different sideband-to-signal region
extrapolations (in 0-jet, 1-jet and 2-jet categories).
 
The uncertainty in the gluon-initiated and the vector-boson-initiated \zzstar\ process is taken into account by
changing the relative composition of the quark-initiated, the gluon-initiated and the vector-boson-scattered \zzstar\ components according to the theoretical uncertainty in the predicted cross-sections and the respective $K$-factors. In addition, the event yield and NN discriminant shapes in each event category are compared with the data
in an $m_{4\ell}$ sideband around the signal region (105 \gev$<m_{4\ell}<$~115~\gev\ or 130~\gev$<m_{4\ell}<$~160~\gev ). Good agreement
between the \progname{Sherpa} predictions and the data is found.
 
For the \tXX\ process, uncertainties due to PDF and QCD scale variations are considered in the relative fraction of events present in the \ttH-like categories, in the \CatSBtXX control region and in the NN discriminant shape. Differences between \MGMCatNLO and \SHERPA are considered as an additional systematic uncertainty.
For all other categories where this process is estimated from simulation, the impact of these uncertainties on the SM cross-section and acceptance are also considered.
 
Uncertainties in the PDF and in missing higher-order corrections in QCD are applied to the \emph{VVV} background estimate, which is fully taken from MC simulation.
 
To probe the tensor structure of the Higgs boson coupling in the EFT approach, theoretical uncertainties due to PDF and QCD scale variations are assigned to the signal predictions based on the simulated highest-order SM signal samples. The same uncertainties are assigned to all corresponding BSM signal predictions, since it is shown using the MC signal samples simulated at LO accuracy that the uncertainties change negligibly as a function of the Wilson coefficients.

\FloatBarrier

\section{Measurement of the Higgs boson production mode cross-sections}
\label{sec:stxs}
\subsection{Observed data}
\label{subsec:stxs_observed}
 
\begin{figure}[!htbp]
\begin{center}
\includegraphics[width=0.5\linewidth]{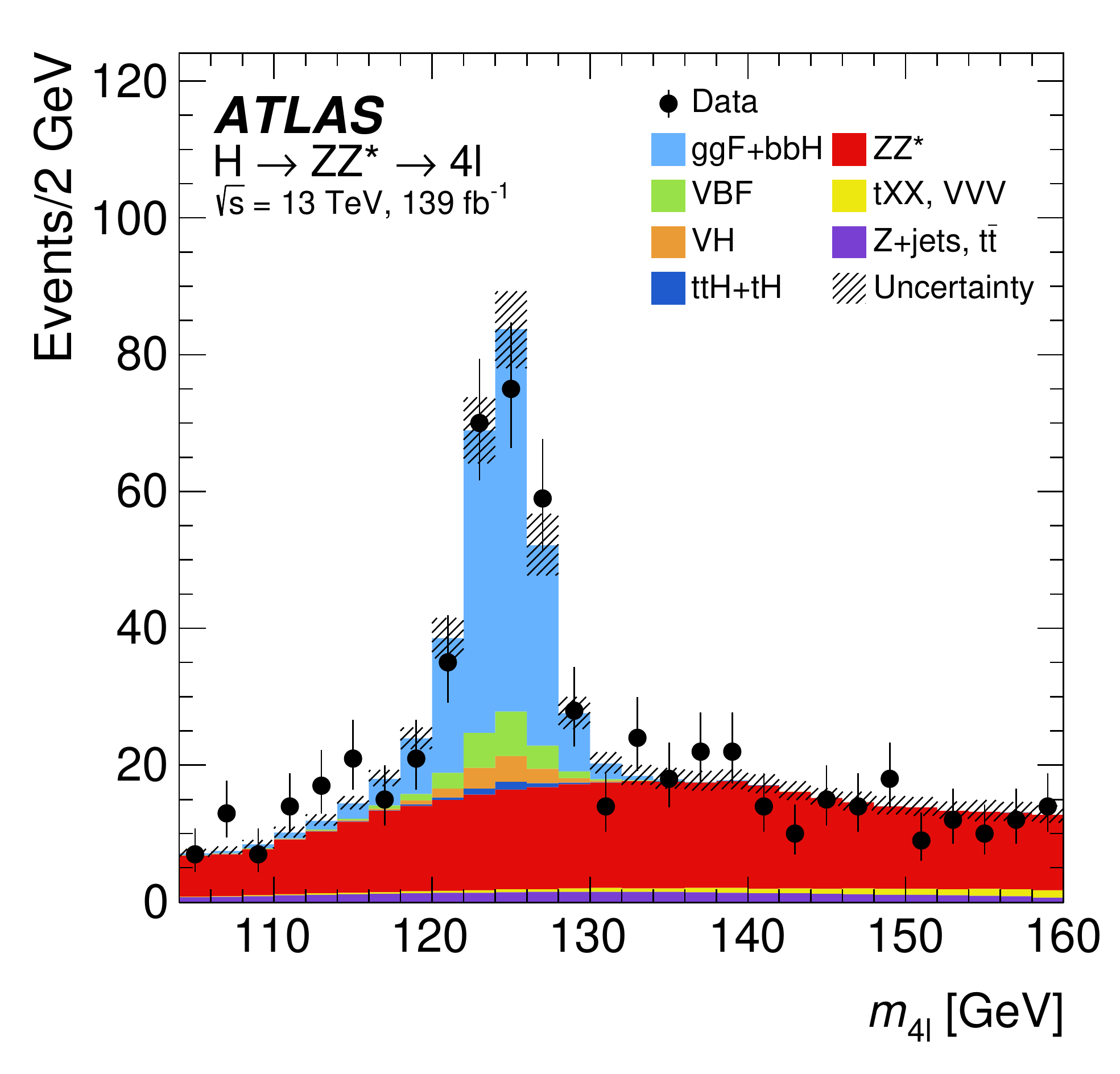}
\end{center}
\vskip-0.2cm
\caption{The observed and expected (post-fit) four-lepton invariant mass distributions for the selected Higgs boson candidates, shown for an integrated luminosity of \LumExact\ at $\sqrt{s}=13$~\TeV. The SM Higgs boson signal is assumed to have a mass $m_{H}$ = 125~\gev. The uncertainty in the prediction is shown by the hatched band, calculated as described in Section~\ref{sec:systematics}. \TheoryBlurb
}
\label{fig:m4lmass}
\end{figure}

The expected and observed four-lepton invariant mass (post-fit) distributions of the selected Higgs boson candidates after the event selection are shown in \figref{fig:m4lmass}.
 
\begin{figure}[!htbp]
\begin{center}
\subfloat[]{
\includegraphics[width=0.37\linewidth]{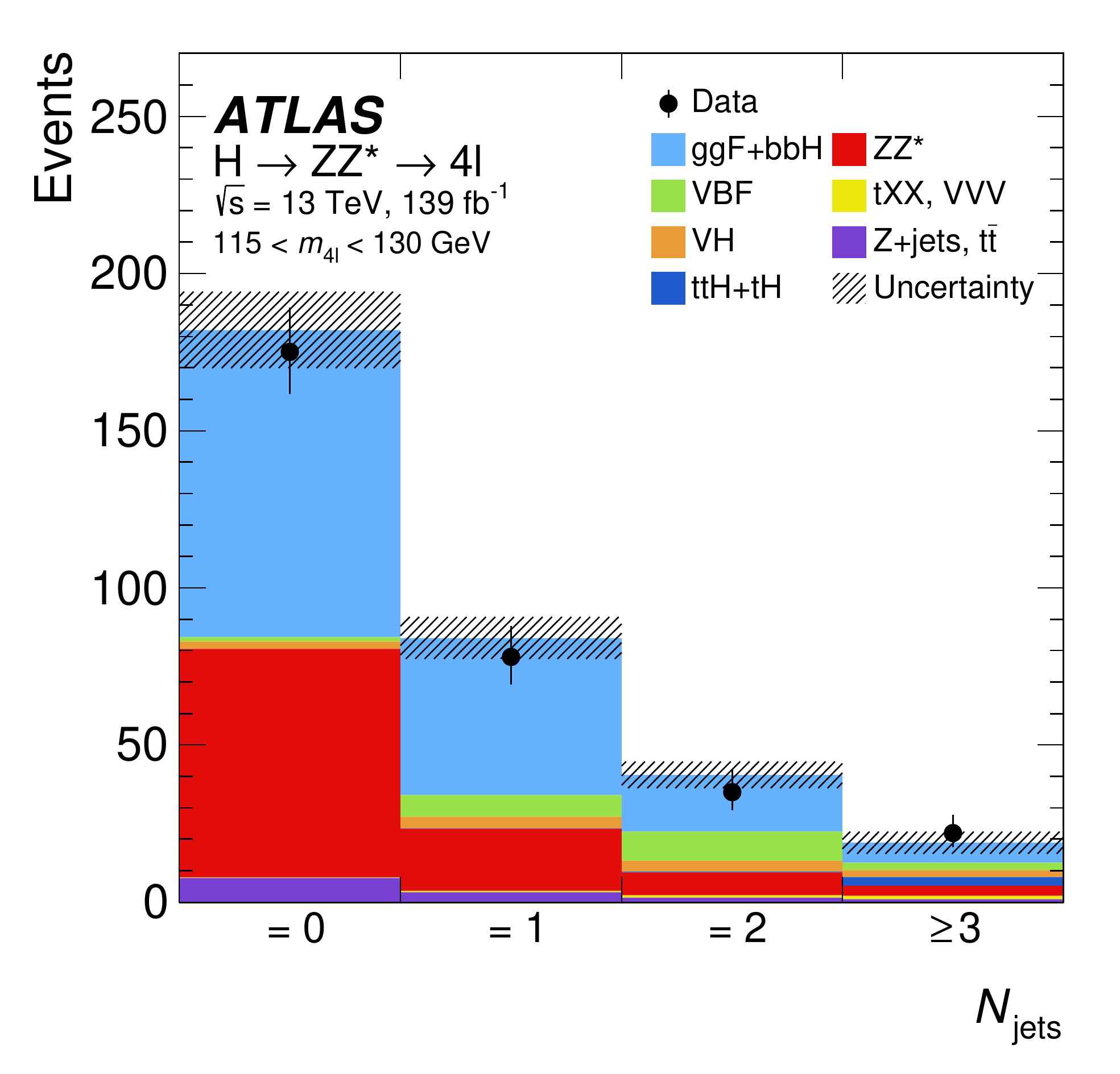}\label{fig:oNjets}}\\
\subfloat[]{
\includegraphics[width=0.37\linewidth]{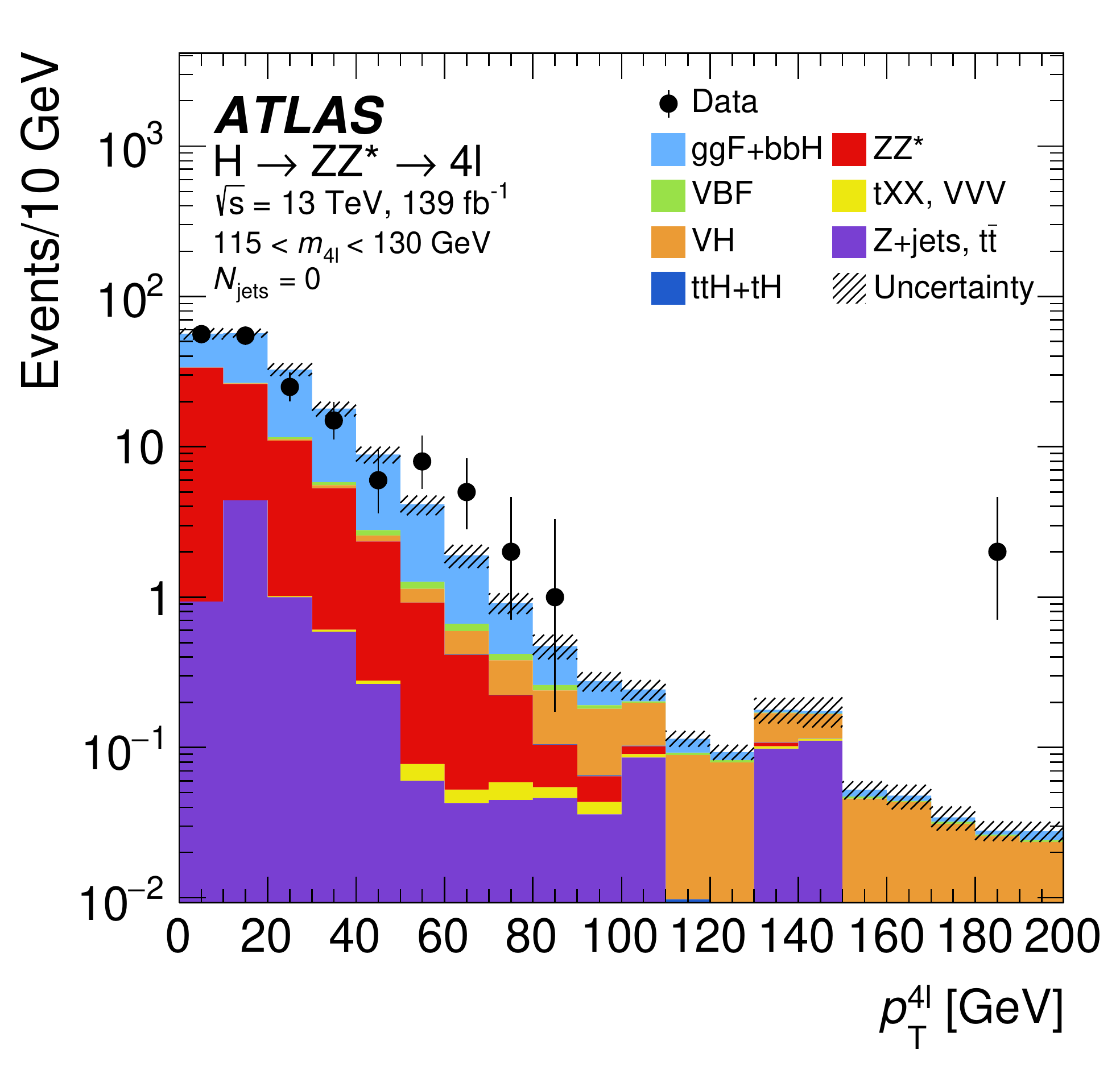}\label{fig:omelpt0j}}
\subfloat[]{
\includegraphics[width=0.37\linewidth]{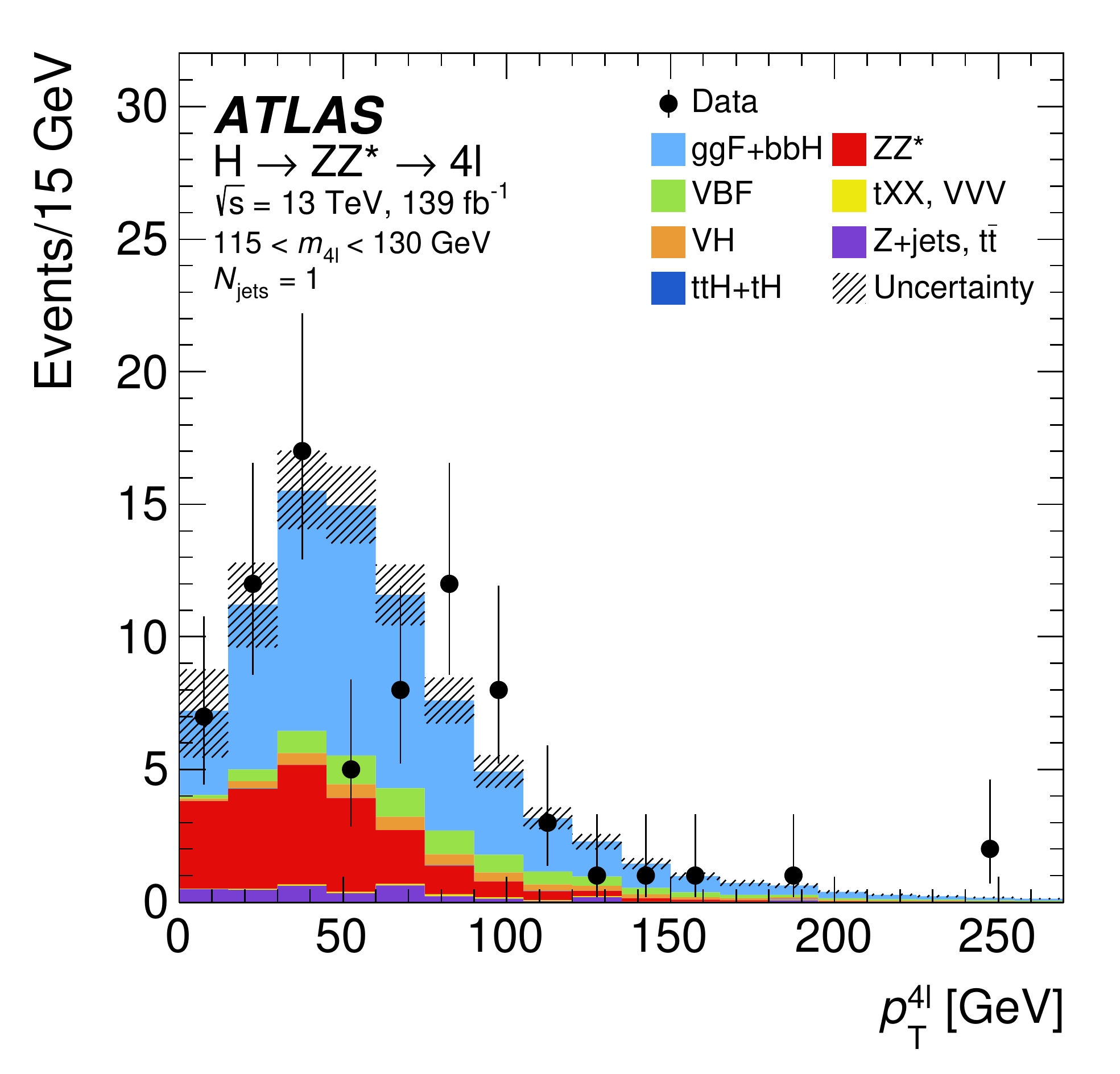}\label{fig:omelpt1j}}\\
\subfloat[]{
\includegraphics[width=0.37\linewidth]{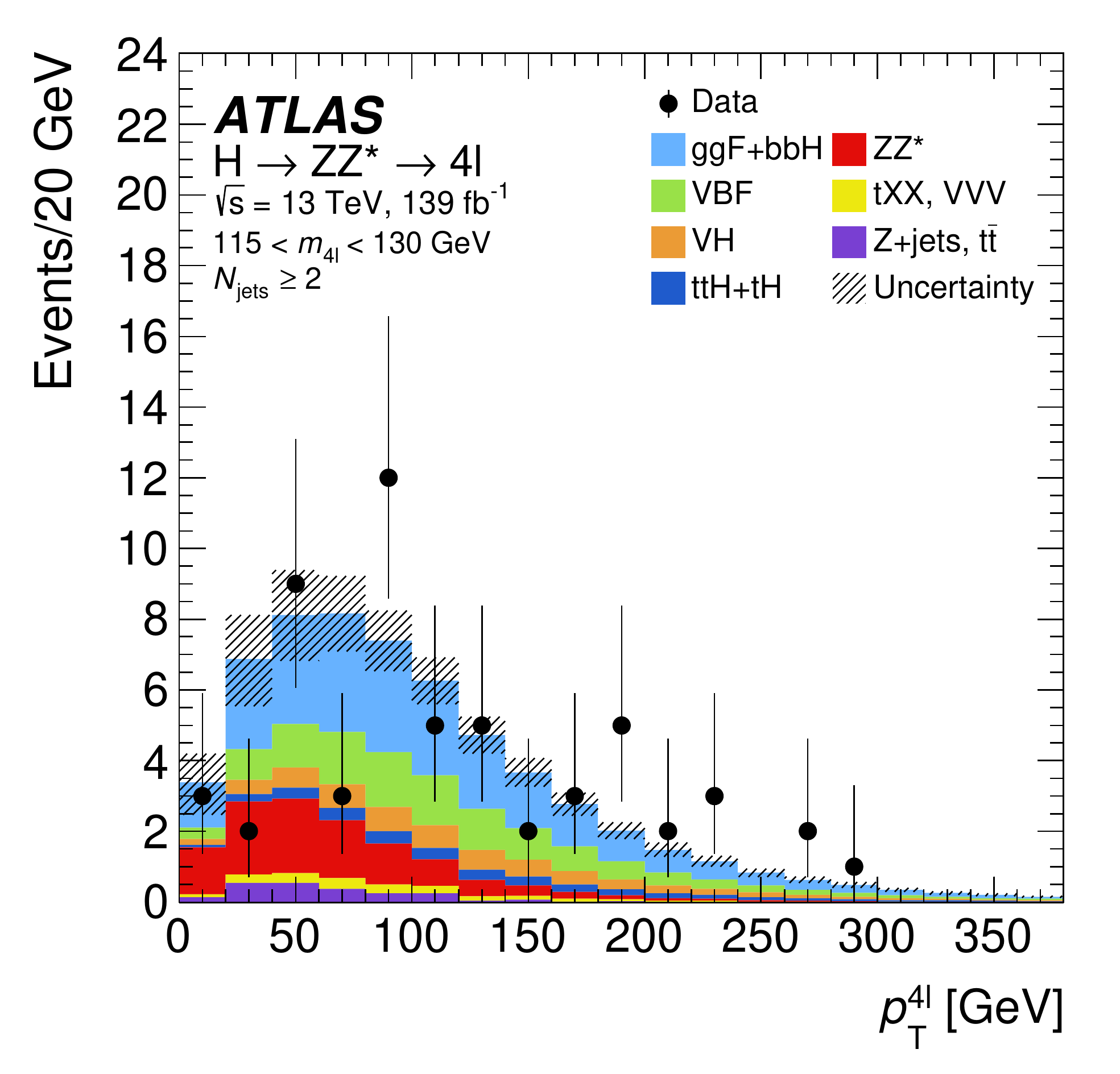}\label{fig:omelpt2j}}
\subfloat[]{
\includegraphics[width=0.37\linewidth]{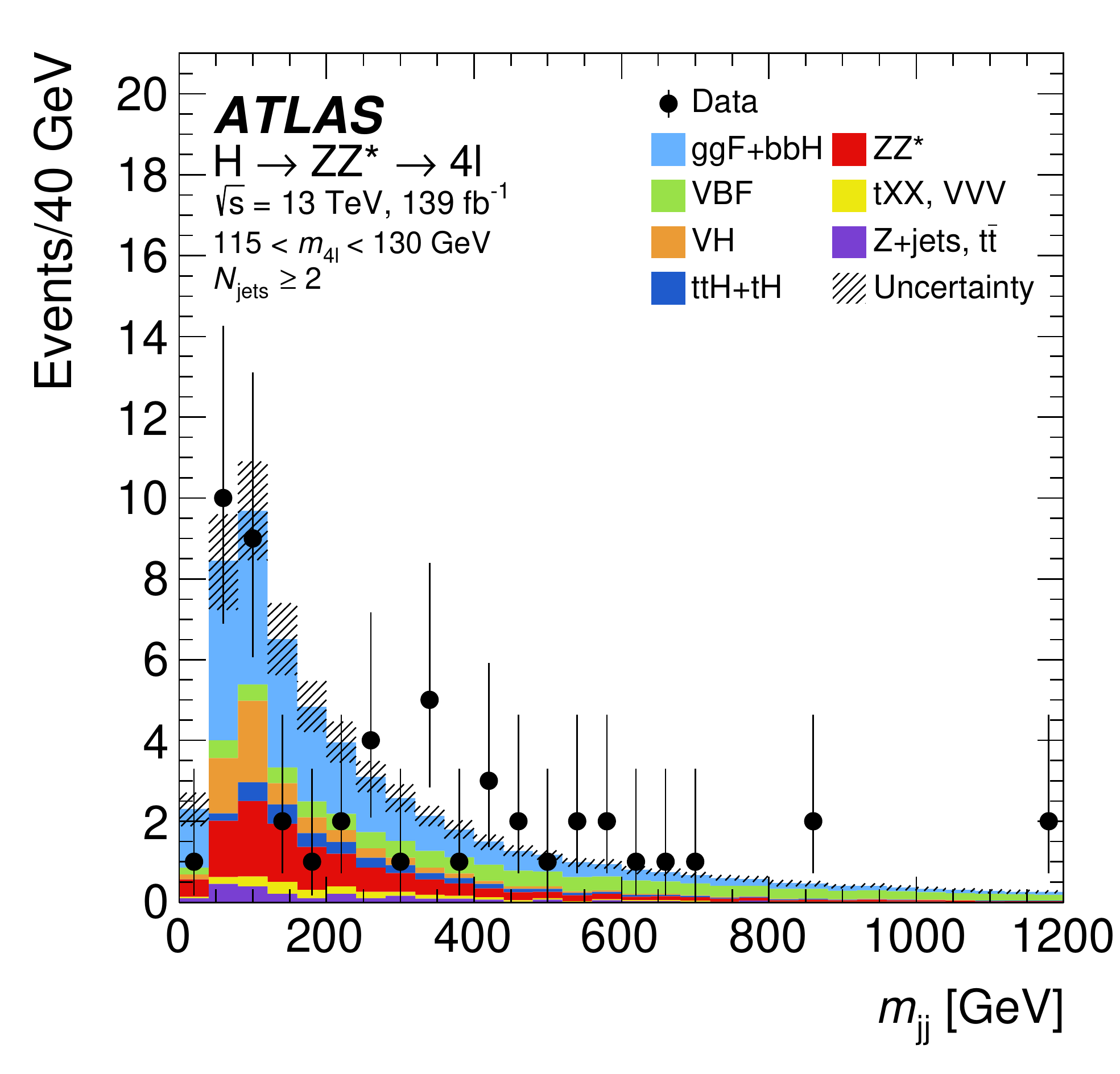}\label{fig:omjj}}
\end{center}
\vskip-0.2cm
\caption{The observed and expected distributions (post-fit) of (a) the jet multiplicity $N_{\textrm{jets}}$ after the inclusive event selection, the four-lepton transverse momenta \ptH\ for events with (b) exactly zero jets, (c) with exactly one jet and (d) with at least two jets and (e) the dijet invariant mass $m_{jj}$ for events with at least two jets. The SM Higgs boson signal is assumed to have a mass $m_{H}$ = 125~\gev. The uncertainty in the prediction is shown by the hatched band, calculated as described in Section~\ref{sec:systematics}. \TheoryBlurb
}
\label{fig:Cat_observables}
\end{figure}

The observed and expected (post-fit) distributions of the jet multiplicity, the dijet invariant mass, and the four-lepton transverse momenta in different $N_{\textrm{jets}}$ bins, which are used for the categorisation of reconstructed events, are shown in \figref{fig:Cat_observables} for different steps of the event categorisation.
 
\begin{figure}[!htbp]
\begin{center}
\subfloat[]{
\includegraphics[width=0.33\linewidth]{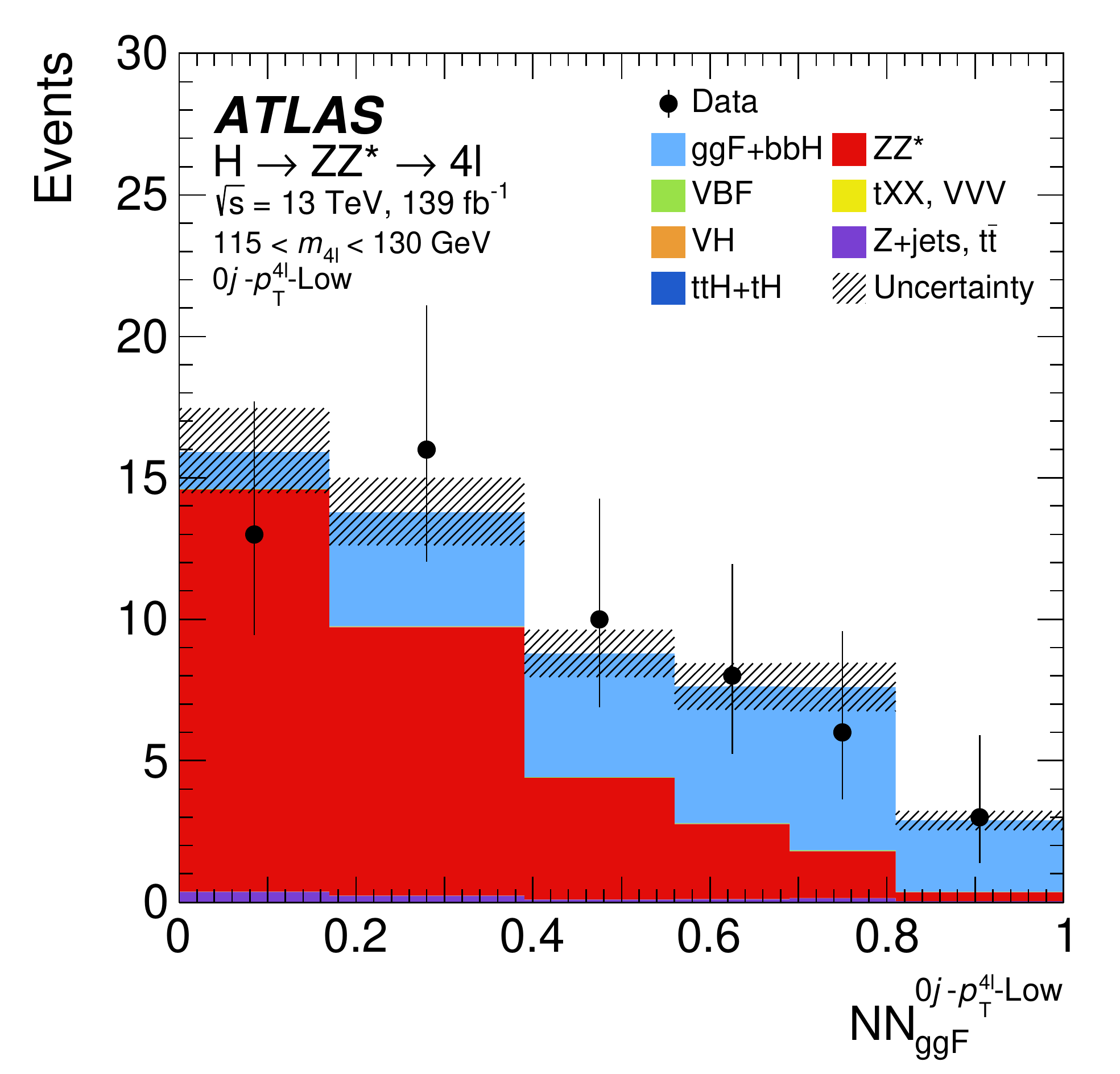}\label{fig:oNN0jptl}}
\subfloat[]{
\includegraphics[width=0.33\linewidth]{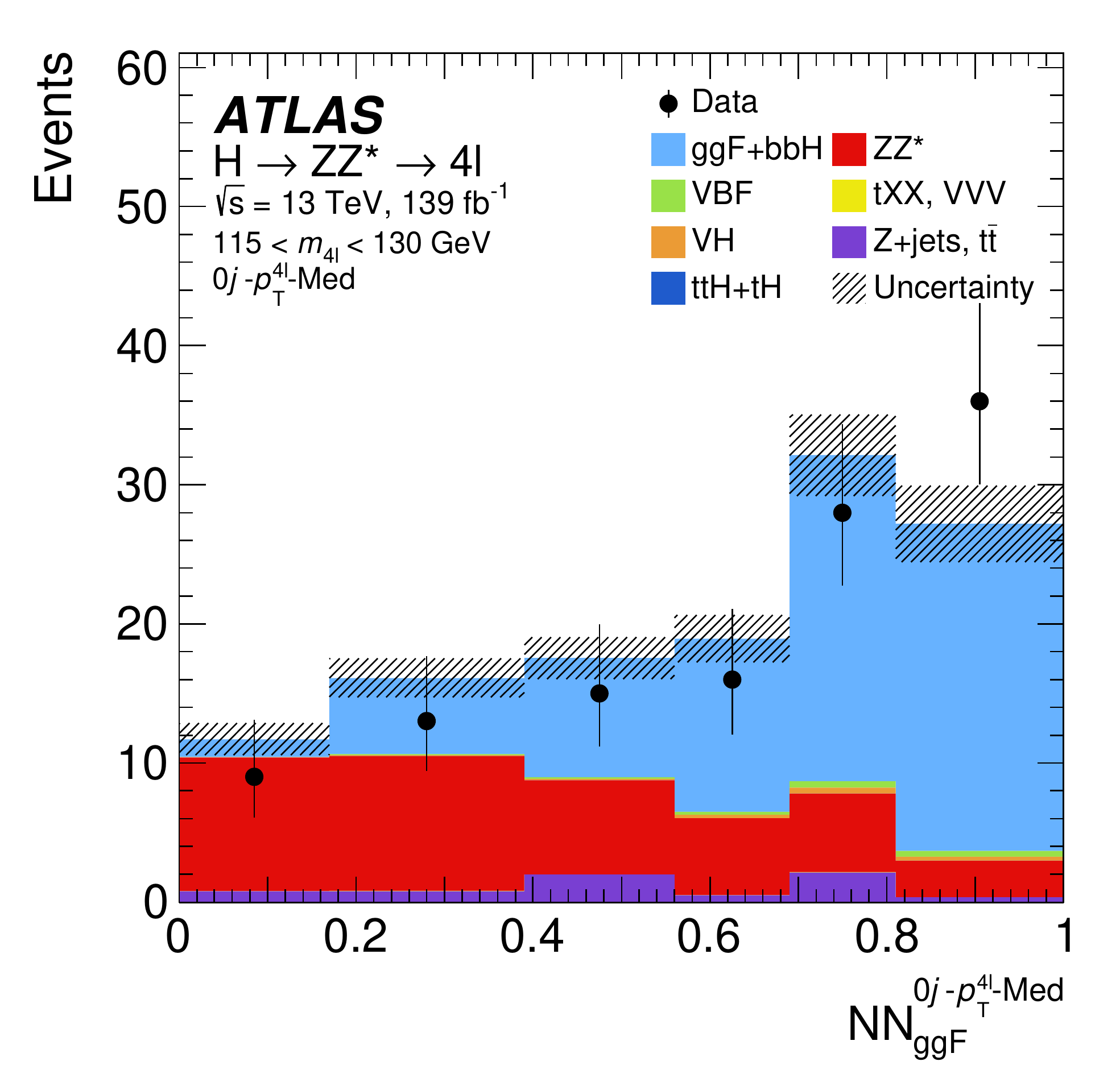}\label{fig:oNN0jpthm}}\\
\subfloat[]{
\includegraphics[width=0.33\linewidth]{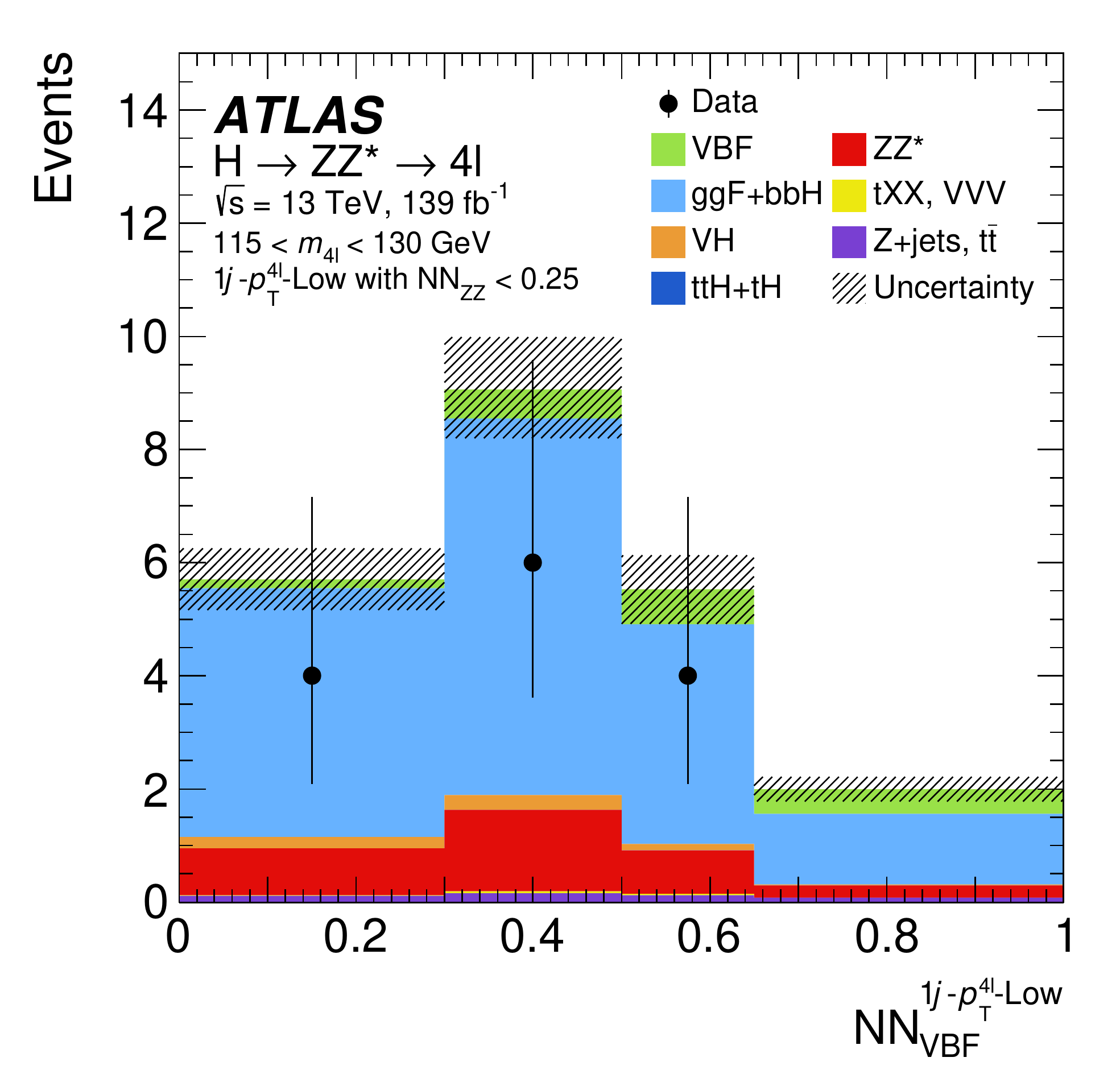}\label{fig:oNN1jpthlvbf}}
\subfloat[]{
\includegraphics[width=0.33\linewidth]{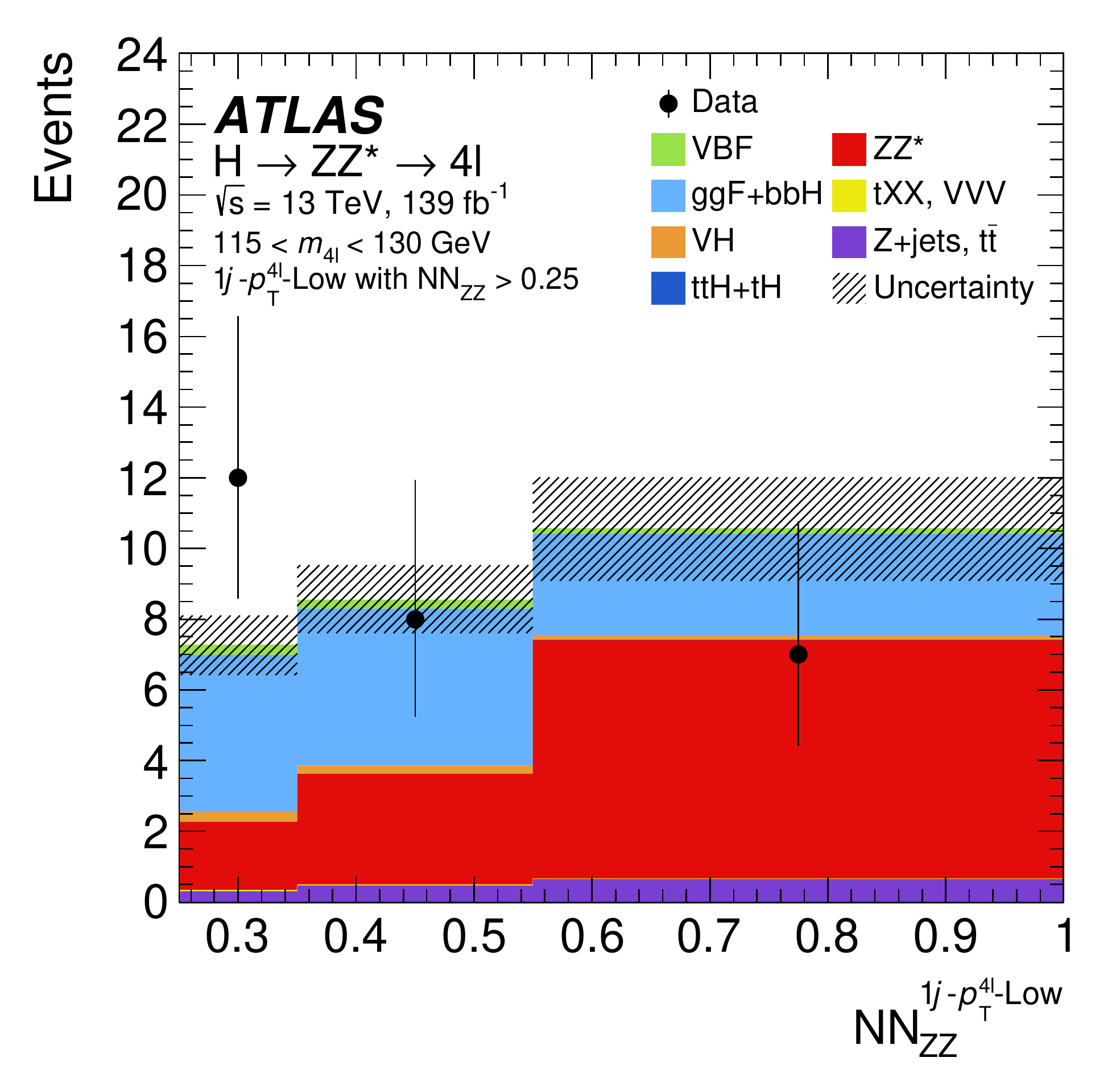}\label{fig:oNN1jptlzz}}\\
\subfloat[]{
\includegraphics[width=0.33\linewidth]{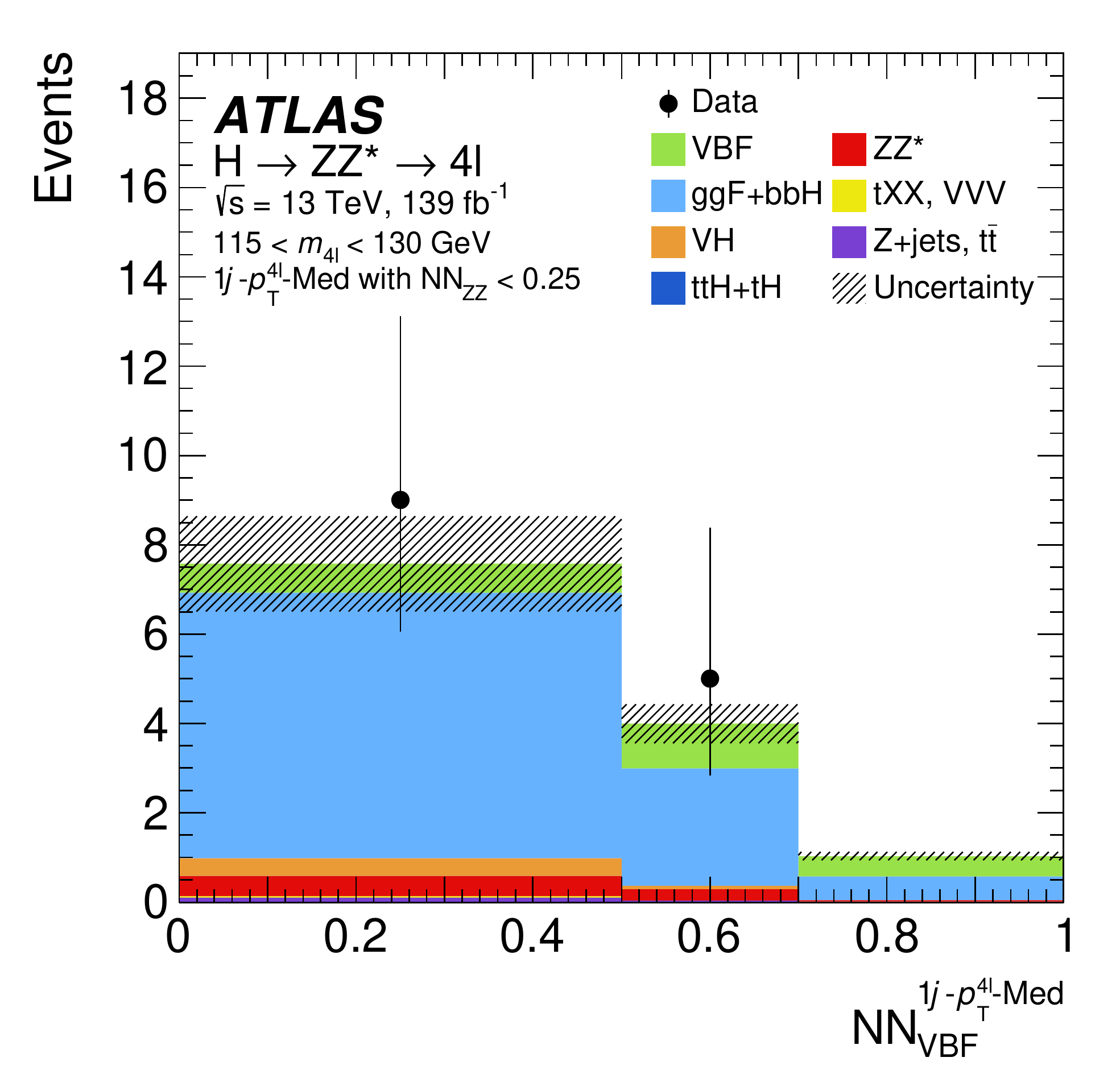}\label{fig:oNN1jpthmvbf}}
\subfloat[]{
\includegraphics[width=0.33\linewidth]{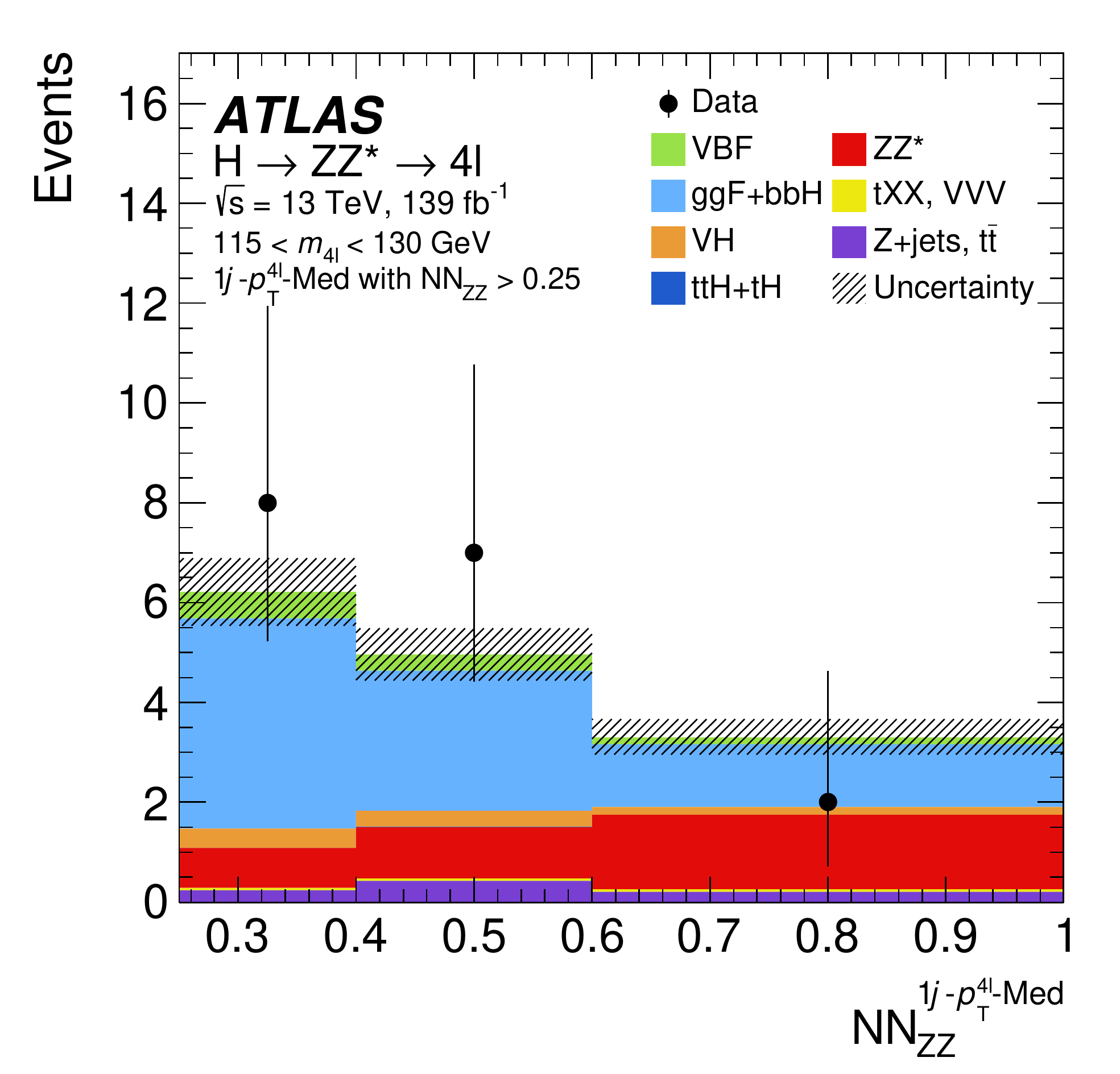}\label{fig:oNN1jpthmzz}}
\subfloat[]{
\includegraphics[width=0.33\linewidth]{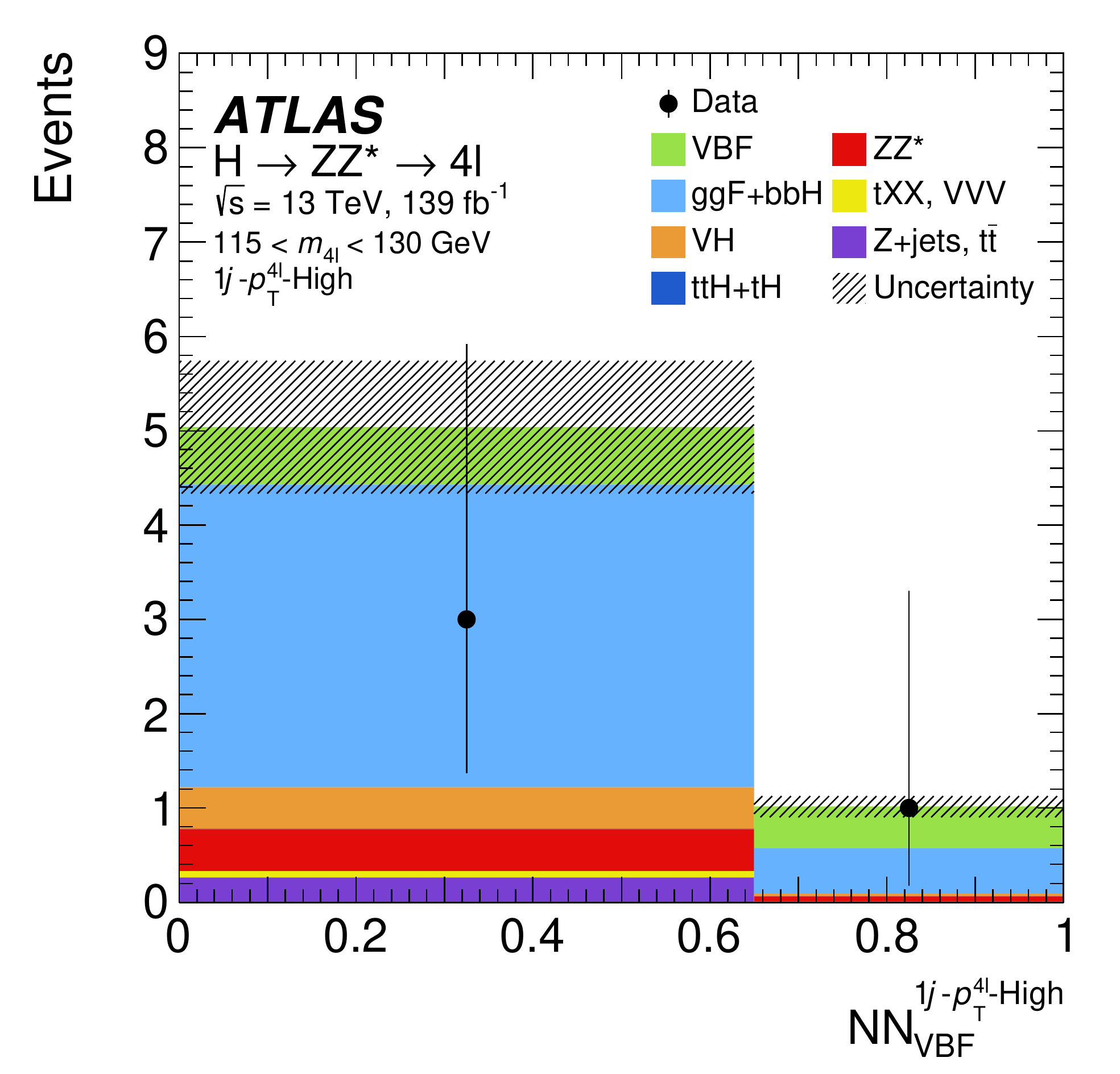}\label{fig:oNN1jpthhzz}}
\end{center}
\vskip-0.2cm
\caption{The observed and expected NN output (post-fit) distributions for an integrated luminosity of \Lum\ at $\sqrt{\mathrm{s}}=$~13~\TeV\ in the different zero- and one-jet categories: (a)  $\mathrm{NN_{ggF}}$ in \CatZeroJL, (b) $\mathrm{NN_{ggF}}$ in \CatZeroJM, (c)  $\mathrm{NN_{VBF}}$  \CatOneJL\ with $\mathrm{NN}_{ZZ}<0.25$, (d)  $\mathrm{NN}_{ZZ}$ in \CatOneJL\ with $\mathrm{NN}_{ZZ}>0.25$, (e) $\mathrm{NN_{VBF}}$ in \CatOneJM\ with $\mathrm{NN}_{ZZ}<0.25$, (f) $\mathrm{NN}_{ZZ}$ in \CatOneJM\ with $\mathrm{NN}_{ZZ}>0.25$ and (g) $\mathrm{NN_{VBF}}$ in \CatOneJH. The SM Higgs boson signal is assumed to have a mass $m_{H}$ = 125~\gev. The uncertainty in the prediction is shown by the hatched band, calculated as described in Section~\ref{sec:systematics}. \TheoryBlurb
The bin boundaries are chosen to maximise the significance of the targeted signal in each category. }
\label{fig:NNoutputs_observed}
\end{figure}

\begin{figure}[!htbp]
\begin{center}
\subfloat[]{
\includegraphics[width=0.33\linewidth]{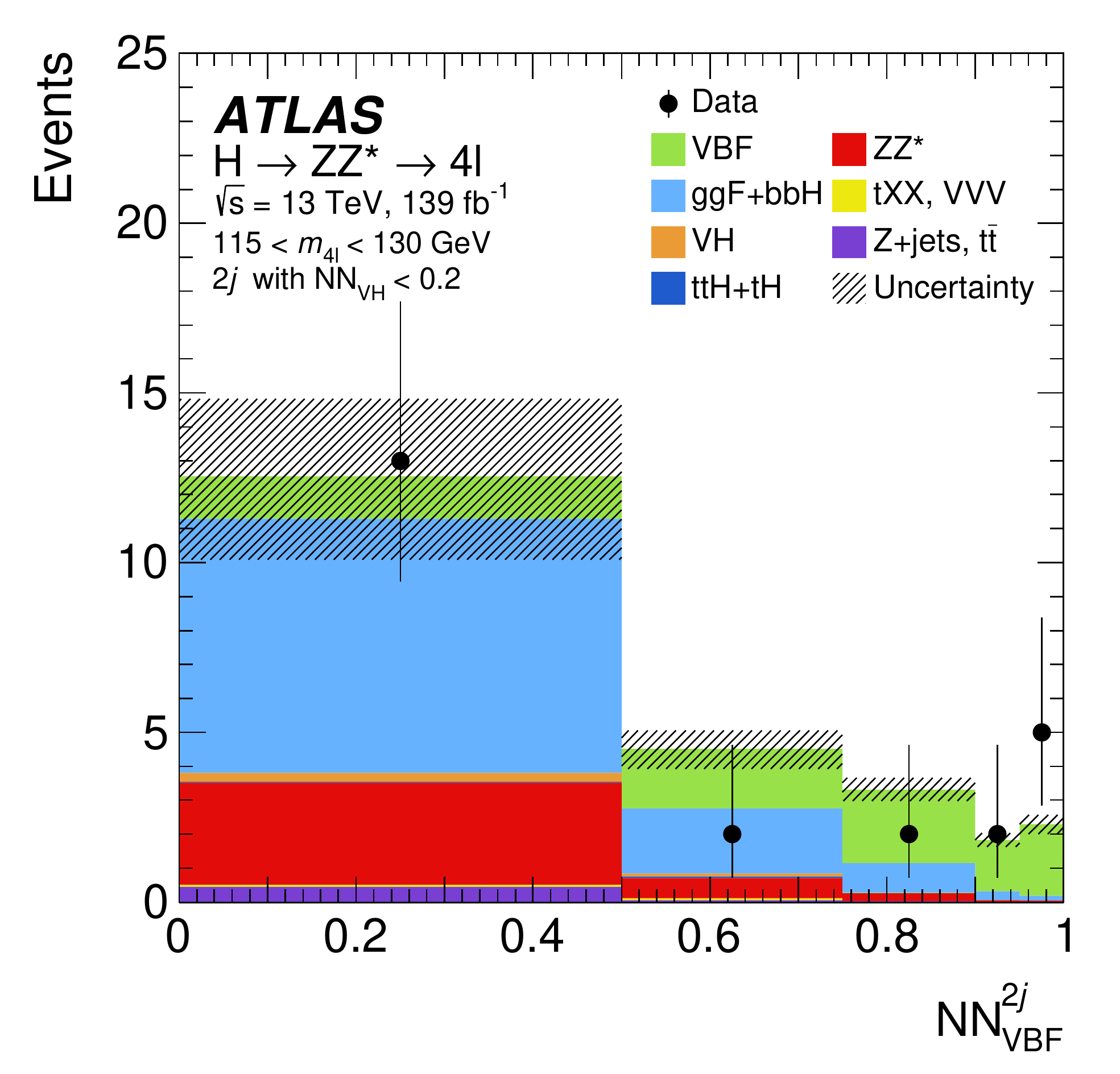}\label{fig:oNN2jvbf}}
\subfloat[]{
\includegraphics[width=0.33\linewidth]{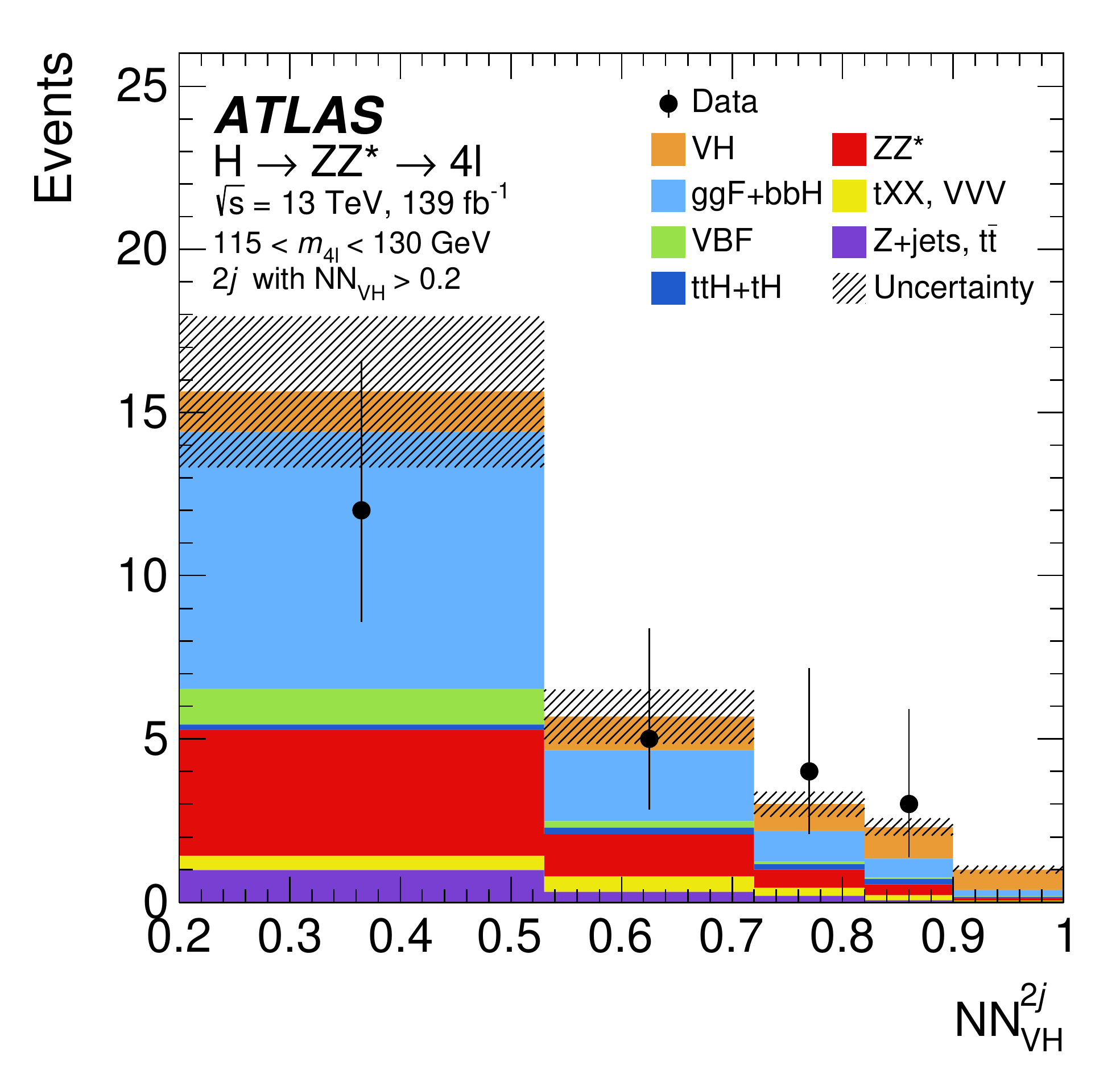}\label{fig:oNN2jvh}}
\subfloat[]{
\includegraphics[width=0.33\linewidth]{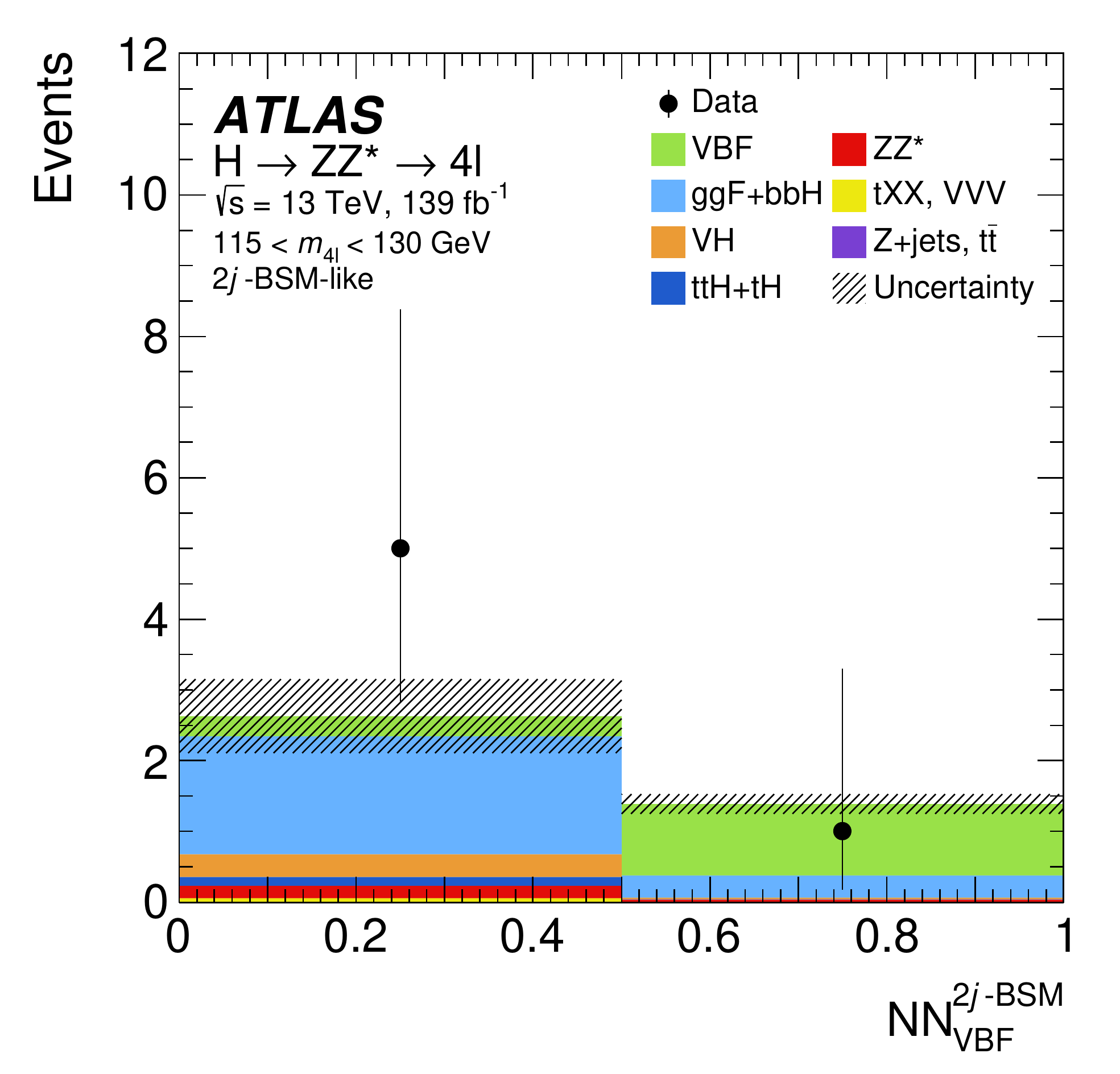}\label{fig:oNNbsmvbf}}\\
\subfloat[]{
\includegraphics[width=0.33\linewidth]{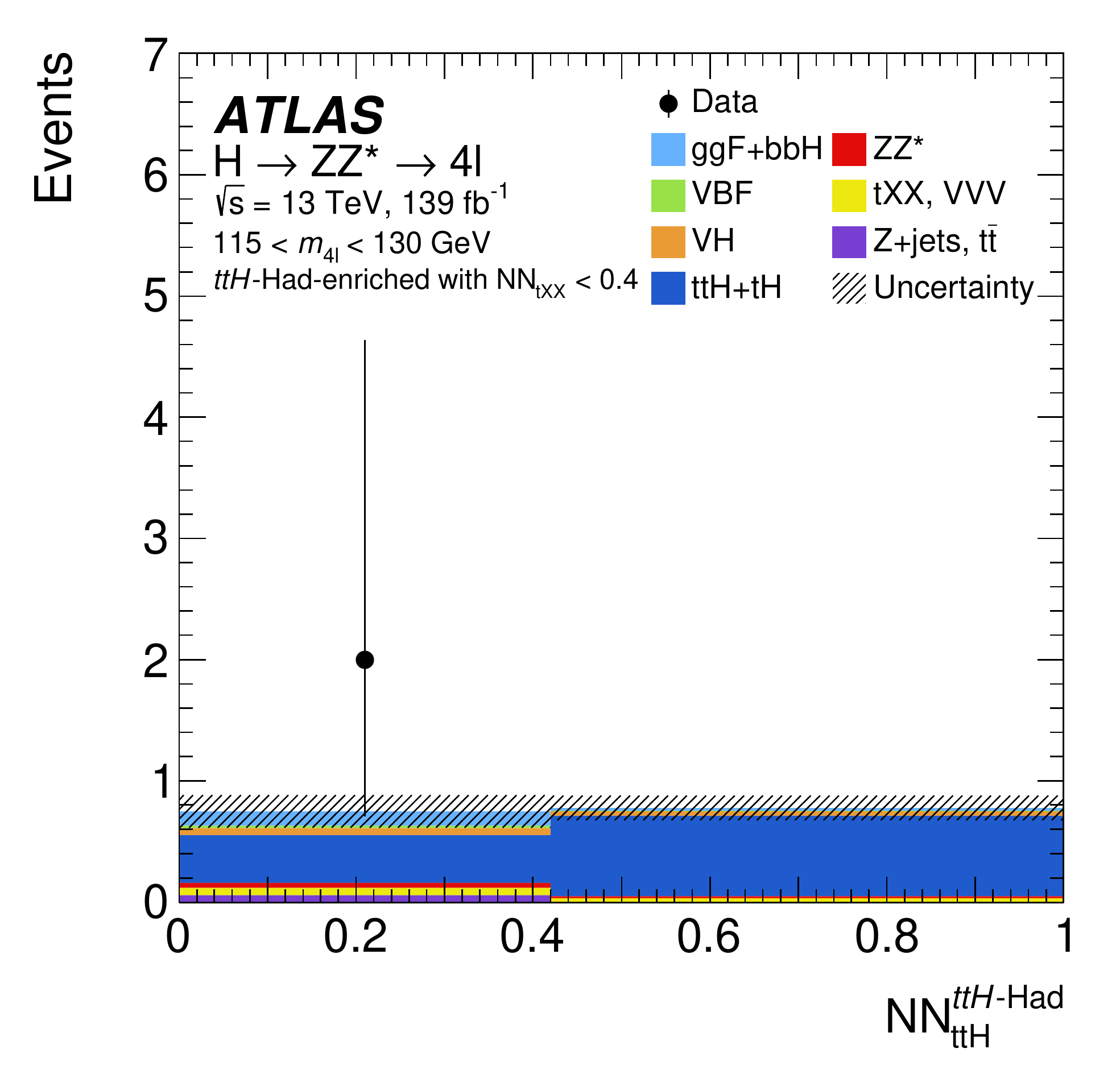}\label{fig:oNNtthhadtxxl}}
\subfloat[]{
\includegraphics[width=0.33\linewidth]{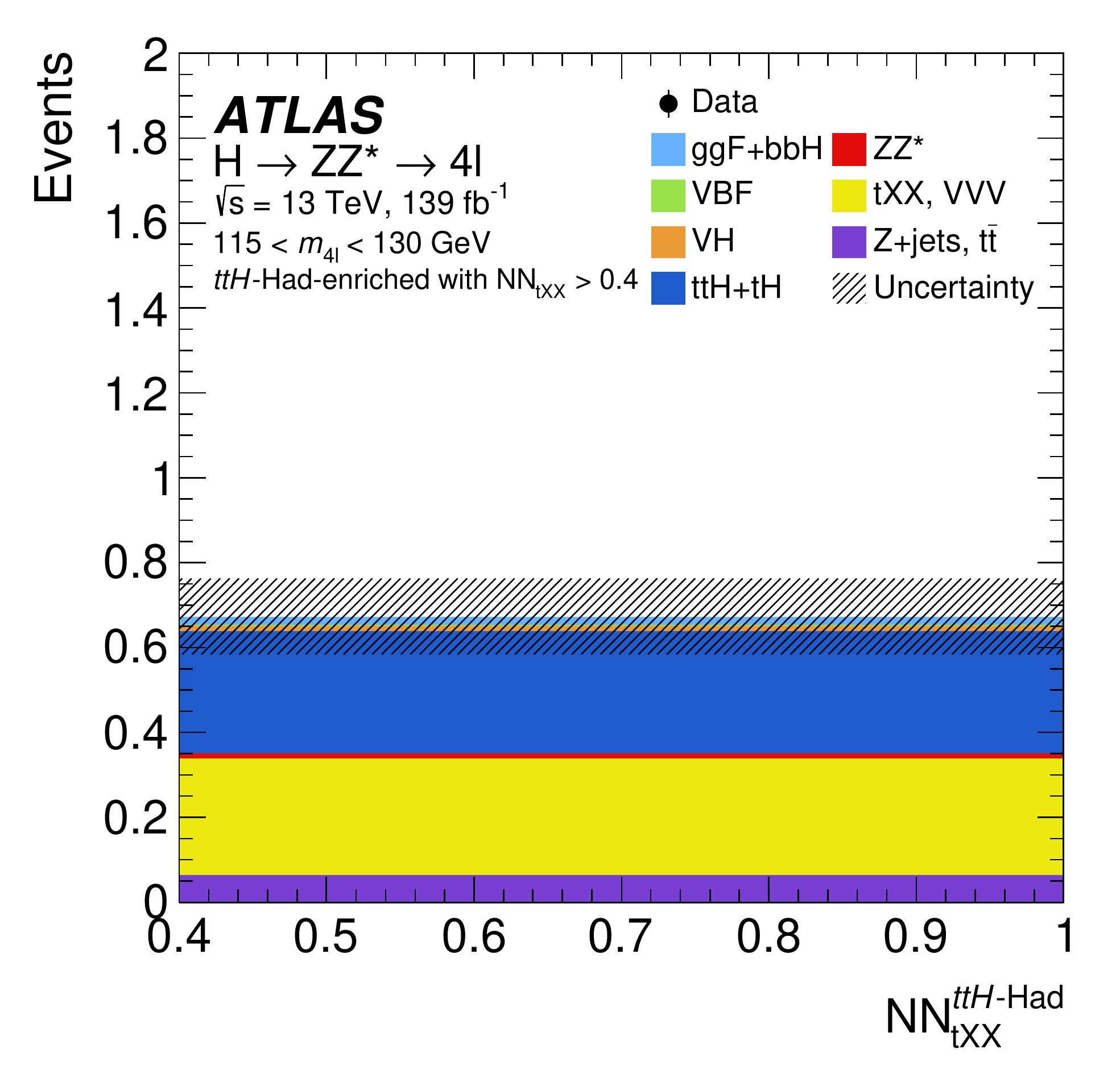}\label{fig:oNNtthhadtxxh}}
\subfloat[]{
\includegraphics[width=0.33\linewidth]{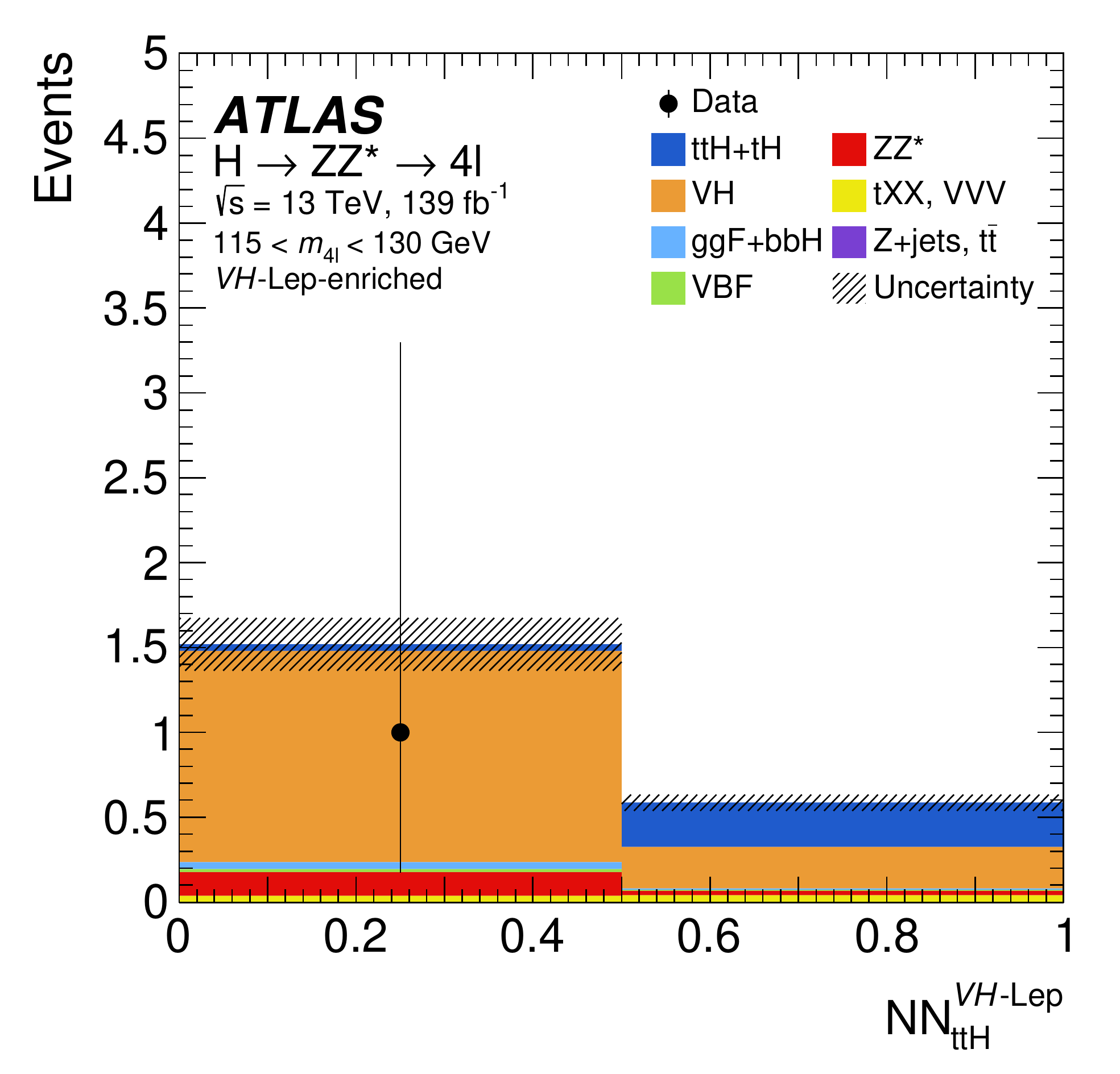}\label{fig:oNNvhleptth}}\\
\subfloat[]{
\includegraphics[width=0.33\linewidth]{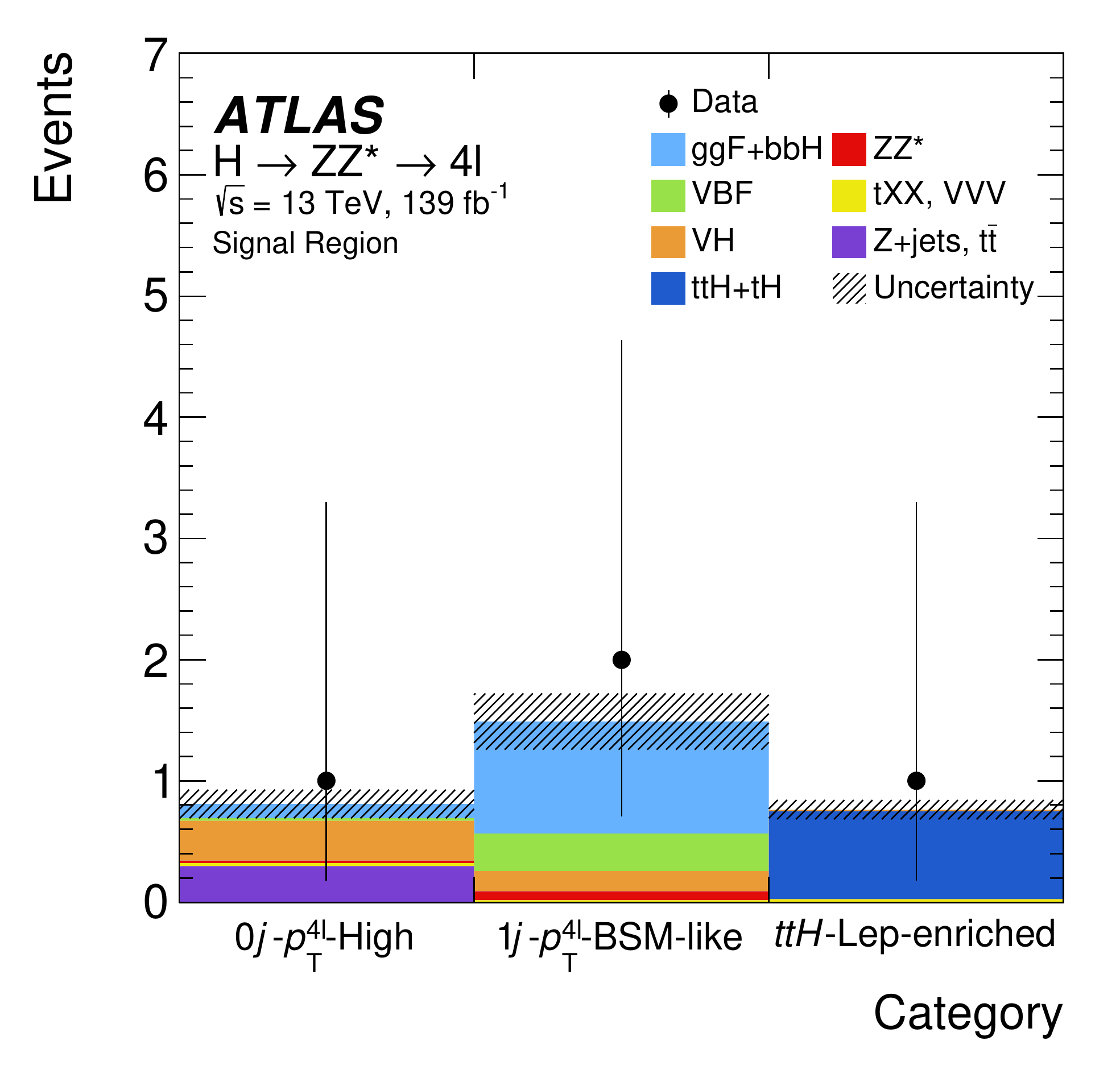}\label{fig:SRcount}}
\subfloat[]{
\includegraphics[width=0.33\linewidth]{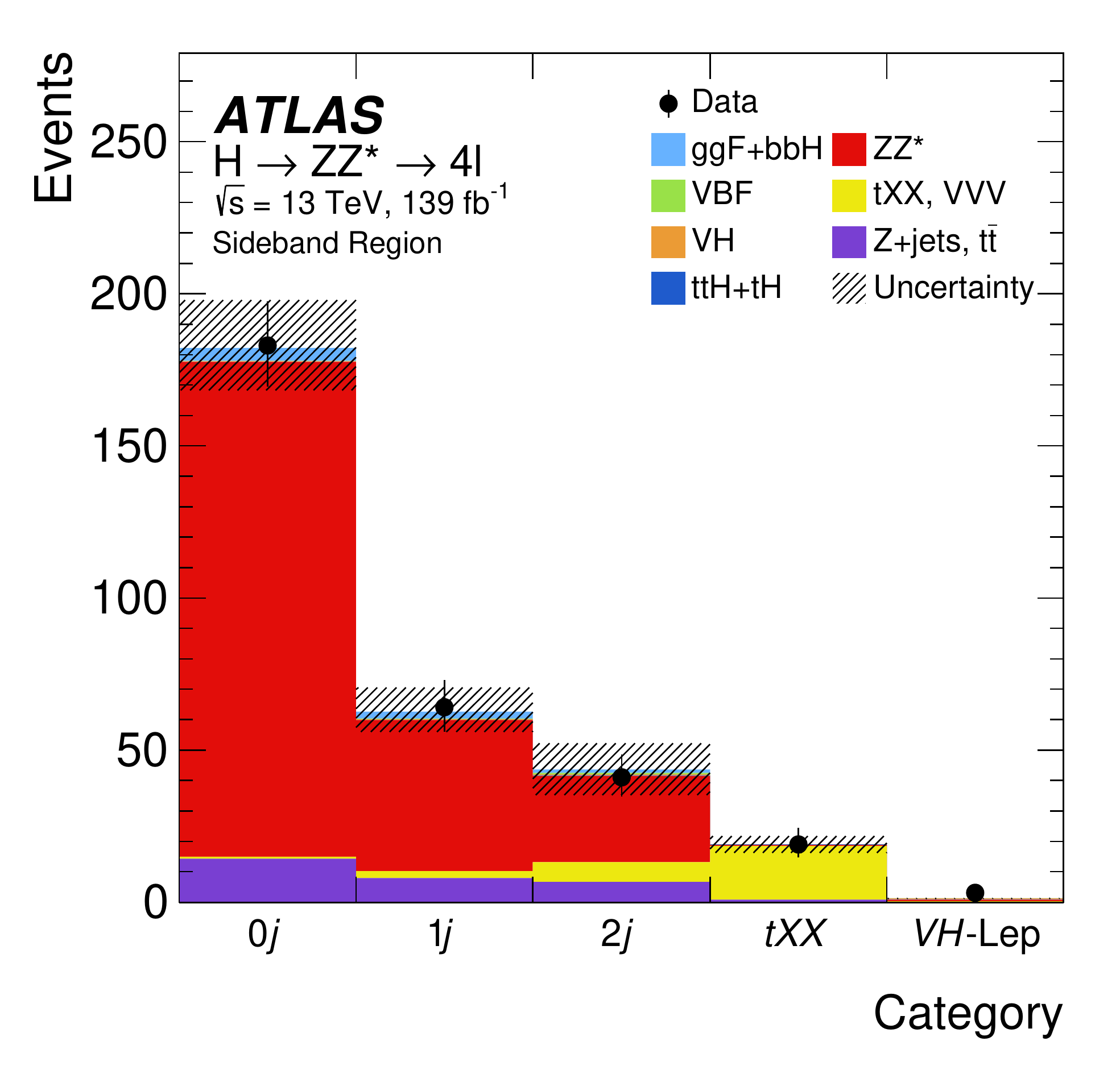}\label{fig:SBcount}}\\
\end{center}
\vskip-0.2cm
\caption{The observed and expected NN output (post-fit) distributions for an integrated luminosity of \Lum\ at $\sqrt{\mathrm{s}}=$~13~\TeV\ in the different categories: (a) $\mathrm{NN_{VBF}}$ in \CatTwoJ\ with $\mathrm{NN}_{VH}<0.2$, (b) $\mathrm{NN}_{VH}$ in \CatTwoJ\ with  $\mathrm{NN}_{VH}>0.2$, (c) $\mathrm{NN_{VBF}}$  in \CatTwoJBSM,  (d) $\mathrm{NN}_{ttH}$ in \CatttHhad\ with $\mathrm{NN}_{tXX}<0.4$,  (e) $\mathrm{NN}_{tXX}$ in \CatttHhad\ with  $\mathrm{NN}_{tXX}>0.4$ and (f) $\mathrm{NN}_{ttH}$ in \CatVHLep.
(g) shows the categories where no NN discriminant is used while (h) shows the sidebands used to constrain the \zzstar\  and $tXX$ backgrounds. The SM Higgs boson signal is assumed to have a mass $m_{H}$ = 125~\gev. The uncertainty in the prediction is shown by the hatched band, calculated as described in Section~\ref{sec:systematics}. \TheoryBlurb
The bin boundaries are chosen to maximise the significance of the targeted signal in each category.}
\label{fig:NNoutputs_observed2}
\end{figure}

The expected numbers of signal and background events in each reconstructed event category are shown in Table~\ref{tab:event_yields_all} together with the corresponding observed number of events. The expected event yields are in good agreement with the observed ones.
\begin{table*}[!htbp]
\centering
\caption{The expected (pre-fit) and observed numbers of events for an integrated luminosity of \Lum\ at $\sqrt{\mathrm{s}}=$~13~\TeV\ in the signal region \mbox{$115<m_{4\ell}<130$~\GeV\ } and sideband region \mbox{$105<m_{4\ell}<115$~\GeV\ } or \mbox{$130<m_{4\ell}<160$~\GeV\ (350~\GeV\ for \tXX-enriched) } in each reconstructed event category assuming the SM Higgs boson signal with a mass $m_{H}$ = 125~\GeV. The sum of the expected number of SM Higgs boson events and the estimated background yields is compared with the data. Combined statistical and systematic uncertainties are included for the predictions. Expected contributions that are below 0.2\% of the total yield in each reconstructed event category are not shown and replaced by `-'.}
\label{tab:event_yields_all}
\vspace{0.1cm}
\vspace{0.1cm}
{\renewcommand{\arraystretch}{1.1}
\resizebox{\linewidth}{!}{
\begin{tabular}{l ccccccc}
\hline\hline
Reconstructed&
\multicolumn{1}{c}{Signal}&
\multicolumn{1}{c}{\zzstar}&
\multicolumn{1}{c}{$tXX$}&
\multicolumn{1}{c}{Other}&
\multicolumn{1}{c}{Total}&
\multicolumn{1}{c}{Observed}\\
 
event category&
&
\multicolumn{1}{c}{background}&
\multicolumn{1}{c}{background}&
\multicolumn{1}{c}{backgrounds}&
\multicolumn{1}{c}{expected}&
\\
 
\hline
 
Signal&
\multicolumn{6}{c}{$115<m_{4\ell}<130$ \GeV}\\  \hline
 
\CatZeroJL
& $   24.2 \pm    3.5~~$
& $     30 \pm      4~~ $
& $               -   $
& $   0.93 \pm   0.13 $
& $     55 \pm      5~~ $
& $~~56$ \\
\CatZeroJM
& $     76 \pm      8~~ $
& $     37 \pm      4~~ $
& $               -   $
& $    6.5 \pm    0.6 $
& $    120 \pm      9~~~~ $
& $117$ \\
\CatZeroJH
& $  0.355 \pm  0.031 $
& $  0.020 \pm  0.012 $
& $ 0.0094 \pm 0.0027 $
& $   0.30 \pm   0.05 $
& $   0.69 \pm   0.06 $
& $~~~~1$ \\
\CatOneJL
& $     34 \pm      4~~ $
& $   15.5 \pm    2.7~~ $
& $               - $
& $   1.91 \pm   0.29 $
& $     52 \pm      5~~ $
& $~~41$ \\
\CatOneJM
& $   20.8 \pm    2.8~~ $
& $    4.0 \pm    0.7 $
& $  0.114 \pm  0.013 $
& $   1.02 \pm   0.19 $
& $   26.0 \pm    2.9~~ $
& $~~31$ \\
\CatOneJH
& $    4.7 \pm    0.8 $
& $   0.48 \pm   0.10 $
& $  0.043 \pm  0.008 $
& $   0.27 \pm   0.04 $
& $    5.5 \pm    0.8 $
& $~~~~4$ \\
\CatOneJBSM
& $   1.23 \pm   0.23 $
& $  0.069 \pm  0.031 $
& $ 0.0067 \pm 0.0031 $
& $  0.062 \pm  0.012 $
& $   1.37 \pm   0.23 $
& $~~~~2$ \\
\CatTwoJ
& $     38 \pm      5~~ $
& $    9.1 \pm    2.7 $
& $   0.95 \pm   0.08 $
& $   2.13 \pm   0.31 $
& $     50 \pm      6~~$
& $~~48$ \\
\CatTwoJBSM
& $    3.3 \pm    0.6 $
& $   0.18 \pm   0.06 $
& $  0.032 \pm  0.005 $
& $  0.091 \pm  0.017 $
& $    3.6 \pm    0.6 $
& $~~~~6$ \\
\CatVHLep
& $   1.29 \pm   0.07 $
& $  0.156 \pm  0.025 $
& $  0.039 \pm  0.009 $
& $ 0.0194 \pm 0.0032 $
& $   1.50 \pm   0.08 $
& $~~~~1$ \\
\CatttHhad
& $   1.02 \pm   0.18 $
& $  0.058 \pm  0.025 $
& $  0.252 \pm  0.032 $
& $  0.119 \pm  0.033 $
& $   1.45 \pm   0.18 $
& $~~~~2$ \\
\CatttHlep
& $   0.42 \pm   0.04 $
& $  0.002 \pm  0.005 $
& $ 0.0157 \pm 0.0023 $
& $ 0.0028 \pm 0.0029 $
& $   0.44 \pm   0.04 $
& $~~~~1$ \\

\hline
Sideband &
\multicolumn{5}{c}{$105<m_{4\ell}<115$ \GeV~or $130<m_{4\ell}<160$ \GeV}\\
\hline
 
\CatSBZeroJ
& $    4.5 \pm    0.5 $
& $    150 \pm     13~~ $
& $                -  $
& $   16.2 \pm    2.2~~ $
& $    171 \pm     13~~$
& $183$ \\
\CatSBOneJ
& $   2.80 \pm   0.30 $
& $     51 \pm      7~~ $
& $   1.29 \pm   0.16 $
& $    8.4 \pm    1.2 $
& $     63 \pm      7~~ $
& $~~64$ \\
\CatSBTwoJ
& $   2.02 \pm   0.27 $
& $     25 \pm      7~~ $
& $    4.4 \pm    0.5 $
& $    6.0 \pm    0.9 $
& $     38 \pm      7~~$
& $~~41$ \\
\CatSBVHL
& $  0.273 \pm  0.015 $
& $   0.48 \pm   0.06 $
& $  0.125 \pm  0.018 $
& $  0.126 \pm  0.019 $
& $   1.00 \pm   0.07 $
& $~~~~3$ \\

\hline
& \multicolumn{5}{c}{$105<m_{4\ell}<115$ \GeV~or $130<m_{4\ell}<350$ \GeV}\\
\hline
 
\CatSBttV
& $  0.071 \pm  0.012 $
& $   0.32 \pm   0.12 $
& $   12.1 \pm    1.3~~ $
& $   0.84 \pm   0.33 $
& $   13.3 \pm    1.4~~ $
& $~~19$ \\

\hline \hline
\noalign{\vspace{0.05cm}}
\label{tab:yields_obs_sum}
\end{tabular}
}}
\end{table*}
The observed  and expected (post-fit) distributions of the NN discriminants are shown in \figref{fig:NNoutputs_observed} and in \figref{fig:NNoutputs_observed2}. In addition, \figref{fig:SRcount} and \figref{fig:SBcount} show the observed and expected yields in the categories where no NN discriminant is used and in the mass sidebands used to constrain the \zzstar\ and $tXX$ background, respectively. All distributions are in good agreement with the data.
 
The statistical interpretation of the results and compatibility with the SM are discussed in the following.
 
\label{subsec:stxs_inclusive}
 
\subsection{Measurement of simplified template cross-sections}
\label{subsec:stxs_stxs}
 
To measure the product $\vec{\sigma}\cdot$\BR of the Higgs boson production cross-section and the branching ratio for $H \rightarrow ZZ^*$ decay for the production bins of the Production Mode Stage or the Reduced Stage 1.1, a fit to the discriminant observables introduced in Section~\ref{subsec:cat_nn} is performed using the likelihood function ${\cal L}( \vec{\sigma}, \vec{\theta})$ that depends on the Higgs boson production cross-section $\vec{\sigma} = \{ \sigma_{1}, \sigma_{2}, \ldots , \sigma_{N} \} $ where $\sigma_p$ is the cross-section in each production bin $p$
and the nuisance parameters $\vec{\theta}$ accounting for the systematic uncertainties. The likelihood function is defined as a product of conditional probabilities over binned distributions of the discriminating observables in each reconstructed signal and sideband event category $j$,
\begin{equation}
\begin{split}
& \mathcal{L}(\vec{\sigma}, \vec{\theta}) =  \displaystyle  \prod_{j}^{N_{\mathrm{categories}}} \prod_{i}^{N_{\mathrm{bins}}}
P\left(N_{i,j} ~| ~L\cdot\vec{{\sigma}} \cdot \BR \cdot  \vec{A}_{i, j}(\vec{\theta}) ~+  B_{i,j}(\vec{\theta}) \right)  \times \prod_{m}^{N_{\mathrm{nuisance}}} \mathcal{C}_{m}(\vec{\theta}) \;,
\end{split}
\label{eq:likelihood_simplified}
\end{equation}
with Poisson distributions $P$ corresponding to the observation of $N_{i,j}$ events in each histogram bin $i$ of the\linebreak discriminating  observable given the expectations for each background process, $B_{i,j}(\vec{\theta})$, and for the signal,\linebreak $S_{i,j}(\vec{\theta}) = L \cdot\vec{{\sigma}} \cdot \BR \cdot \vec{A}_{i,j}(\vec{\theta})$, where $L$ is the integrated luminosity and $\vec{A}_{i,j} = \{ A_{i,j}^{1}, A_{i,j}^{2}, \ldots , A_{i,j}^{N} \} $ is the set of signal acceptances from each production bin. The signal acceptance $A_{i,j}^{p}$ is defined as the fraction of generated signal events in the production bin $p$ that satisfy the event reconstruction and selection criteria in the histogram bin $i$ of the reconstructed event category $j$. For a given production bin $p$, the acceptance consists of $A_{i,j}^{p} = a^{p} \cdot \epsilon_{i,j}^{p}$, where $a^{p}$ is the particle-level acceptance in the fiducial region defined from requirements listed in Sections~\ref{sub:selection} and~\ref{sec:cat} and $\epsilon_{i,j}^{p}$ is the reconstruction efficiency of these particle-level events. Constraints on the nuisance parameters corresponding to systematic uncertainties described in Section~\ref{sec:systematics} are
represented by the functions $\mathcal{C}_{m}(\vec{\theta})$. The cross-sections are treated as independent parameters for each production bin and correlated among the different reconstructed event categories.
The test statistic used to perform the measurements is the ratio of profile likelihoods~\cite{Cowan:2010js},
\[
q(\vec{\sigma}) = -2 \ln \frac{\mathcal{L}(\vec{\sigma}, \hat{\hat{\vec{\theta}}}(\vec{\sigma})) } {\mathcal{L}(\hat{{\vec{\sigma}}}, \hat{{\vec{\theta}}} )} = -2 \ln \lambda(\vec{\sigma})\;,
\]
where $\vec{\sigma}$ represents only the cross-section(s) considered as parameter(s) of interest in a given fit. The likelihood in the numerator is the estimator of a conditional fit, i.e.\ with parameter(s) of interest ${\sigma_i}$ fixed to a given value, while the remaining cross-sections and nuisance parameters are free-floating parameters in the fit. The values of the nuisance parameters $\hat{\hat{\vec{\theta}}}(\vec{\sigma}))$ maximise the likelihood on the condition that the parameters of interest are held fixed to a given value. The likelihood in the denominator is the estimator of an unconditional fit in which all $\vec{\sigma}$ and $\vec{\theta}$ parameters are free parameters of the fit.
The parameter of interest $\sigma$ in each production bin is alternatively replaced by $\mu \cdot \sigma_{\mathrm{SM}}(\vec{\theta})$, allowing an interpretation in terms of the signal strength $\mu$ relative to the SM prediction $\sigma_{\mathrm{SM}}(\vec{\theta})$. 
 
Assuming that the relative signal fractions in each production bin are given by the predictions for the SM Higgs boson, the inclusive $H \to ZZ^{*}$ production cross-section for $|y_H|<2.5$ is measured to be:
\begin{equation*}
\sBR \equiv \sBRZZ~=~{\color{black}{
1.34 \pm 0.11(\mathrm{stat.})\pm 0.04(\mathrm{exp.})\pm 0.03(\mathrm{th.})~\mathrm{pb}~=
~1.34 \pm 0.12~ \mathrm{pb}}},
\end{equation*}
where the uncertainties are either statistical (stat.) or of experimental (exp.) or theoretical (th.) systematic nature.
 
The SM prediction is \sBRsm~$\equiv~$\sBRZZsm~$= 1.33 \pm 0.08$~pb.
The data are also interpreted in terms of the global signal strength, yielding
\begin{equation*}
\mu = 1.01\pm 0.08(\mathrm{stat.}) \pm0.04(\mathrm{exp.})\pm0.05(\mathrm{th.}) = 1.01\pm0.11.
\end{equation*}
The measured cross-section and signal strength are in an excellent agreement with the SM prediction, with a $p$-value of 98.6\% for both compatibility tests.
 
\begin{figure}[!htbp]
\begin{center}
\subfloat[]{
\includegraphics[width=0.45\linewidth]{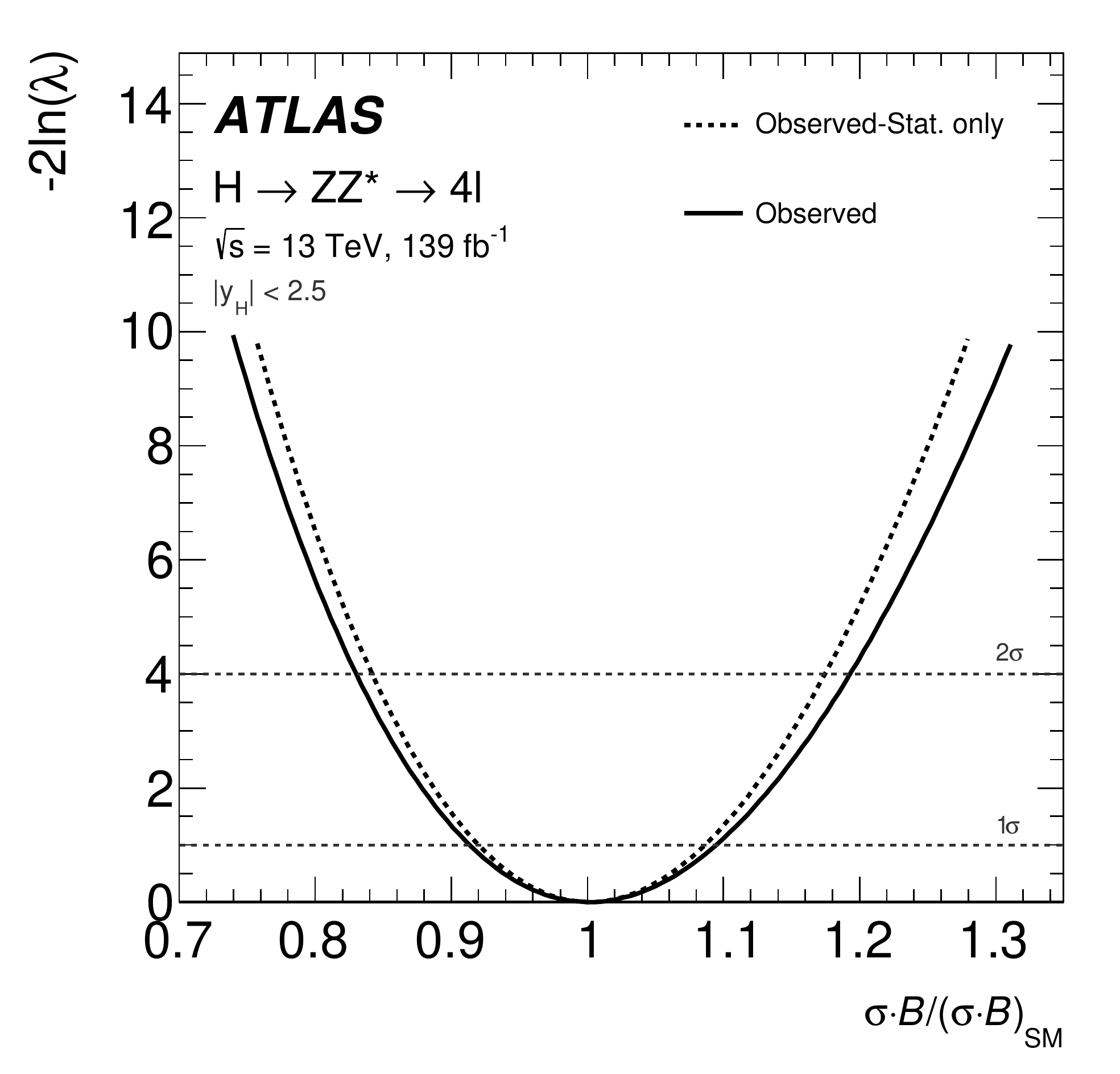}\label{fig:totXS}}
\subfloat[]{
\includegraphics[width=0.45\linewidth]{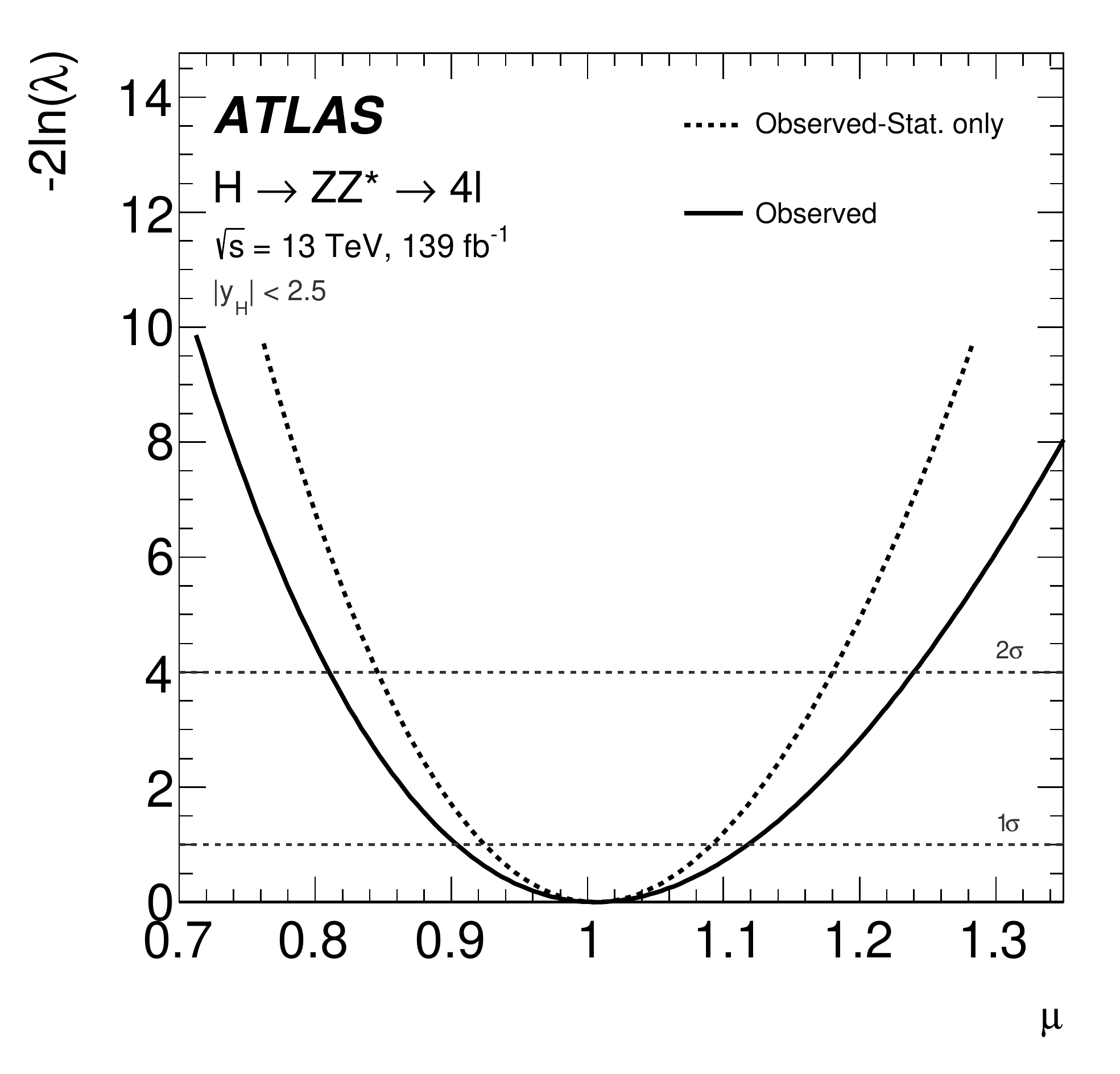}\label{fig:totMU}}
\end{center}
\vskip-0.4cm
\caption{Observed profile likelihood as a function of (a) \sBRZZ normalised by the SM expectation and (b) the inclusive signal strength $\mu$; the scans are shown both with (solid line) and without (dashed line) systematic uncertainties. }
\label{fig:incl_results}
\end{figure}

The corresponding likelihood functions are shown in \figref{fig:incl_results}. The dominant systematic uncertainty in the cross-section measurement is the experimental uncertainty in the lepton efficiency and integrated luminosity measurements and theoretical uncertainties related to parton shower modelling affecting the acceptance. The signal-strength measurement is also affected by the theoretical uncertainty in the \ggF cross-section due to missing higher-order corrections in QCD.
 
The expected SM cross-section, the observed values of \sBRZZ and their ratio for the inclusive production and in each production bin of the Production Mode Stage and the Reduced Stage 1.1 are shown in Table~\ref{tab:XS_cat_result}.
\begin{table*}[!htbp]
\def\arraystretch{1.8}
\centering
\caption{The expected SM cross-section \sBRsm, the observed value of \sBR, and their ratio \sBRratio for the inclusive production and for each Production Mode Stage and Reduced Stage-1.1 production bin for the $H\to ZZ^*$ decay for an integrated luminosity of \Lum\ at $\sqrt{\mathrm{s}}=13$~\TeV. The \bbH ($tH$) contribution is included in the  \STXSggF (\STXSttH) production bins. The uncertainties are given as (stat.)+(exp.)+(th.) for the inclusive cross-section and the Production Mode Stage, and as (stat.)+(syst.) for the Reduced Stage 1.1. The Reduced Stage-1.1 results are dominated by the statistical uncertainty and the impact of theory uncertainties is smaller than for the Production Mode Stage. The impact of the theory uncertainties for the Reduced Stage 1.1 is smaller than the least significant digit.
}
\label{tab:XS_cat_result}
\vspace{0.1cm}
{\renewcommand{\arraystretch}{1.5}
 
\begin{tabular}{l | c c | c }
\hline\hline
Production bin &
\multicolumn{2}{c|}{ Cross-section (\sBR) [pb]} &
\sBRratio
\\
&
\multicolumn{1}{c}{SM expected} & \multicolumn{1}{c|}{Observed} &  \multicolumn{1}{c}{Observed} \\
\hline
 
\multicolumn{4}{c}{Inclusive production, $|y_{H}|<2.5$}\\
\hline
&
$1.33 \pm 0.08$ &
$1.34 \pm 0.11 \pm 0.04 \pm 0.03$ &
$1.01 \pm 0.08 \pm 0.03 \pm 0.02$ \\
\hline
 
\multicolumn{4}{c}{Production Mode Stage bins, $|y_{H}|<2.5$}\\
\hline
\STXSggF	  &
$1.17 \pm 0.08$ &
$1.12 \pm 0.12 \pm 0.04 \pm 0.03$ &
$0.96 \pm 0.10 \pm 0.03 \pm 0.03$\\
\STXSVBF	  &
$0.0920 \pm 0.0020$ &
$0.11 \pm 0.04 \pm 0.01 \pm 0.01$ &
$1.21 \pm 0.44$ $^{+0.13}_{-0.08}$ $^{+0.07}_{-0.05}$\\
\STXSVH	  &
$0.0524^{+0.0027}_{-0.0049}$ &
$0.075^{+0.059}_{-0.047}$ $^{+0.011}_{-0.007}$ $^{+0.013}_{-0.009}$&
$1.44^{+1.13}_{-0.90}$ $^{+0.21}_{-0.14}$ $^{+0.24}_{-0.17}$ \\
\STXSttH	  &
$0.0154^{+0.0010}_{-0.0013}$ &
$0.026^{+0.026}_{-0.017} \pm 0.002 \pm 0.002$ &
$1.7^{+1.7}_{-1.2} \pm 0.2 \pm 0.2$\\
\hline
 
\multicolumn{4}{c}{Reduced Stage-1.1 bins, $|y_{H}|<2.5$}\\
\hline
\STXSggToHZeroJL    &
$0.176 \pm 0.025$ &
$0.17 \pm 0.05 \pm 0.02$ &
$0.96 \pm 0.30 \pm 0.09$\\
\STXSggToHZeroJH    &
$0.55 \pm 0.04$ &
$0.63 \pm 0.09 \pm 0.06$ &
$1.15 \pm 0.17 \pm 0.11$\\
\STXSggToHOneJL     &
$0.172 \pm 0.025$ &
$0.05 \pm 0.07$ $^{+0.04}_{-0.06}$ &
$0.3 \pm 0.4$ $^{+0.2}_{-0.3}$\\
\STXSggToHOneJM     &
$0.119 \pm 0.018$ &
$0.17 \pm 0.05$ $^{+0.02}_{-0.01} $ &
$1.4 \pm 0.4 \pm 0.1$\\
\STXSggToHOneJH     &
$0.020 \pm 0.004$ &
$0.009^{+0.016}_{-0.011} \pm 0.002$ &
$0.5^{+0.8}_{-0.6} \pm 0.1$\\
\STXSggToHTwoJ      &
$0.127 \pm 0.027$ &
$0.04 \pm 0.07 \pm 0.04 $&
$0.3 \pm 0.5 \pm 0.3 $\\
\STXSggToHHigh      &
$0.015 \pm 0.004$ &
$0.038^{+0.021}_{-0.016}$ $^{+0.003}_{-0.002}$ &
$2.5^{+1.3}_{-1.0}$ $^{+0.2}_{-0.1}$\\
\STXSqqtoHqqVHLike       &
$0.0138^{+0.0004}_{-0.0006}$ &
$0.021^{+0.037}_{-0.029}$ $^{+0.009}_{-0.006}$ &
$1.5^{+2.7}_{-2.1}$ $^{+0.6}_{-0.4}$\\
\STXSqqtoHqqRest      &
$0.1076^{+0.0024}_{-0.0035}$  &
$0.15 \pm 0.05$ $^{+0.02}_{-0.01} $ &
$1.4 \pm 0.5$ $^{+0.2}_{-0.1}$\\
\STXSqqtoHqqBSM        &
$0.00420 \pm 0.00018$ &
$0.0005^{+0.0079}_{-0.0047} \pm 0.008$ &
$0.1^{+1.9}_{-1.1} \pm 0.2$\\
\STXSVHLep        &
$0.0164 \pm 0.0004$ &
$0.022^{+0.028}_{-0.018}$ $^{+0.003}_{-0.001}$ &
$1.3^{+1.7}_{-1.1}$ $^{+0.2}_{-0.1}$\\
\STXSttH          &
$0.0154^{+0.0010}_{-0.0013}$ &
$0.025^{+0.026}_{-0.017}$ $^{+0.005}_{-0.003}$ &
$1.6^{+1.7}_{-1.1}$ $^{+0.3}_{-0.2}$ \\
 
\hline\hline
\end{tabular}
}
\end{table*}
 
\begin{figure}[!htbp]
\centering
\subfloat[]{
\includegraphics[width=0.50\linewidth]{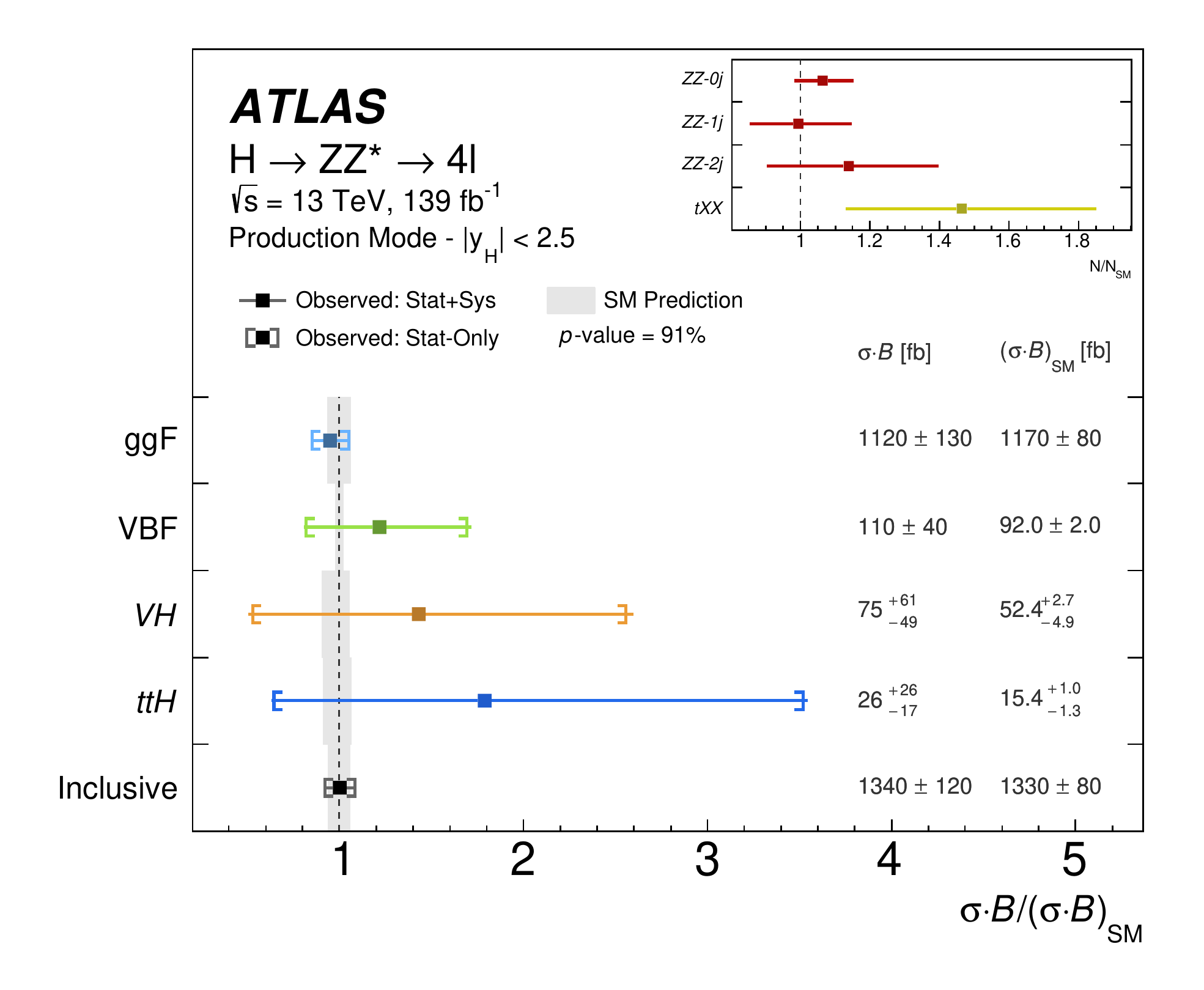}\label{fig:stage0sigxb}}
\subfloat[]{
\includegraphics[width=0.45\linewidth]{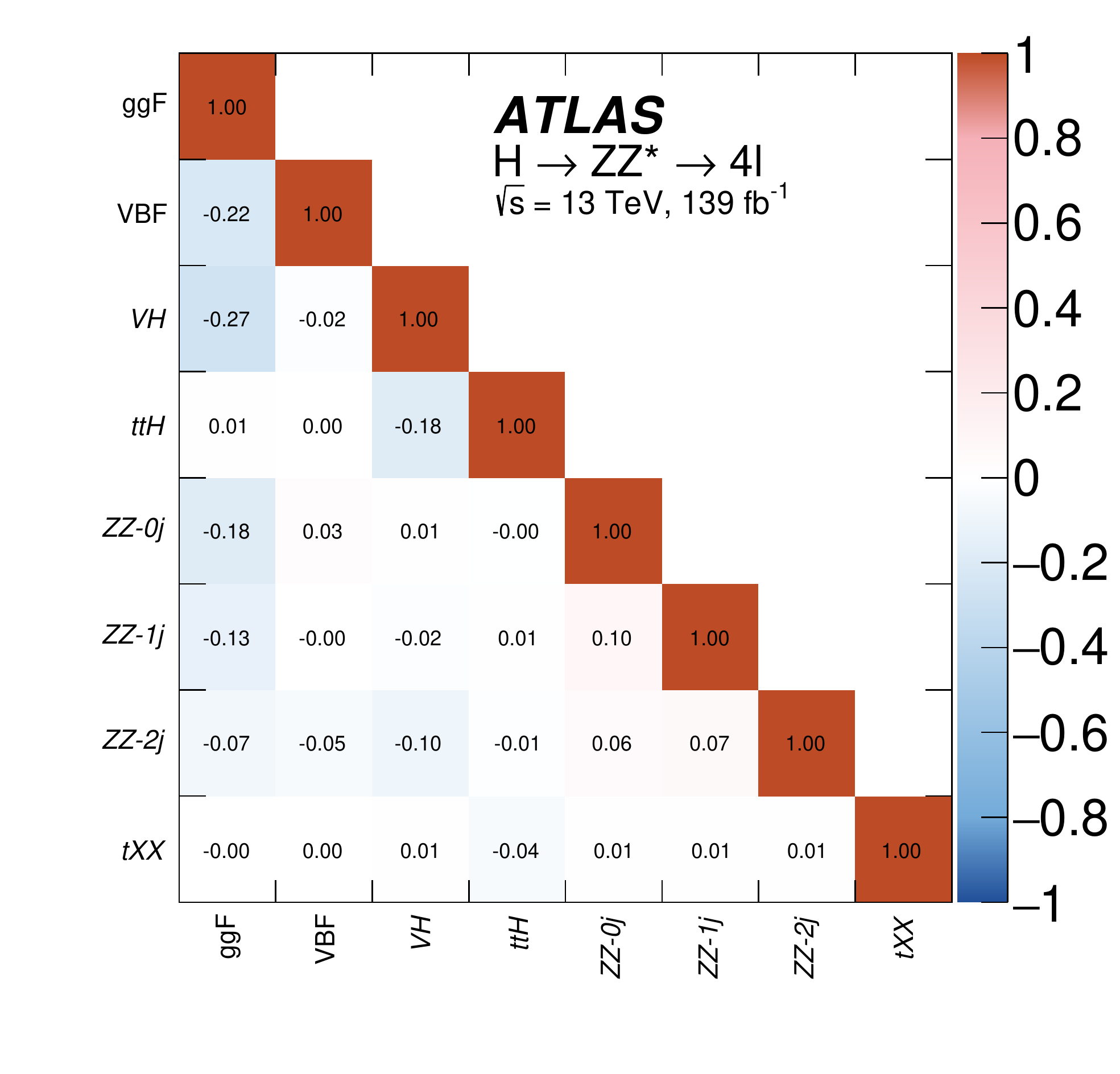}\label{fig:stage0corr}}\\
\subfloat[]{
\includegraphics[width=0.50\linewidth]{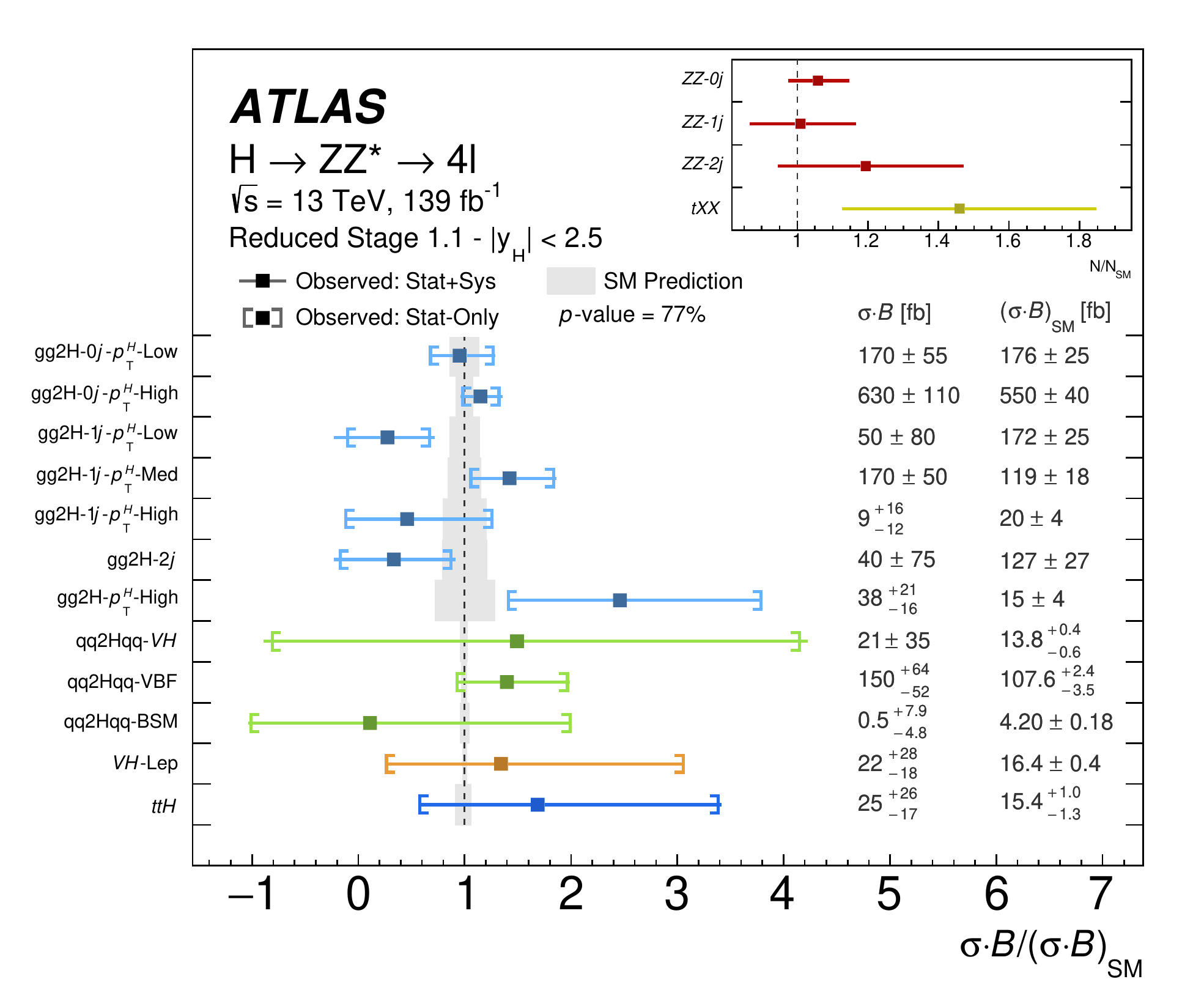}\label{fig:stage1sigxb}}
\subfloat[]{
\includegraphics[width=0.45\linewidth]{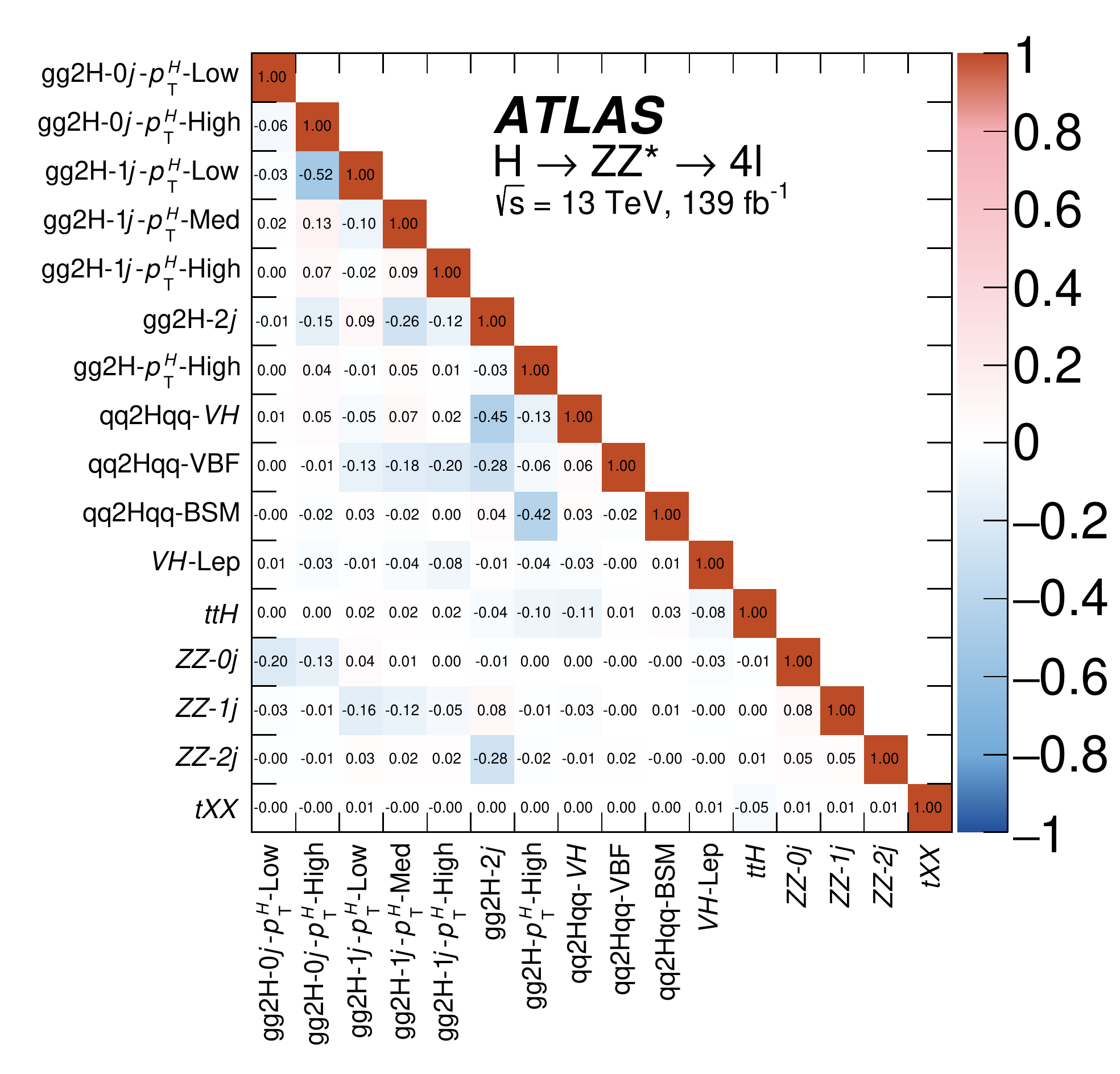}\label{fig:stage1corr}}
\caption{The observed and expected SM values of the cross-sections \sBR normalised by the SM expectation \sBRsm for (a) the inclusive production and in the  Production Mode Stage and (c) the Reduced Stage-1.1 production bins for an integrated luminosity of \Lum\ at $\sqrt{\mathrm{s}}=13$~\TeV. The fitted normalisation factors for the $ZZ$ and $tXX$ background are shown in the inserts. Different colours indicate different Higgs boson production modes (or background sources). The vertical band represents the theory uncertainty in the signal prediction. The correlation matrices between the measured cross-sections and the $ZZ$ and $tXX$ normalisation factors are shown  for (b) the Production Mode Stage and (d) the Reduced Stage 1.1.
}
\label{fig:Stage01_results}
\end{figure}

The corresponding values are summarised in \figref{fig:Stage01_results}.
In the ratio calculation, uncertainties in the SM expectation are not taken into account. The Production Mode Stage and Reduced Stage-1.1 measurements agree with the predictions for the SM Higgs boson. The $p$-values of the corresponding compatibility tests are 91\% and 77\%, respectively.

For the \STXSqqtoHqqRest bin, most of the sensitivity to the \VBF production mode comes from the phase space with $m_{jj} > 350$~\GeV\ and $\pt^H < 200$~\GeV. To probe the \VBF contribution more directly, the cross-sections in this and in the remaining phase space region of the \STXSqqtoHqqRest bin are fitted separately to the data, simultaneously with the other Reduced Stage 1.1 bins, using the reconstruction categories described in \secref{sec:cat}. The cross-section in the $m_{jj} > 350$~\GeV\ and $\pt^H < 200$~\GeV~phase space is measured to be $0.060^{+0.025}_{-0.020}\ \textrm{pb}$  compared with the predicted cross-section of  $0.0335 ^{+0.0007} _{-0.0011}\ \textrm{pb}$. This measurement has a correlation of 20\% with the measurement in the \STXSggToHTwoJ bin, while correlations with other bins are up to 50\%.

The dominant contribution to the measurement uncertainty in the \ggF Production Mode Stage bin originates from the same sources as in the inclusive measurement. For the \VBF production bin, the dominant systematic uncertainties are related to parton showering modelling and jet energy scale and resolution uncertainties. The \STXSVBF, \STXSVH and \STXSttH production bins are also affected by the theoretical uncertainties related to the modelling of the \ggF process. For the Reduced Stage-1.1
bins, the dominant cross-section uncertainties are the jet energy scale and resolution, and  parton shower uncertainties.
 
\begin{figure}[!htbp]
\begin{center}
\subfloat[]{\includegraphics[width=0.45\linewidth]{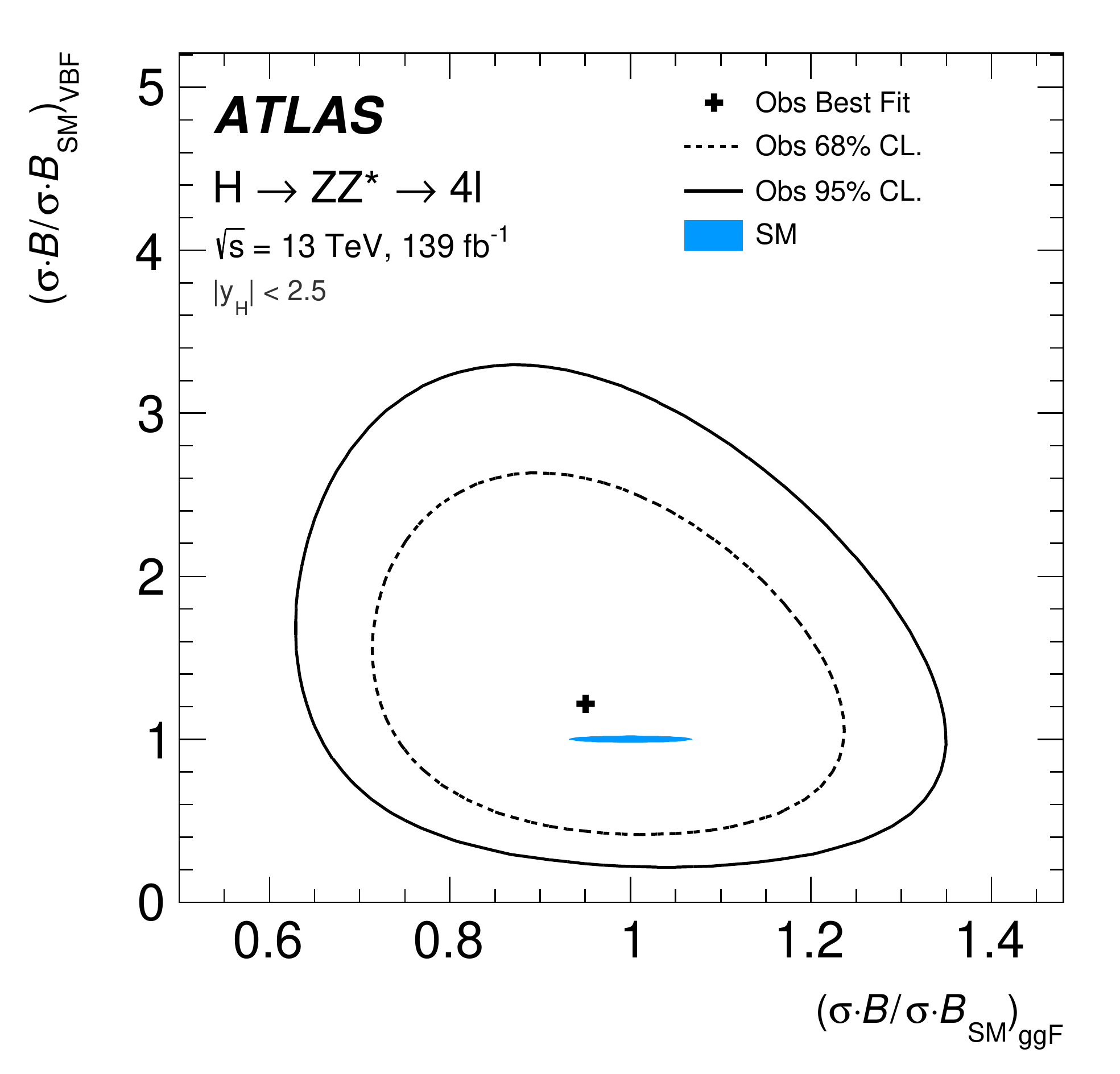}\label{fig:sigGGFVBF}}
\subfloat[]{\includegraphics[width=0.45\linewidth]{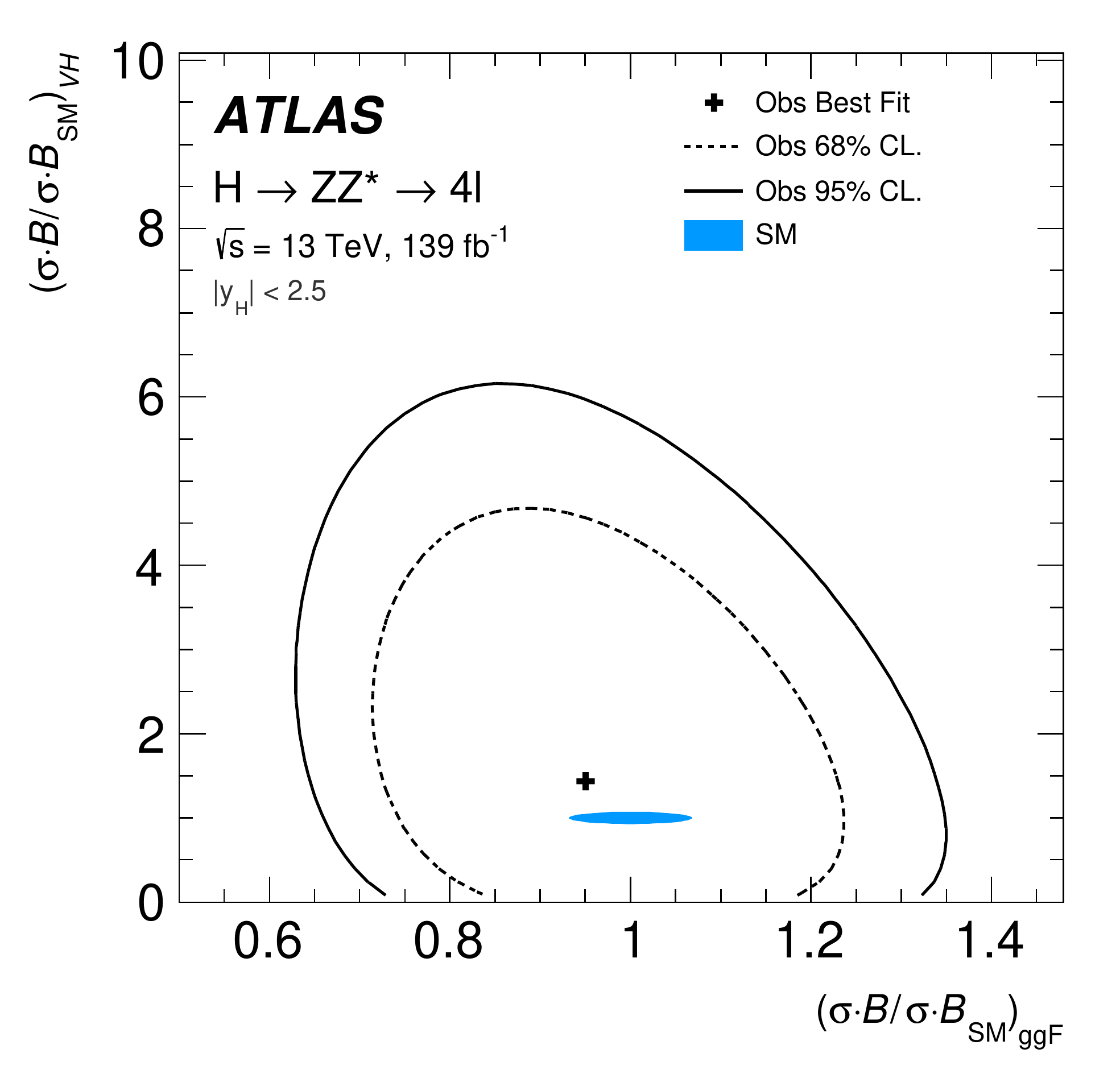}\label{fig:sigVBFVH}}\\
\subfloat[]{\includegraphics[width=0.45\linewidth]{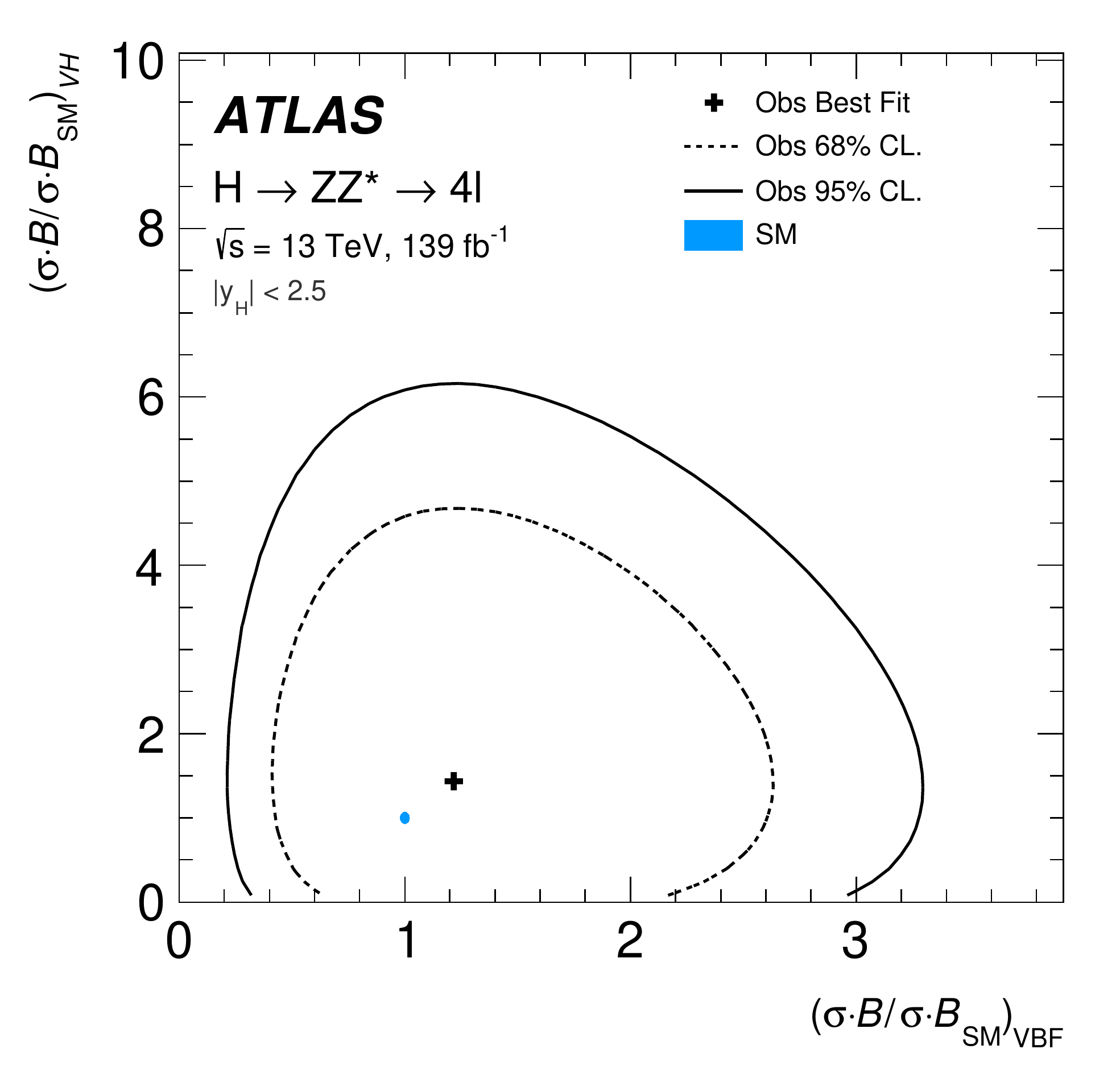}\label{fig:sigRS1p1}}
\subfloat[]{\includegraphics[width=0.45\linewidth]{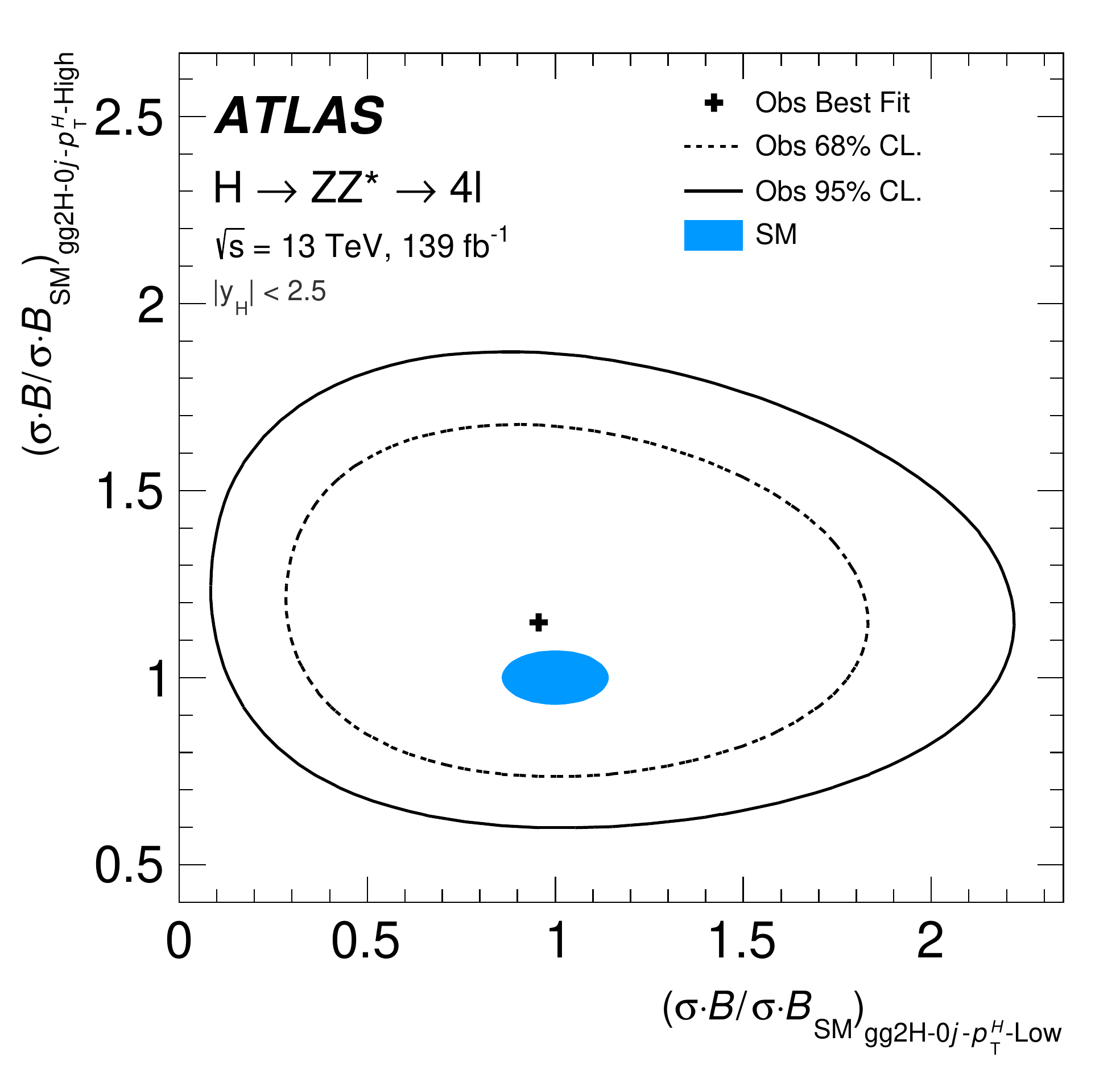}\label{fig:sigRS1p1ZeroJLH}}
 
\end{center}
\caption{ Likelihood contours at 68\% CL (dashed line) and 95\% CL (solid line) in the (a)  (\STXSggF, \STXSVBF), (b) (\STXSggF, \STXSVH), (c) (\STXSVBF, \STXSVH) and (d) (\STXSggToHZeroJL, \STXSggToHZeroJH) plane. The SM prediction is shown together with its theory uncertainty (filled ellipse). The \STXSVH parameter of interest is constrained to positive values.}
\label{fig:Cont2D}
\end{figure}

\figref{fig:Cont2D} shows the likelihood contours in the (\STXSggF, \STXSVBF), (\STXSggF, \STXSVH), (\STXSVBF, \STXSVH) and (\STXSggToHZeroJL, \STXSggToHZeroJH) planes.
The other cross-section parameters are left free in the fit, i.e.\ they are not treated as parameters of interest.
The compatibility with the SM expectation is
at the level of 0.22, 0.25, 0.19 and 0.33 standard deviations, respectively.

\FloatBarrier

\section{Constraints on the Higgs boson couplings in the $\kappa$-framework}
\label{sec:kappaframework}
\begin{figure}[!htbp]
\begin{center}
\includegraphics[width=0.5\linewidth]{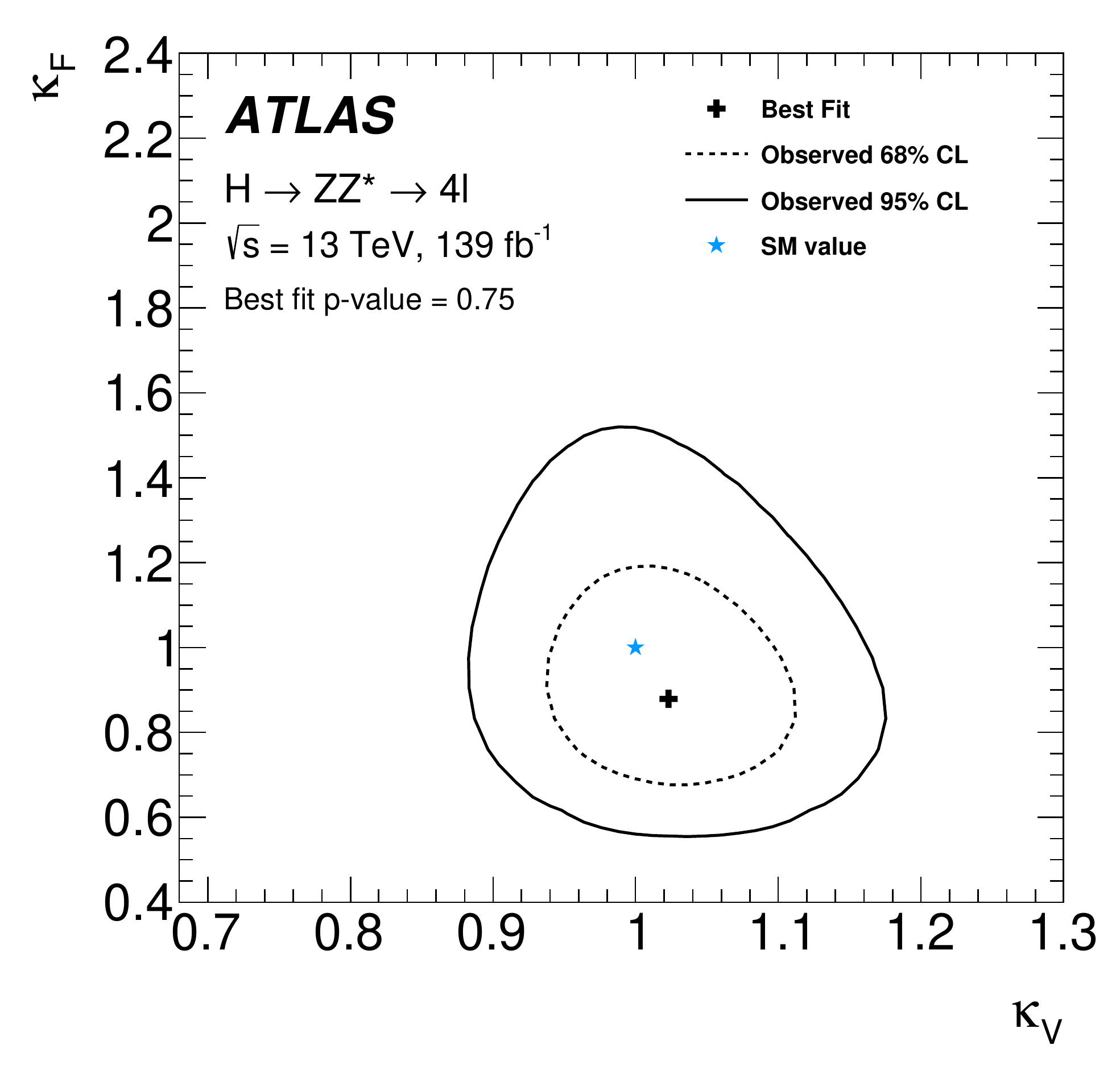}
\end{center}
\caption{ Likelihood contours at 68\% CL (dashed line) and 95\% CL (solid line) in the $\kappa_{V}$--$\kappa_{F}$  plane. The best fit to the data (solid cross) and the SM prediction (star) are also indicated. }
\label{fig:kvkf}
\end{figure}

The cross-sections measured at the Production Mode Stage are interpreted in the $\kappa$-framework described in Section~\ref{subsec:intro_kappa}. The relevant cross-sections and the branching ratio of Eq.~(\ref{eq:likelihood_simplified}) are parameterised in terms of the coupling-strength modifiers $\vec{\kappa}$. One interesting benchmark allows two different Higgs boson coupling-strength modifiers to fermions and bosons, reflecting the different structure of the interactions of the SM Higgs sector with gauge bosons and fermions. The universal coupling-strength modifiers $\kappa_{F}$ for fermions and $\kappa_{V}$ for vector bosons are defined as $\kappa_{V}= \kappa_{W}=\kappa_{Z}$ and $\kappa_{F}=\kappa_{t}=\kappa_{b}=\kappa_{c}=\kappa_{\tau}=\kappa_{\mu}$. It is assumed that there are no undetected or invisible Higgs boson decays. The observed likelihood contours in the $\kappa_{V}$--$\kappa_{F}$ plane are shown in \figref{fig:kvkf} (only the quadrant $\kappa_{F}>0$ and $\kappa_{V}>0$ is shown since this channel is not sensitive to the relative sign of the two coupling modifiers). The best-fit value is $\hat{\kappa}_V=1.02\pm 0.06$ and $\hat{\kappa}_F=0.88\pm0.16$, with the correlation of $-$0.17.
The probability of compatibility with the Standard Model expectation is at the level of 75\%.\\
 
\FloatBarrier

\section{Constraints on the tensor coupling structure in the EFT approach}
\label{sec:eft}
To interpret the observed data in the framework of an effective field theory, an EFT signal model is built by parameterising the production cross-sections in each production bin of the Reduced~Stage~1.1, as well as the branching ratio and the signal acceptances, as a function of the SMEFT Wilson coefficients introduced in Section~\ref{subsec:intro_eft}. The constraints on the Wilson coefficients are then obtained from the simultaneous fit to the data in all reconstructed signal and sideband event categories. Due to the  statistical precision of the data sample, the constraints are always set on one or at most two of the Wilson coefficients at a time, while the values of the remaining coefficients are assumed to be equal to zero.
 
\subsection{EFT signal model}
\label{subsec:eft_predictions}
 
The EFT parameterisation of the production cross-sections in each production bin of the Reduced~Stage~1.1 is obtained from Eq.~(\ref{eq:sigma_parametrisation}) using simulated BSM samples introduced in Section~\ref{sec:simulation}. The contribution from the $gg\to Z(\to \ell\ell)H$ process is taken from the SM simulation and assumed to scale with BSM parameters in the same way as the $qq\to Z(\to \ell\ell)H$ processes. As in the case of simplified template cross-section measurements, \ttH and $tH$ processes are combined into a single \STXSttH production bin. The cut-off scale is set to $\Lambda = 1$~\TeV. Only LO computation of QCD and SM electroweak processes is provided, with LO effective couplings for the SM Higgs boson to gluon and to photon vertices. An assumption is made that higher-order corrections, applied in a multiplicative way, are the same for both the SM and the BSM LO predictions and therefore no changes in the parameterisation are expected due to higher-order effects~\cite{Degrande_2017}. With the current amount of data, the constraints from the \VBF, \VH and \ttH production modes on the relevant Wilson coefficients still allow a rather large range of parameter values in which the quadratic term (the last term in Eq.~(\ref{eq:sigma_parametrisation})) cannot be neglected even though its contribution is suppressed by $\Lambda^4$. Such dimension-six quadratic terms are therefore included in the EFT parameterisation. Since the linear terms from dimension-eight operators are suppressed by the same factor, they could in general also give similar non-negligible contributions. Dimension-eight terms are currently not available in the SMEFT model and are thus not taken into account.
 
The branching ratio for the $H\to ZZ^{*}\to 4\ell$ decay is parameterised in terms of Wilson coefficients following Eq.~(\ref{eq:br_EFT}). The partial and total decay widths are calculated in \MGMCatNLO. The total decay width is calculated by taking into account the dominant Higgs boson decay modes: $\gamma\gamma$, $Z\gamma$, \bbprod, $gg$, $WW$ and $ZZ$. Other decay modes are not affected by the probed Wilson coefficients. Their contribution to the total decay width is therefore given by the corresponding SM predictions.

\begin{figure}[!htbp]
\centering
\subfloat[]{
\includegraphics[width=0.3\linewidth]{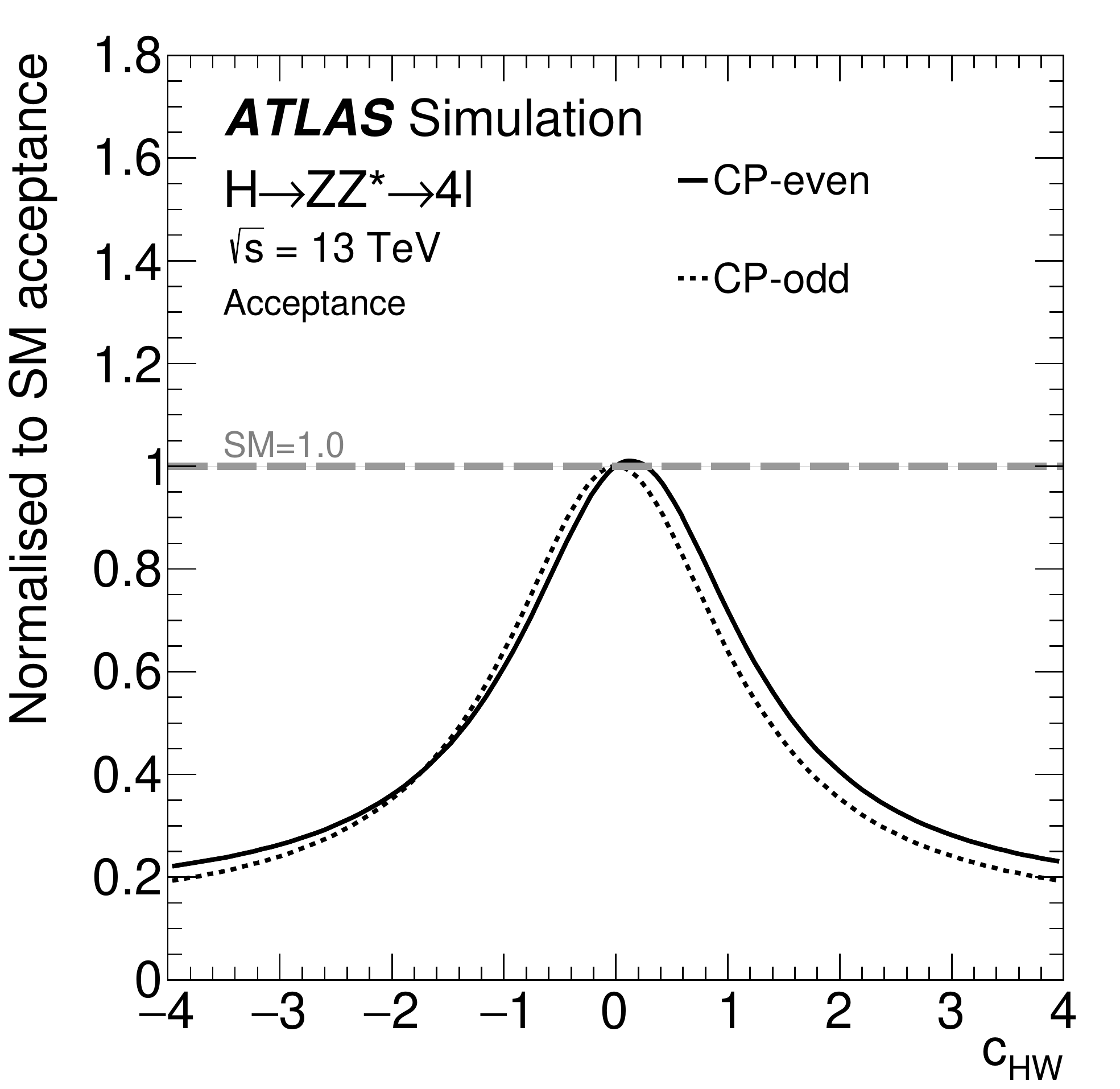}\label{fig:acc_chw}}
\subfloat[]{
\includegraphics[width=0.3\linewidth]{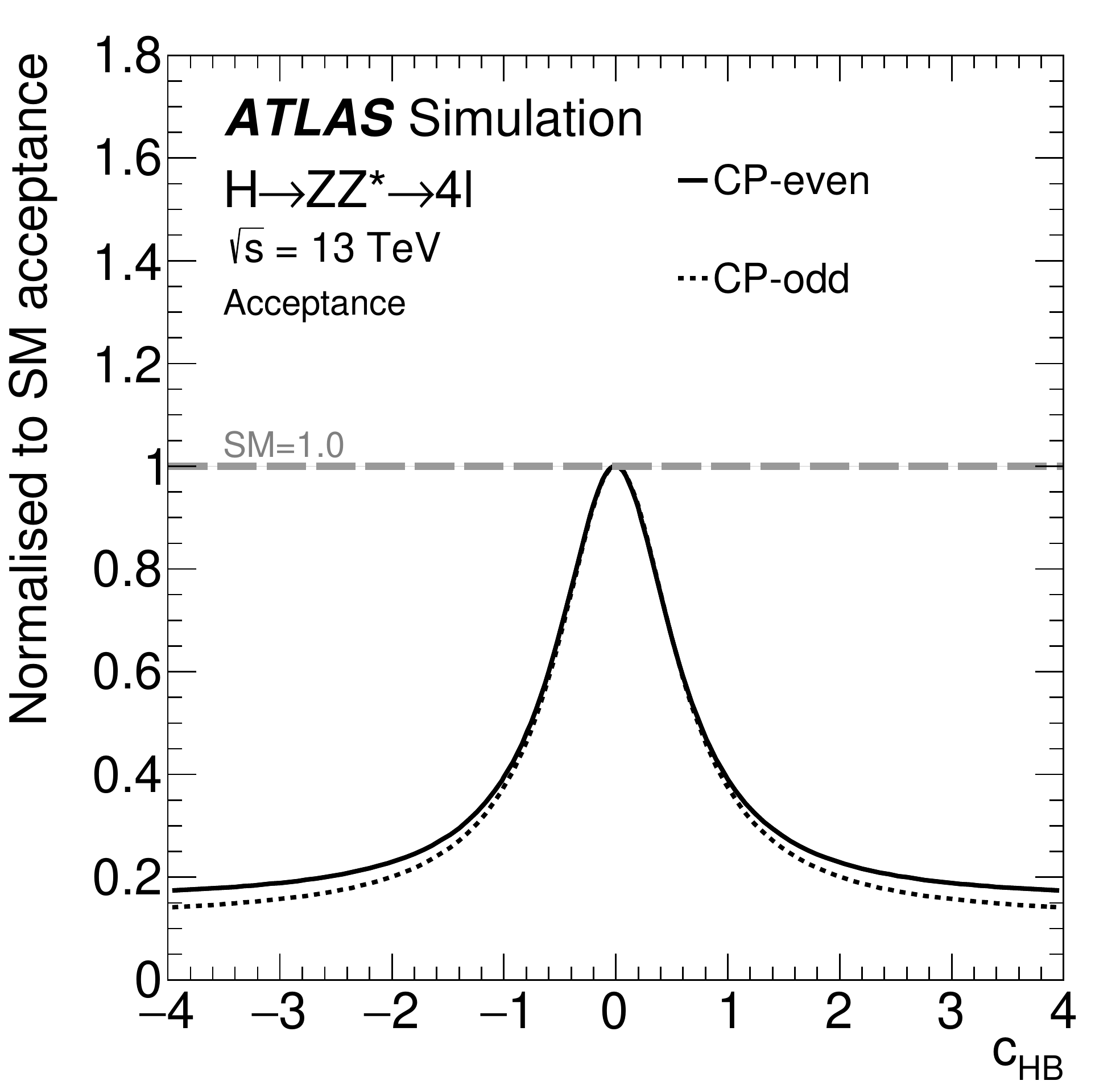}\label{fig:acc_chb}}
\subfloat[]{
\includegraphics[width=0.3\linewidth]{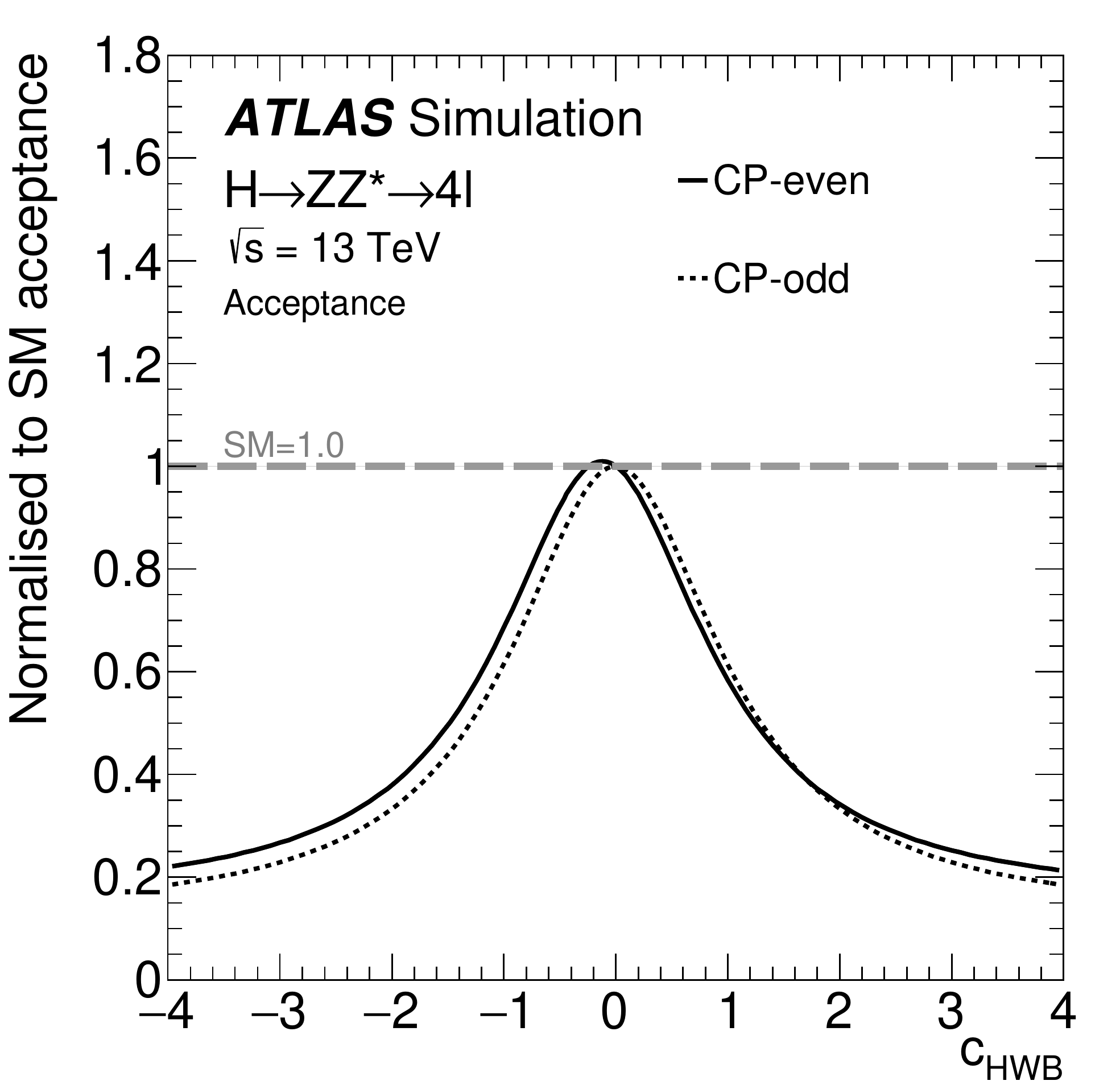}\label{fig:acc_chwb}}\\
\caption{The dependence of the signal acceptance normalised to the SM acceptance on the Wilson coefficients (a) $c_{HW}$ and $c_{H\tilde{W}}$, (b) $c_{HB}$ and $c_{H\tilde{B}}$,  (c) $c_{HWB}$ and $c_{H\tilde{W}B}$ after setting all other coefficients to zero.}
\label{fig:EFT_Acceptance}
\end{figure}

The selection criteria for the four-lepton Higgs boson candidates, in particular the requirements on the minimum invariant mass $m_{34}$ of the subleading lepton pair, introduce an additional dependence of the signal acceptance on the BSM coupling parameters.  The particle-level signal acceptance $A$, defined as the fraction of signal events satisfying the Higgs boson candidate selection criteria applied at particle-level,
has therefore been simultaneously parameterised in terms of the three Wilson coefficients $c_{HW}$, $c_{HB}$ and $c_{HWB}$ ($c_{H\widetilde{W}}$, $c_{H\widetilde{B}}$ and $c_{H\widetilde{W}B}$) assuming that the values of CP-odd (CP-even) parameters vanish. The dependence of the acceptance on other EFT coupling parameters
is shown to be negligible as these parameters have negligible or no impact on the $H\to ZZ^*$ decay. The acceptance correction relative to the SM prediction is described by a three-dimensional Lorentzian function with free acceptance parameters $\alpha_0$, $\alpha_1$, $\alpha_2$, $\beta_i$, $\delta_i$, $\delta_{(i,j)}$ and $\delta_{(i,j,k)}$,
\begin{equation}
\resizebox{0.85\textwidth}{!}{ 
$\frac{A(\vec{c})}{A_{\textrm{SM}}} = \alpha_0 + (\alpha_1)^2 \cdot \left[ \alpha_2 +  \sum\limits_{i}\delta_i\cdot(c_i+\beta_i)^2 + \sum\limits_{\substack{ij \\ i\neq j}} \delta_{(i,j)} \cdot c_i c_j + \underset{i\neq j\neq k}{\delta_{(i,j,k)}} \cdot c_{i} c_{j} c_{k}\right]^{-1},$}
\label{eq:eft_acceptance}
\end{equation}
where indices $i$, $j$ and $k$ run over $(HW$, $HB$, $HWB)$ in case of the acceptance correction for the set of CP-even parameters and over $(H\widetilde{W}$, $H\widetilde{B}$, $H\widetilde{W}B)$ in case of the CP-odd parameters.
A common parameterisation is used for all production bins since the differences between production bins are shown to be negligible. In addition, the reconstructed event categorisation criteria imposed on the selected Higgs boson candidates and the classification in bins of multivariate NN discriminant values do not impact the acceptance parameterisation. The impact of reconstruction efficiencies on the parameterisation is also negligible, such that Eq.~(\ref{eq:eft_acceptance}) also holds for the ratio $A(\vec{c})/A_\textrm{SM}$ of reconstruction-level acceptances defined in Section~\ref{sec:stxs}. The resulting acceptance parameterisation curves are shown in \figref{fig:EFT_Acceptance} for the cases in which all but one of the Wilson coefficients are set to zero.  For all cases, the acceptance correction is equal to one at the SM point. In the case of the $c_{HW}$ and $c_{HWB}$ Wilson coefficients,
the acceptance corrections reach a maximum value slightly larger than one, leading to the shift of the maximum position from the SM point. This shift is compatible with the statistical accuracy of the fit and the impact of linear EFT terms which are not symmetric around the SM point.
 
\begin{figure}[!htbp]
\centering
\subfloat[]{
\includegraphics[width=0.33\linewidth]{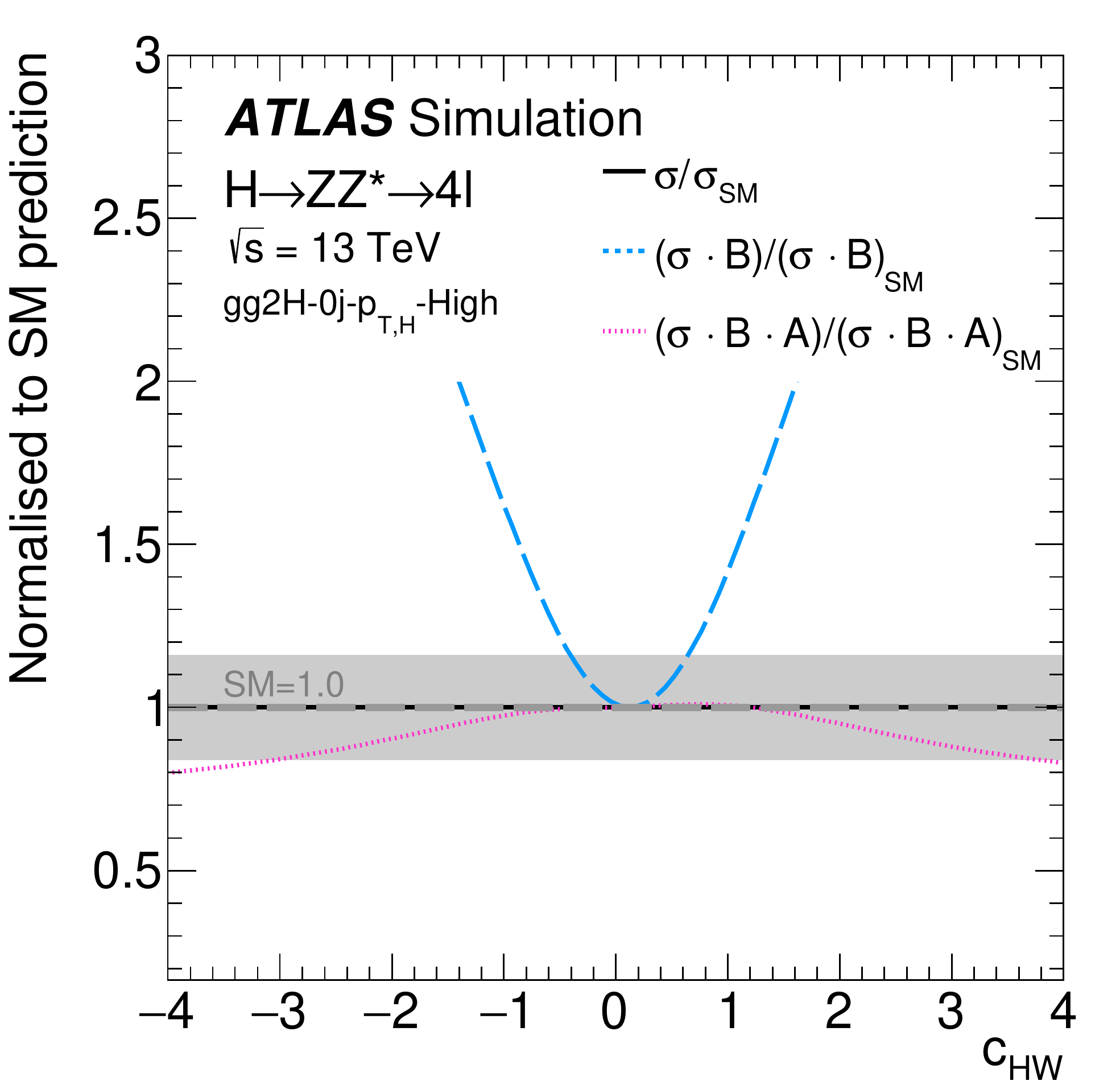}\label{fig:xs_chw_ggF}}
\subfloat[]{
\includegraphics[width=0.33\linewidth]{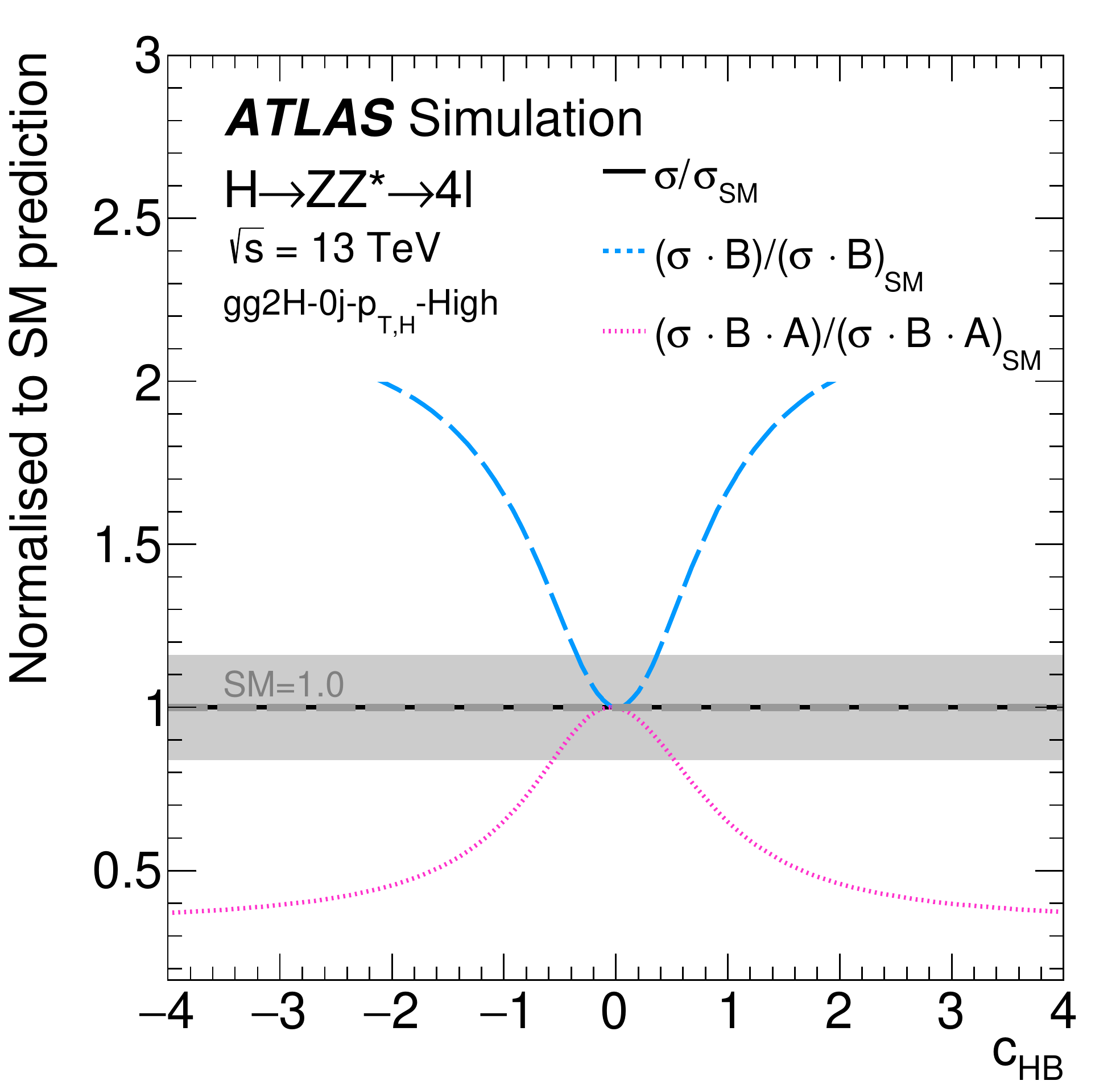}\label{fig:xs_chb_ggF}}
\subfloat[]{
\includegraphics[width=0.33\linewidth]{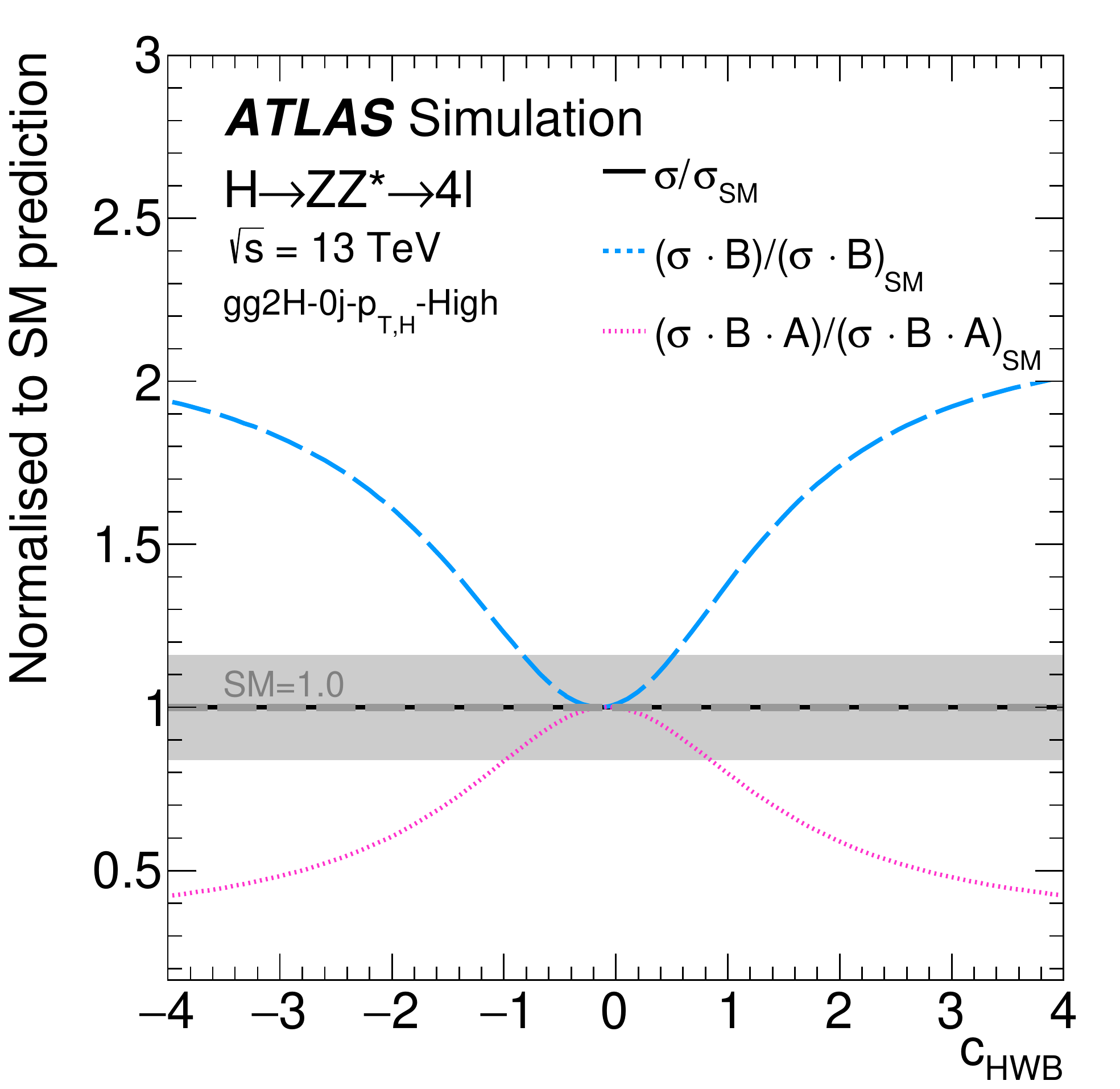}\label{fig:xs_chwb_ggF}}\\
\subfloat[]{
\includegraphics[width=0.33\linewidth]{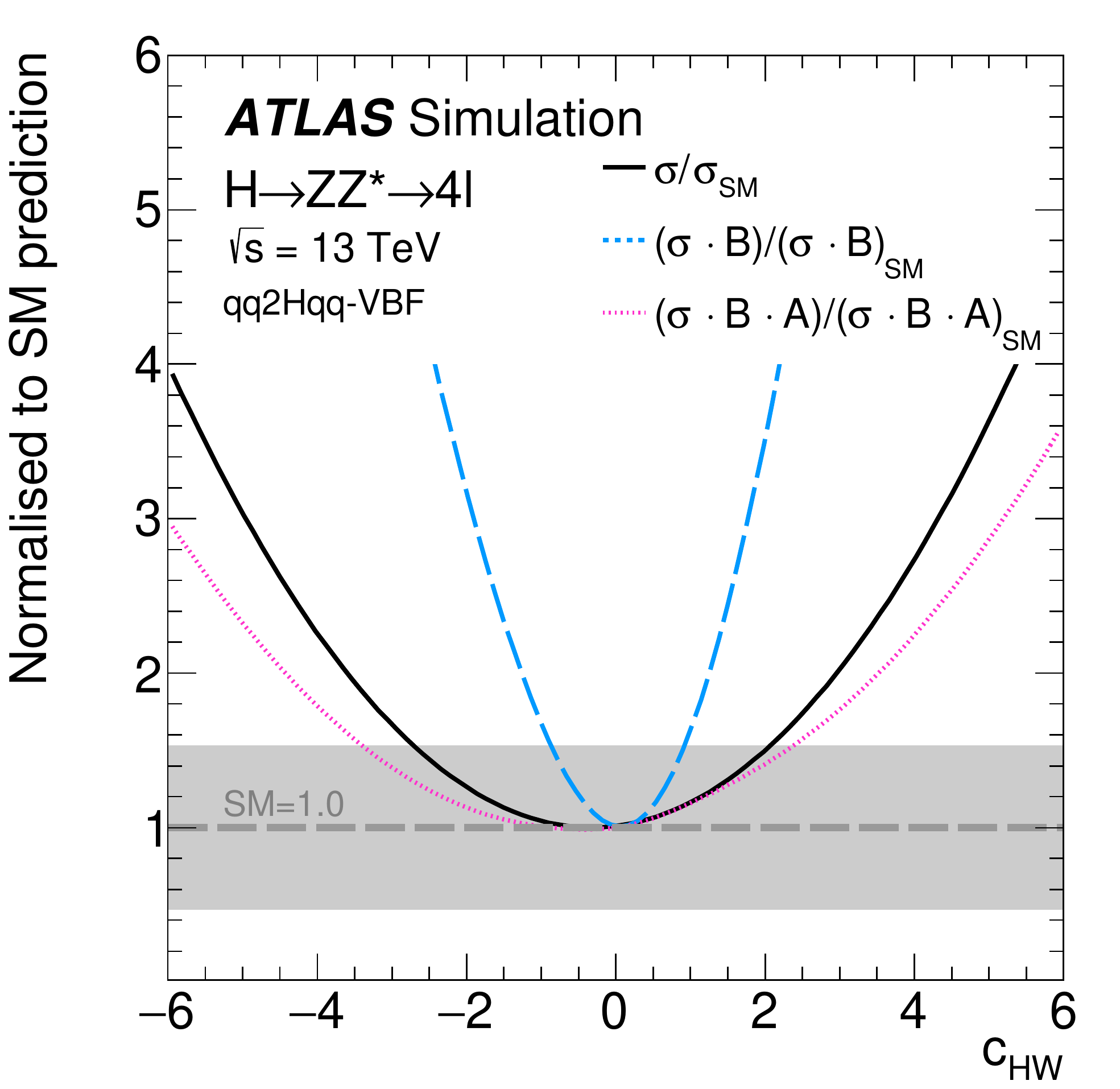}\label{fig:xs_chW_VBF}}
\subfloat[]{
\includegraphics[width=0.33\linewidth]{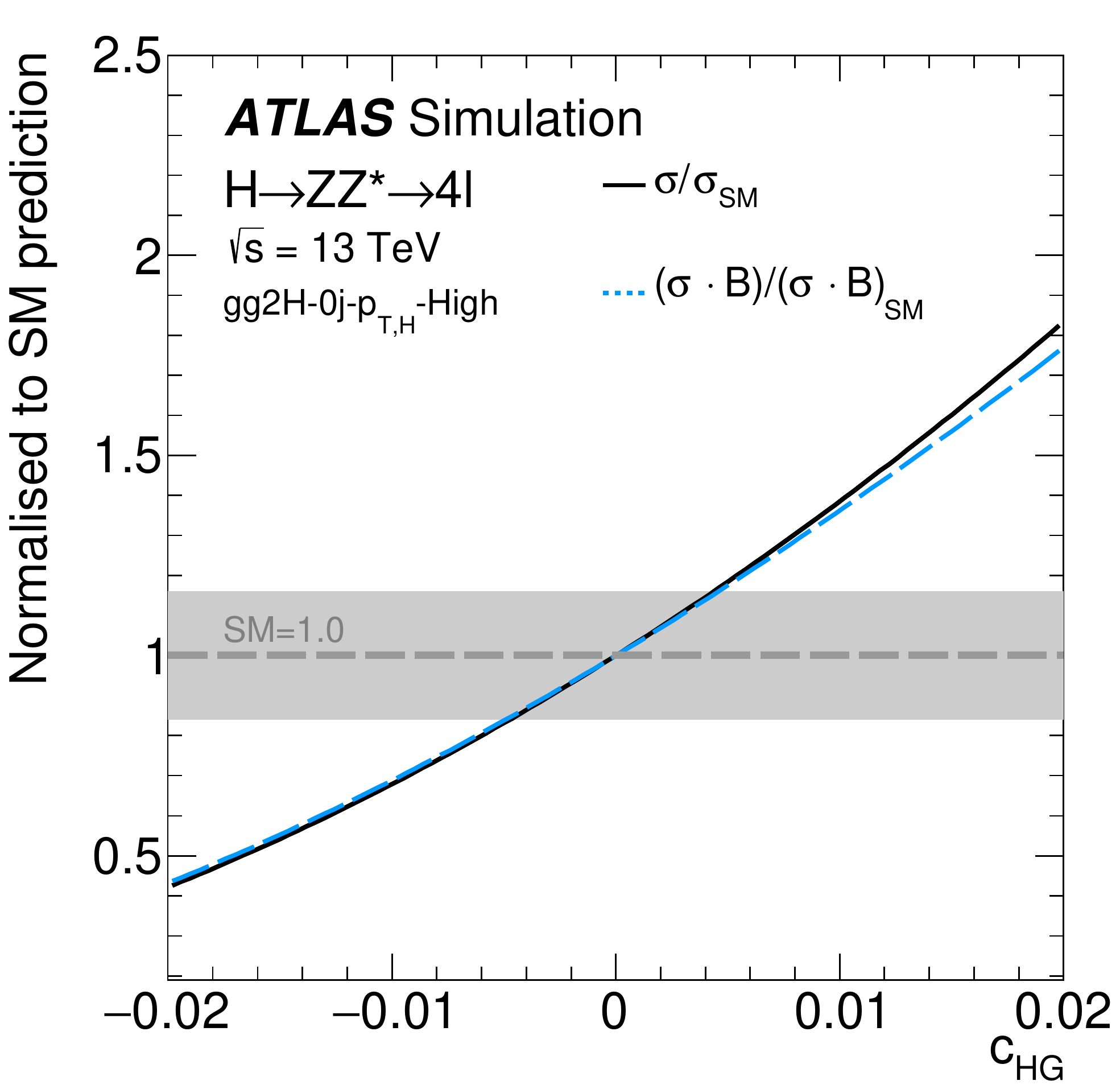}\label{fig:xs_chg_ggF}}
\subfloat[]{
\includegraphics[width=0.33\linewidth]{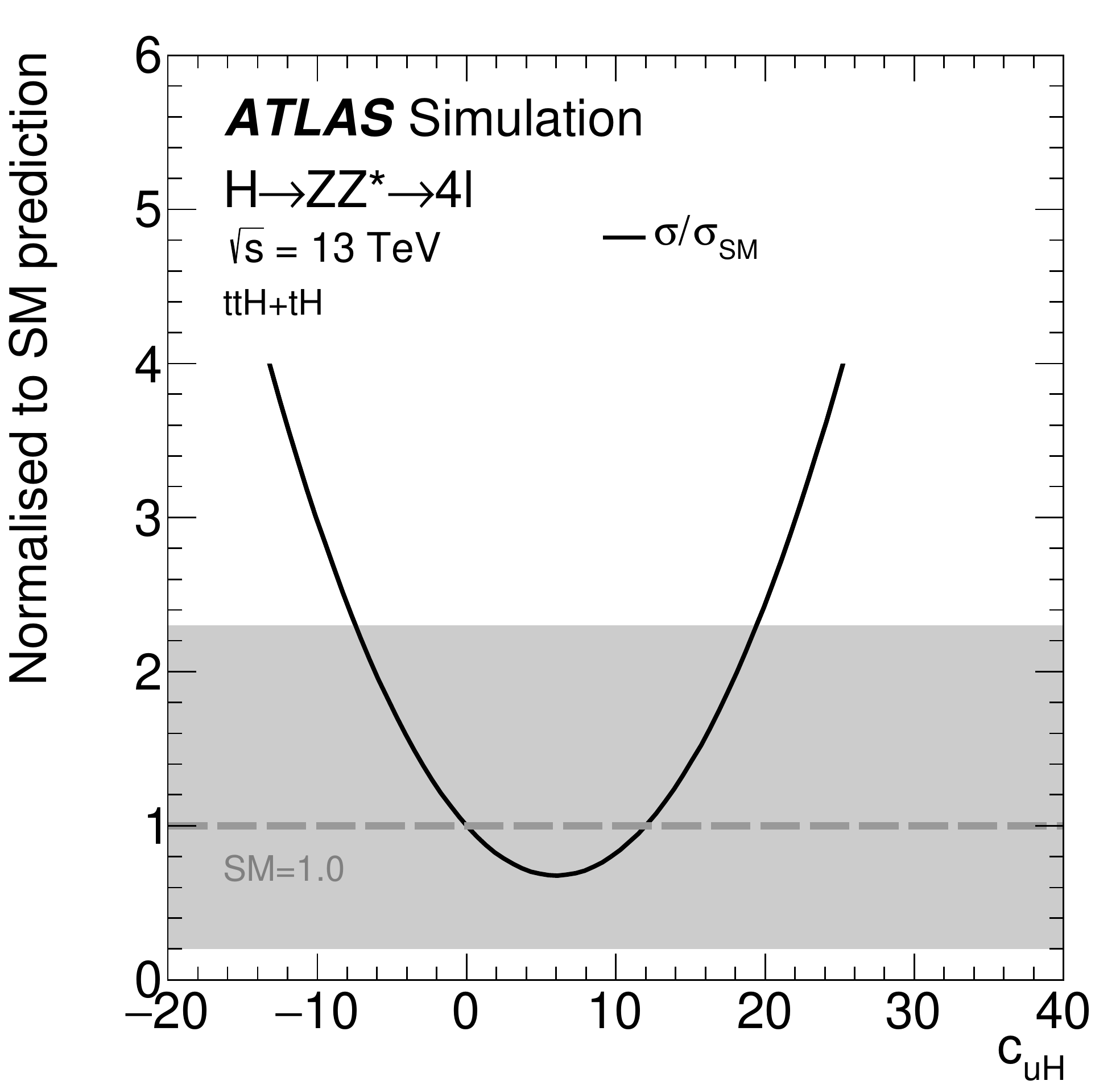}\label{fig:xs_cuh_ttH}}\\
\caption{The expected event yields ($\sigma \cdot B \cdot A$) relative to the SM prediction as a function of the Wilson coefficient (a) $c_{HW}$, (b) $c_{HB}$ and (c) $c_{HWB}$ in the \STXSggToHZeroJH production bin, (d) $c_{HW}$ in the \STXSqqtoHqqRest production bin, (e) $c_{HG}$ in the \STXSggToHZeroJH production bin and (f) $c_{uH}$ in the \STXSttH production bin. The dependence on only one Wilson coefficient is shown on each plot while setting all others to zero. For comparison, the predictions are also shown for the parameterisation without the acceptance corrections ($\sigma \cdot B$) and for the production cross-section only ($\sigma$) without the acceptance and the branching ratio corrections. The $\sigma$ parameterisations in (a), (b) and (c) coincide with the SM expectation at 1 as the coefficients $c_{HW}$, $c_{HB}$ and $c_{HWB}$ are not present in the \ggF production vertex. Since the acceptance does not depend on the $c_{HG}$ and $c_{uH}$ parameters, no corresponding ($\sigma \cdot B \cdot A$) expectation is shown in (e) and (f). Similarly, no ($\sigma \cdot B$) expectation is shown in (f), since the $c_{uH}$ parameter has a negligible impact on the branching ratio. The bands indicate the expected precision of the cross-section measurement in a given production bin at the one standard deviation level.}
\label{fig:EFTpredictions}
\end{figure}

The final parameterisation of signal yields relative to the SM prediction in each production bin of the Reduced~Stage~1.1 is obtained as the product of the corresponding cross-section, branching ratio and acceptance parameterisations. The expected event yields normalised to the SM prediction are shown in Figure~\ref{fig:EFTpredictions} for each of the CP-even Wilson coefficients after setting all other coefficients to zero.
Only production bins with the highest sensitivity to a given Wilson coefficient are shown. The impact of the quadratic terms in the EFT parameterisation can clearly be seen as a non-linear dependence on all but the $c_{HG}$ Wilson coefficient. For comparison, the predictions without the acceptance corrections ($\sigma \cdot \BR$), and without both the acceptance and branching ratio corrections ($\sigma$) are also shown. Both the acceptance and the branching ratio parameterisations have a strong impact on the sensitivity to different Wilson coefficients, especially for the $c_{HW}$, $c_{HB}$ and $c_{HWB}$ parameterisations in $gg2H$ production bins (Figures~\ref{fig:xs_chw_ggF}, \ref{fig:xs_chb_ggF} and~\ref{fig:xs_chwb_ggF}). Since these coefficients do not enter the \ggF production vertex, the corresponding sensitivity is entirely driven by their impact on the decay and the acceptance of selected signal events. The acceptance corrections significantly degrade the sensitivity to the $c_{HW}$ coefficient (see Figure~\ref{fig:xs_chw_ggF}). Additional sensitivity to this coefficient can be gained from the $qq2Hqq$ production bins as shown in Figure~\ref{fig:xs_chW_VBF}. The Wilson coefficients $c_{HG}$ and $c_{uH}$, on the other hand, do not affect the acceptance since they are not present in the decay vertex (Figures~\ref{fig:xs_chg_ggF} and \ref{fig:xs_cuh_ttH}). The coefficient $c_{HG}$ still has a non-vanishing impact on the branching ratio through its contributions to the total decay width. Similar effects are also seen for the Wilson coefficients of CP-odd operators.

\subsection{EFT interpretation results}
\label{subsec:eft_results}
\begin{figure}[!htbp]
\centering
\subfloat[]{
\includegraphics[width=0.50\linewidth]{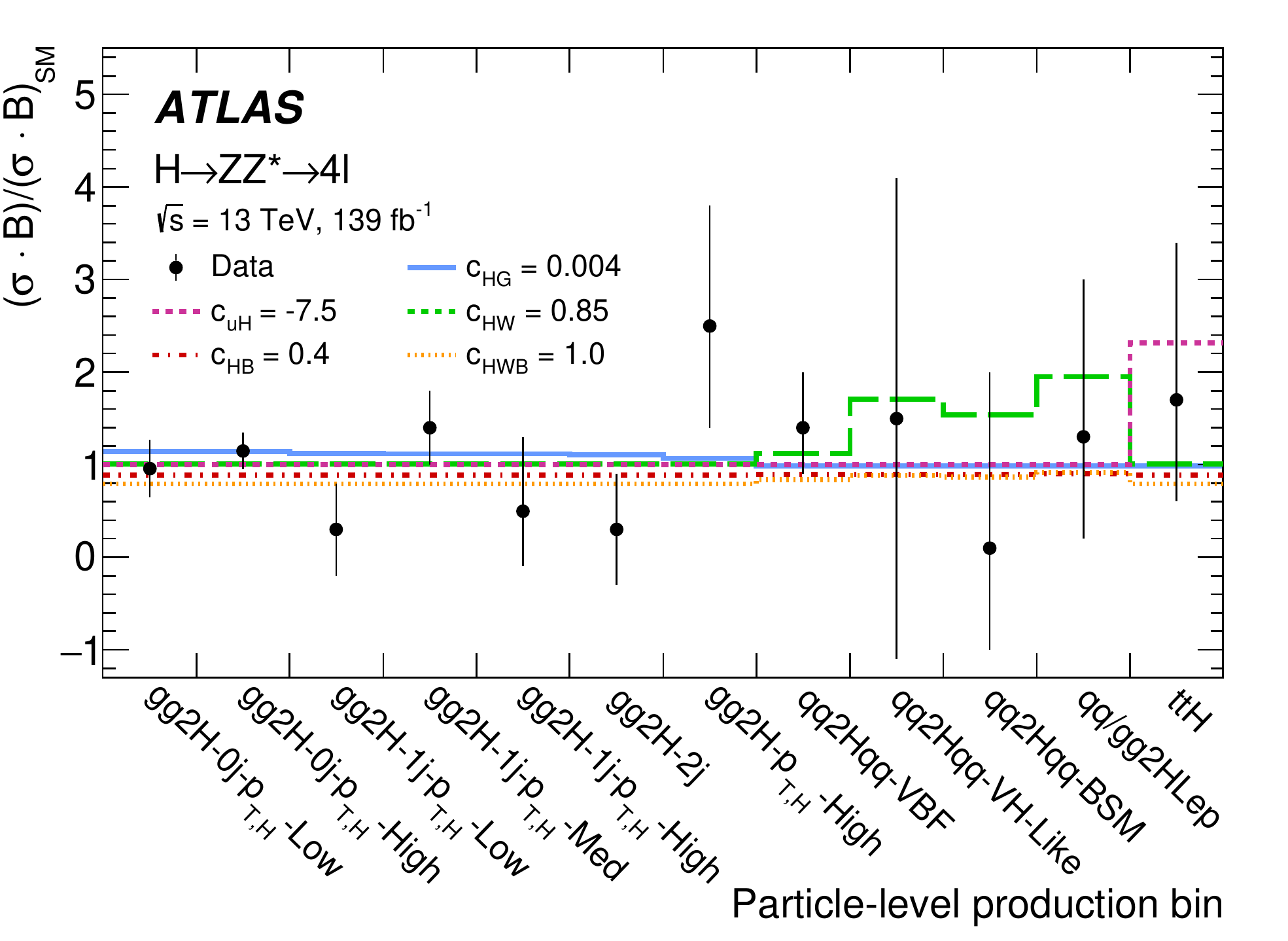}\label{fig:CPEvenObs}}
\subfloat[]{
\includegraphics[width=0.50\linewidth]{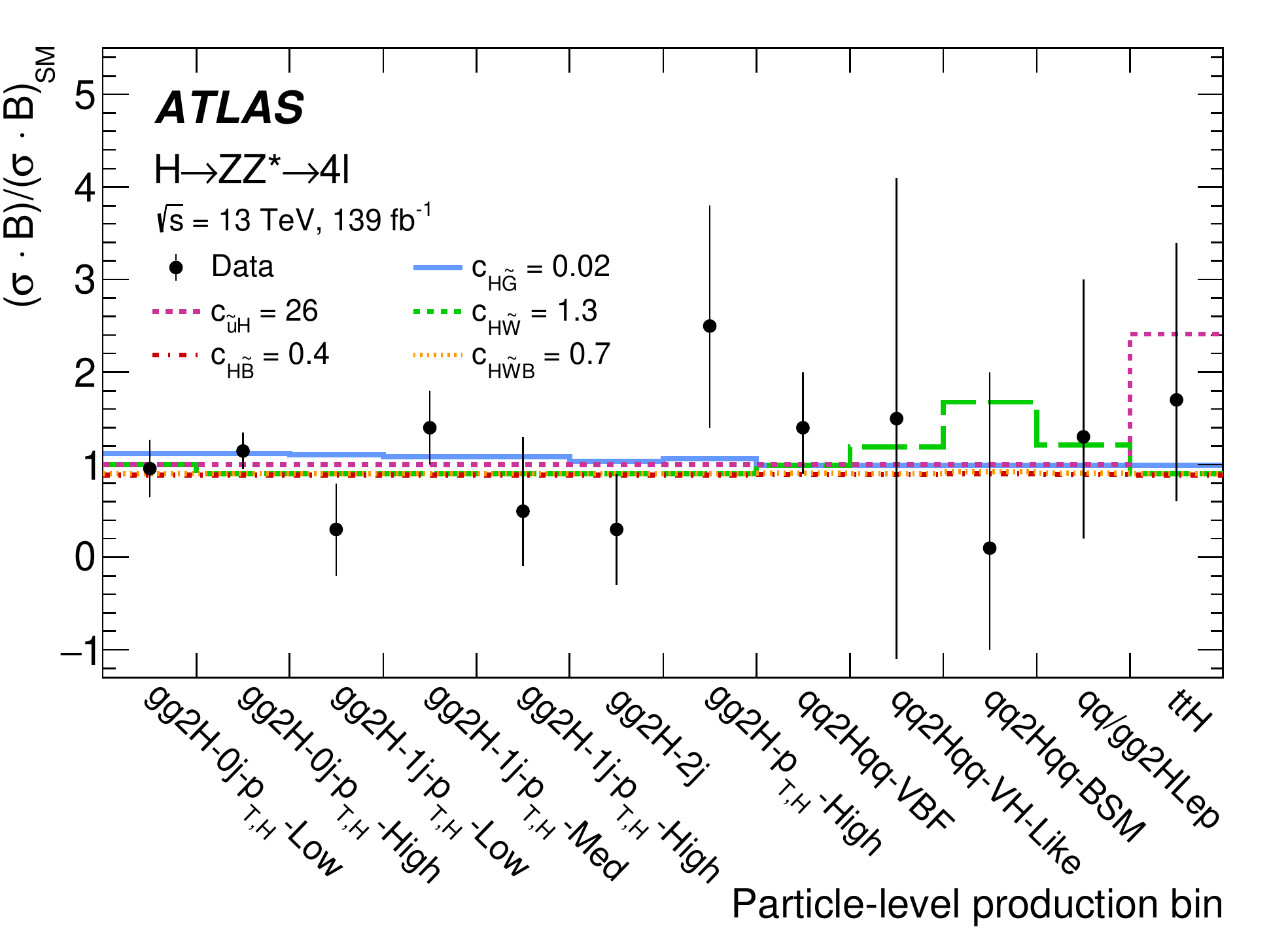}\label{fig:CPOddObs}}\\
\caption{The expected signal yield ratio for chosen (a) CP-even and (b) CP-odd EFT parameter values together with the corresponding cross-section measurement in each production bin of Reduced~Stage~1.1. The parameter values correspond approximately to the expected confidence intervals at the 68\% CL obtained from the statistical interpretation of data.}
\label{fig:ObsXSEFT}
\end{figure}

The ratios of the expected signal yield for a chosen EFT parameter value to its SM prediction are shown in \figref{fig:ObsXSEFT} in each production bin of the Reduced~Stage~1.1, together with the corresponding measurement.
 
The EFT parameterisation of signal yields is implemented in the likelihood function of Eq.~(\ref{eq:likelihood_simplified}) using the BSM-dependent signal-strength parameters $\mu^p(\vec{c})$ for each given production bin $p$,
\begin{equation*}
\mu^p(\vec{c}) = \frac{\sigma^p(\vec{c})}{\sigma_{\textrm{SM}}} \cdot
\frac{\BR^{4\ell}(\vec{c})}{\BR^{4\ell}_{\textrm{SM}}} \cdot
\frac{A(\vec{c})}{A_{\textrm{SM}}}.
\label{eq:eft_mu}
\end{equation*}
This is then fitted to the observed event yields. Default SM predictions at the highest available order are employed for the cross-sections and branching ratios multiplying the signal strengths in the likelihood function.
Modifications of background contributions due to EFT effects are not taken into account.

\begin{figure}[!htbp]
\centering
\subfloat[]{
\includegraphics[width=0.50\linewidth]{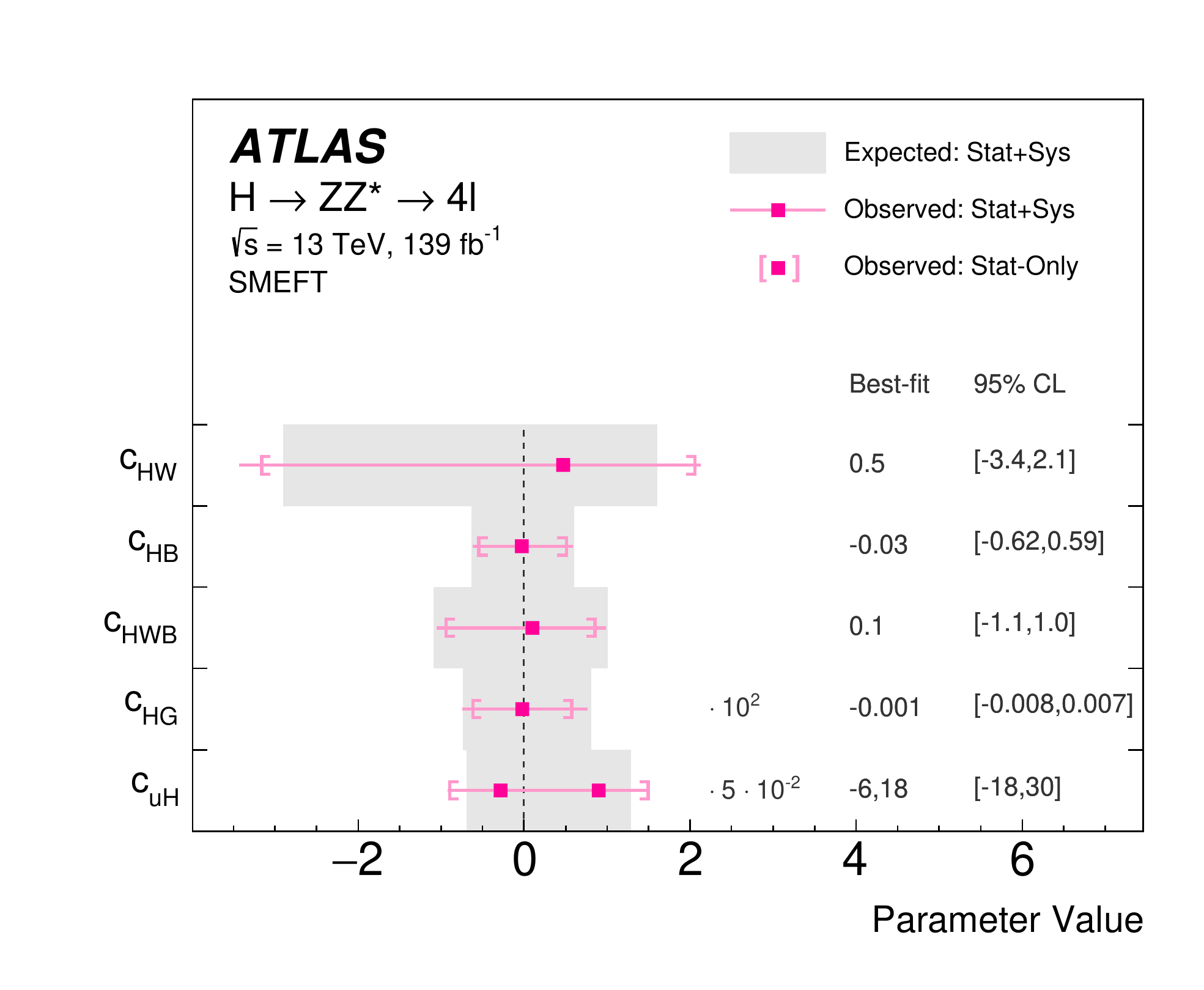}\label{fig:EFTCPEvenResults}}
\subfloat[]{
\includegraphics[width=0.50\linewidth]{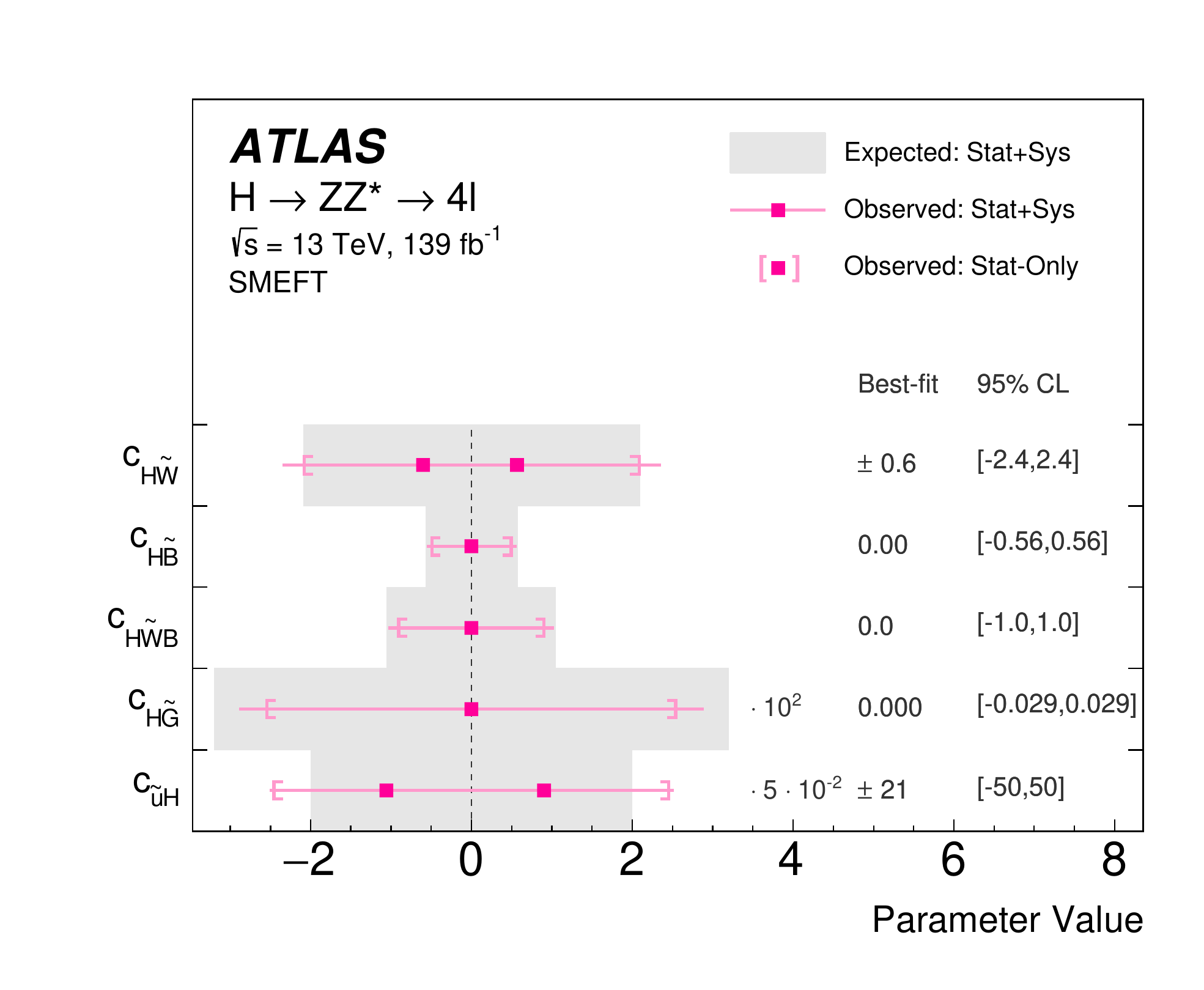}\label{fig:EFTCPOddResults}}\\
\caption{The observed and expected values of SMEFT Wilson coefficients from (a) CP-even and (b) CP-odd operators obtained for an integrated luminosity of \LumExact\ at $\sqrt{s}=13$~\TeV. Only one Wilson coefficient is fitted at a time while all others are set to zero. The values for the $c_{HG}$ and $c_{H\tilde{G}}$ coefficients are scaled by a factor of 100, and for the $c_{uH}$ and $c_{\tilde{u} H}$ coefficients by a factor of 0.05. The horizontal bands represent the expected measurement uncertainty. }
\label{fig:EFT_results}
\end{figure}

The fit results with only one Wilson coefficient fitted at a time are summarised in Figure~\ref{fig:EFT_results} and in Table~\ref{fig:EFT_1DResults}. The results are in good agreement with the SM predictions. The measurements are dominated by the statistical uncertainty. In the case of the CP-odd coupling parameters, each fit gives two degenerate minima since the corresponding EFT parameterisation contains only quadratic terms which are not sensitive to the sign of the fitted parameter. The fit of the CP-even coupling parameter $c_{uH}$ also results in two minima since the corresponding EFT parameterisation curve in the only sensitive \STXSttH production bin crosses the expected SM cross-section value at two different values of the $c_{uH}$ parameter (see Figure~\ref{fig:xs_cuh_ttH}). The same is true also for the observed \STXSttH cross-section.  The small degeneracies for other CP-even coupling parameters are removed by the combination of several sensitive production bins.

The strongest constraint, driven mostly by the \ggF reconstructed event categories, is obtained on the $c_{HG}$ coefficient related to the CP-even Higgs boson interactions with gluons. The highest sensitivity to this parameter is reached by the measurements in the \STXSggToHZeroJL\ and \STXSggToHZeroJH\ production bins due to the highest statistical precision. The sensitivity in the \STXSggToHHigh\ production bin, which is designed to target the BSM physics effects, is limited due to the small number of events observed in the corresponding reconstructed event category. Additional sensitivity in this bin may be provided by the two-loop interactions which are not implemented in the current simulation of the $ggH$ vertex. The constrained range is stringent enough for the linear approximation to hold, i.e.\ the quadratic terms in the signal parameterisation are small compared with the linear ones (see Figure~\ref{fig:xs_chg_ggF}). The constraint on the $c_{H\widetilde{G}}$ parameter of the related CP-odd operator is worse by about a factor of three since the linear terms from CP-odd operators do not contribute to the total production cross-section. The constraints on the remaining EFT parameters are weaker, such that both the CP-even and CP-odd signals become dominated by the quadratic terms and are therefore comparable in size.  The next-strongest constraints are obtained on the $c_{HB}$, $c_{HWB}$, $c_{HW}$, $c_{H\widetilde{B}}$, $c_{H\widetilde{W}B}$ and $c_{H\widetilde{W}}$ coefficients that mostly affect the $H\to ZZ^*$ decays. Due to the larger number of events in the $0$-jet reconstructed event categories, the corresponding $gg2H$ production bins provide the highest sensitivity to these decays. Additional smaller sensitivity is obtained from the production vertex of the \VBF and \VH production modes, with the dominant contribution from \STXSqqtoHqqRest\ and \STXSqqtoHqqBSM\ bins. The latter one is designed to enhance the sensitivity to BSM physics. The $qq2Hqq$ production bins improve in particular the sensitivity to the $c_{HW}$ and $c_{H\widetilde{W}}$ parameters that is otherwise significantly degraded by the acceptance corrections. Finally, looser constraints are set on the top-Yukawa coupling parameters $c_{uH}$ and $c_{\widetilde{u}H}$, driven by the measurements in the \ttH production bin.
 
\begin{table}[!htbp]
\centering
\renewcommand{\arraystretch}{1.3}
\caption{The expected and observed confidence intervals at 68\% and 95\% CL on the SMEFT Wilson coefficients for an integrated luminosity of $139$~fb$^{-1}$ at $\sqrt{s}=13 \ \TeV$. Only one Wilson coefficient is fitted at a time while all others are set to zero. 
}
\label{tab:ResEFT}
\vspace{0.1cm}
\resizebox{\textwidth}{!}{
\begin{tabular}{*{5}{c}rc}
\hline\hline
\noalign{\vspace{0.05cm}}
EFT coupling  &
\multicolumn{2}{c}{Expected} &
\multicolumn{2}{c}{Observed} &
\multicolumn{1}{c}{Best-fit} &
Best-fit \\
parameter &
68\% CL &  95\% CL &
68\% CL &  95\% CL &
\multicolumn{1}{c}{value} &
$p$-value  \\
\hline
$c_{HG}$  &
[$-0.004$, 0.004]  & [$-0.007$, 0.008]&
[$-0.005$, 0.003]&  [$-0.008$, 0.007]&
$-0.001$ &
$0.79$ \\
$c_{uH}$ &
\phantom{0.00}[$-8$, 20]\phantom{0.0} &  [$-14$, 26]&
\phantom{0.0}[$-12$, 6]\phantom{0.00} & [$-18$, 30]&
$-6$, $18$ &
$0.50$\\
$c_{HW}$ &
[$-1.6$, 0.9] & [$-2.9$, 1.6]&
[$-1.5$, 1.3] & [$-3.4$, 2.1] &
$0.5$ &
$0.66$\\
$c_{HB}$ &
[$-0.43$, 0.38] & [$-0.62$, 0.60] &
[$-0.42$, 0.37] & [$-0.62$, 0.59]&
$-0.03$&
$0.98$ \\
$c_{HWB}$ &
[$-0.75$, 0.63] & [$-1.09$, 0.99] &
[$-0.71$, 0.63] &  [$-1.06$, 0.99]&
$0.1$ &
$0.93$\\
\hline
$c_{H\widetilde{G}}$  &
[$-0.022$, 0.022] & [$-0.031$, 0.031] &
[$-0.019$, 0.019] & [$-0.029$, 0.029] &
$0.000$ &
$1.00$\\
$c_{\widetilde{u}H}$ &
[$-26$, 26]  &  [$-40$, 40] &
[$-37$, 37] & [$-50$, 50]&
$\pm21$ &
$0.48$\\
$c_{H\widetilde{W}}$ &
[$-1.3$, 1.3]&  [$-2.1$, 2.1] &
[$-1.5$, 1.5] & [$-2.4$, 2.4] &
$\pm0.6$ &
$0.84$\\
$c_{H\widetilde{B}}$ &
[$-0.39$, 0.39]  & [$-0.57$, 0.57] &
[$-0.37$, 0.37] & [$-0.56$, 0.56]&
$0.00$ &
$1.00$\\
$c_{H\widetilde{W}B}$ &
[$-0.71$, 0.71]  &  [$-1.05$, 1.05]&
[$-0.69$, 0.69] & [$-1.03$, 1.03] &
$0.0$ &
$1.00$\\
\noalign{\vspace{0.05cm}}
\hline\hline
\label{fig:EFT_1DResults}
\end{tabular}
}
\end{table}
 
\begin{figure}[!htbp]
\centering
\subfloat[]{
\includegraphics[width=0.33\linewidth]{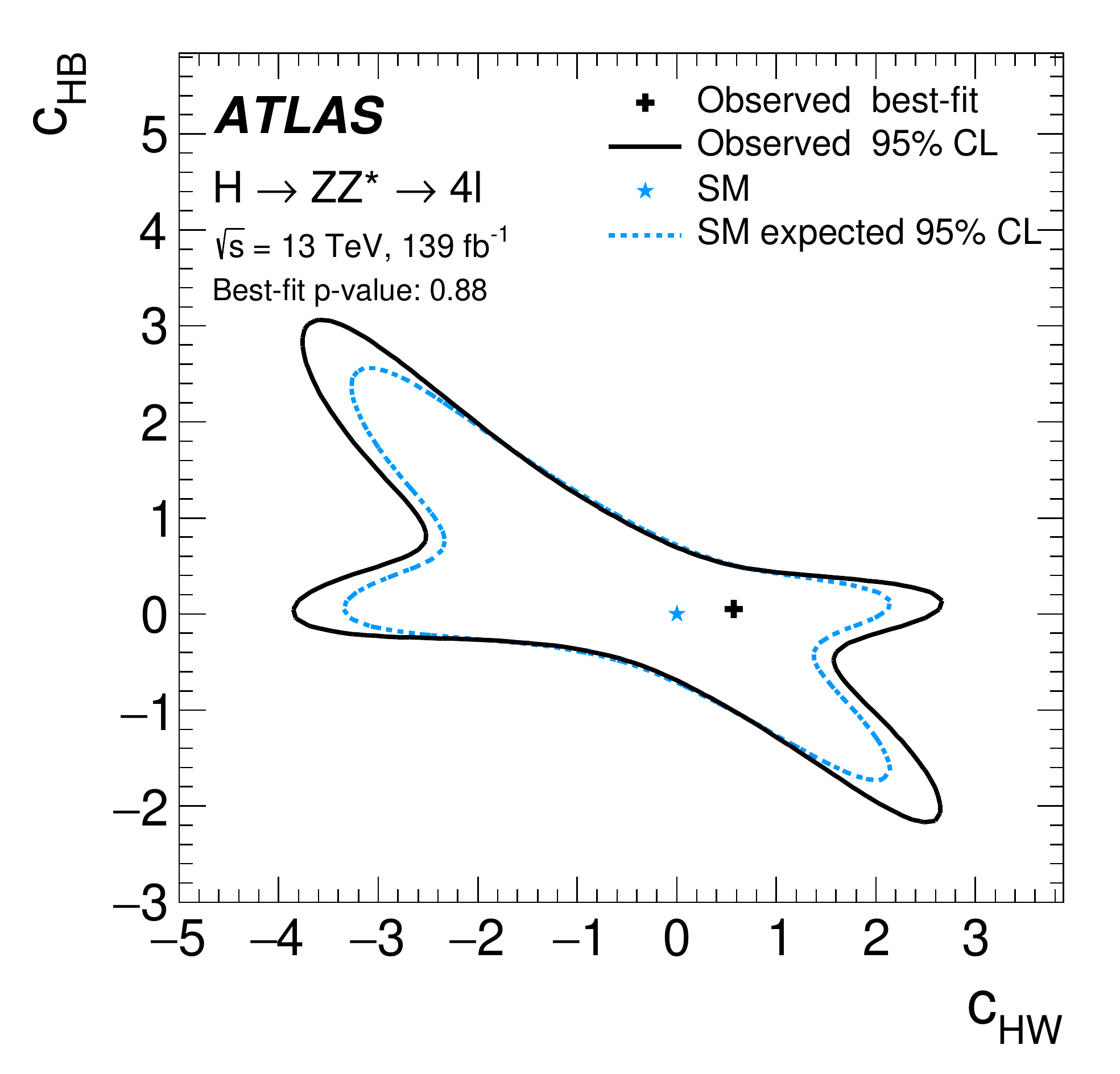}}
\subfloat[]{
\includegraphics[width=0.33\linewidth]{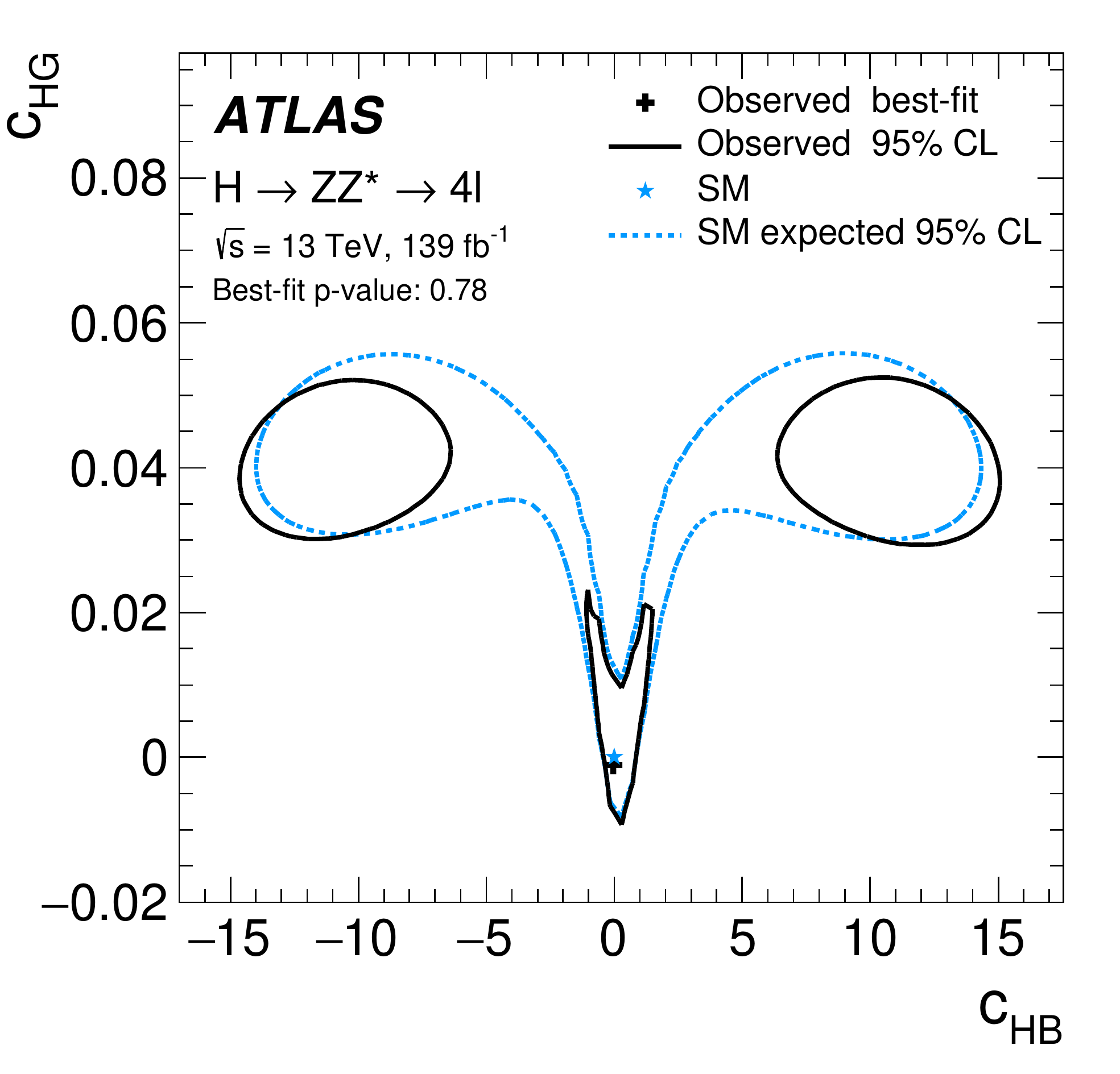}}
\subfloat[]{
\includegraphics[width=0.33\linewidth]{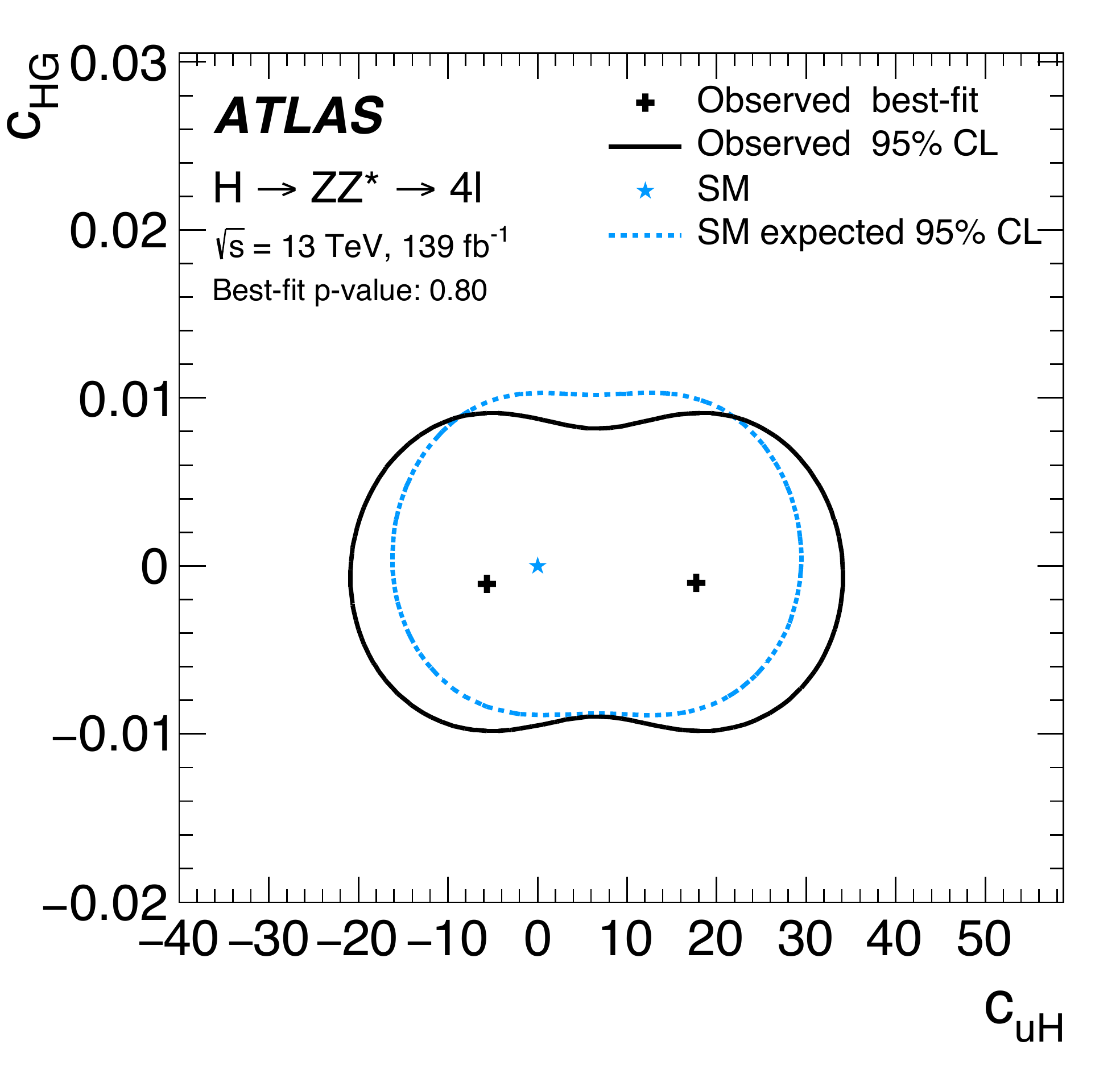}}
\caption{Expected (dashed line) and observed (full line) 2D-fit likelihood curves at the 95\%~CL for the SMEFT Wilson coefficients of CP-even operators at an integrated luminosity of \LumExact\ and  $\sqrt{s}=13$~\TeV. The best fit to the data (solid cross) and the SM prediction (star) are also indicated. Except for the two fitted Wilson coefficients, all others are set to zero.}
\label{fig:mainEFT_results_2da}
\end{figure}

\begin{figure}[!htbp]
\centering
\subfloat[]{
\includegraphics[width=0.33\linewidth]{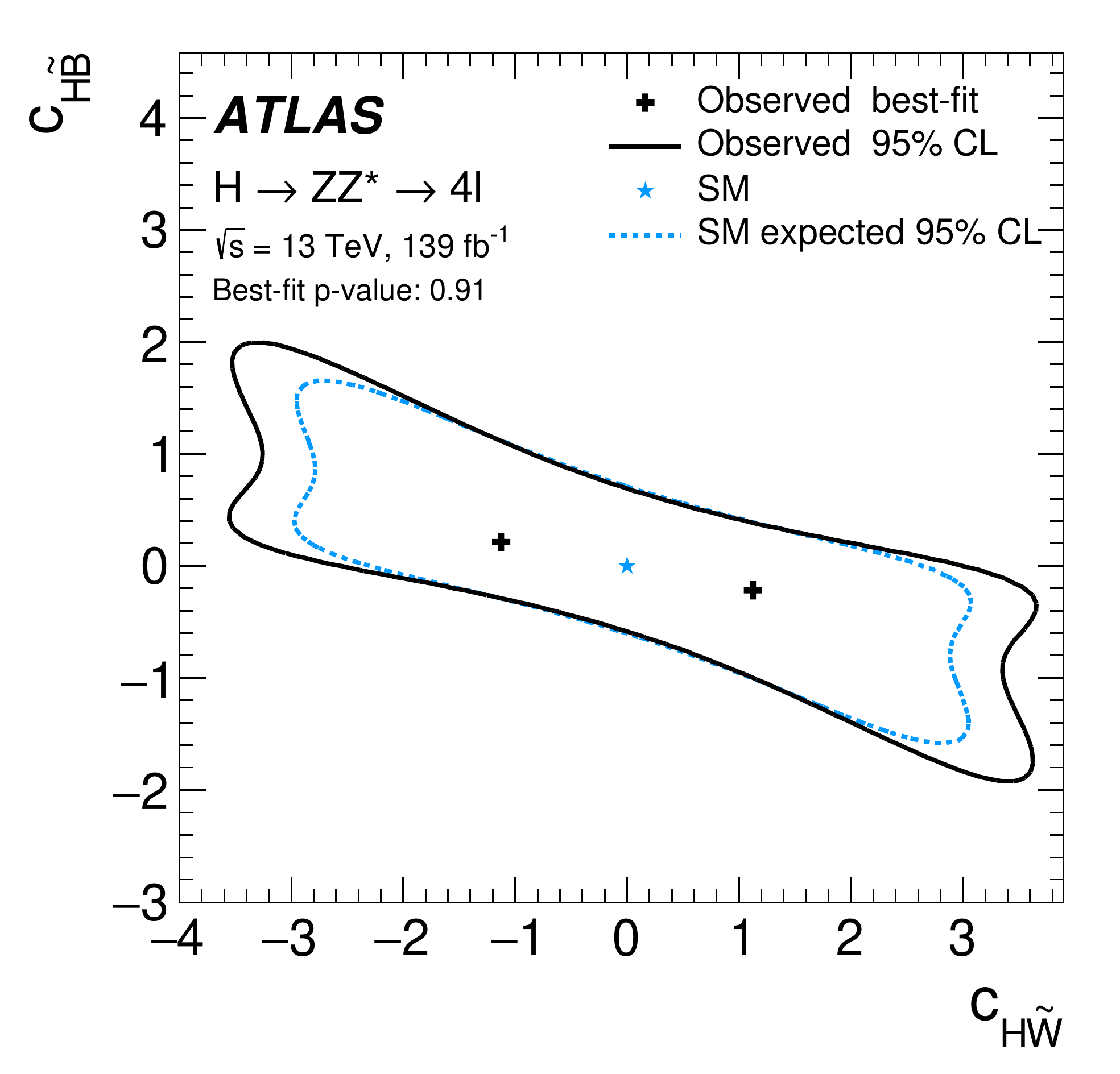}}
\subfloat[]{
\includegraphics[width=0.33\linewidth]{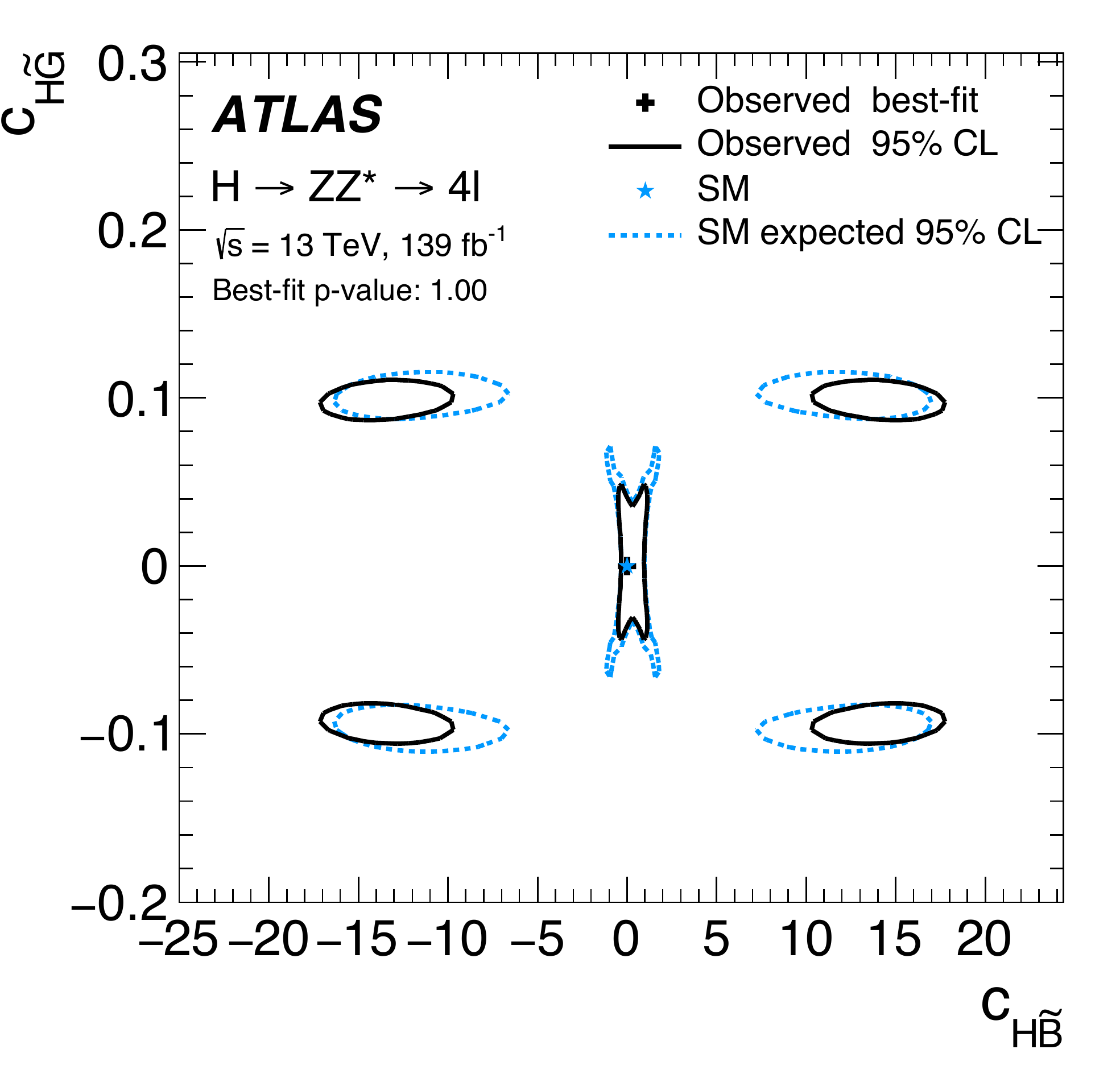}}
\subfloat[]{
\includegraphics[width=0.33\linewidth]{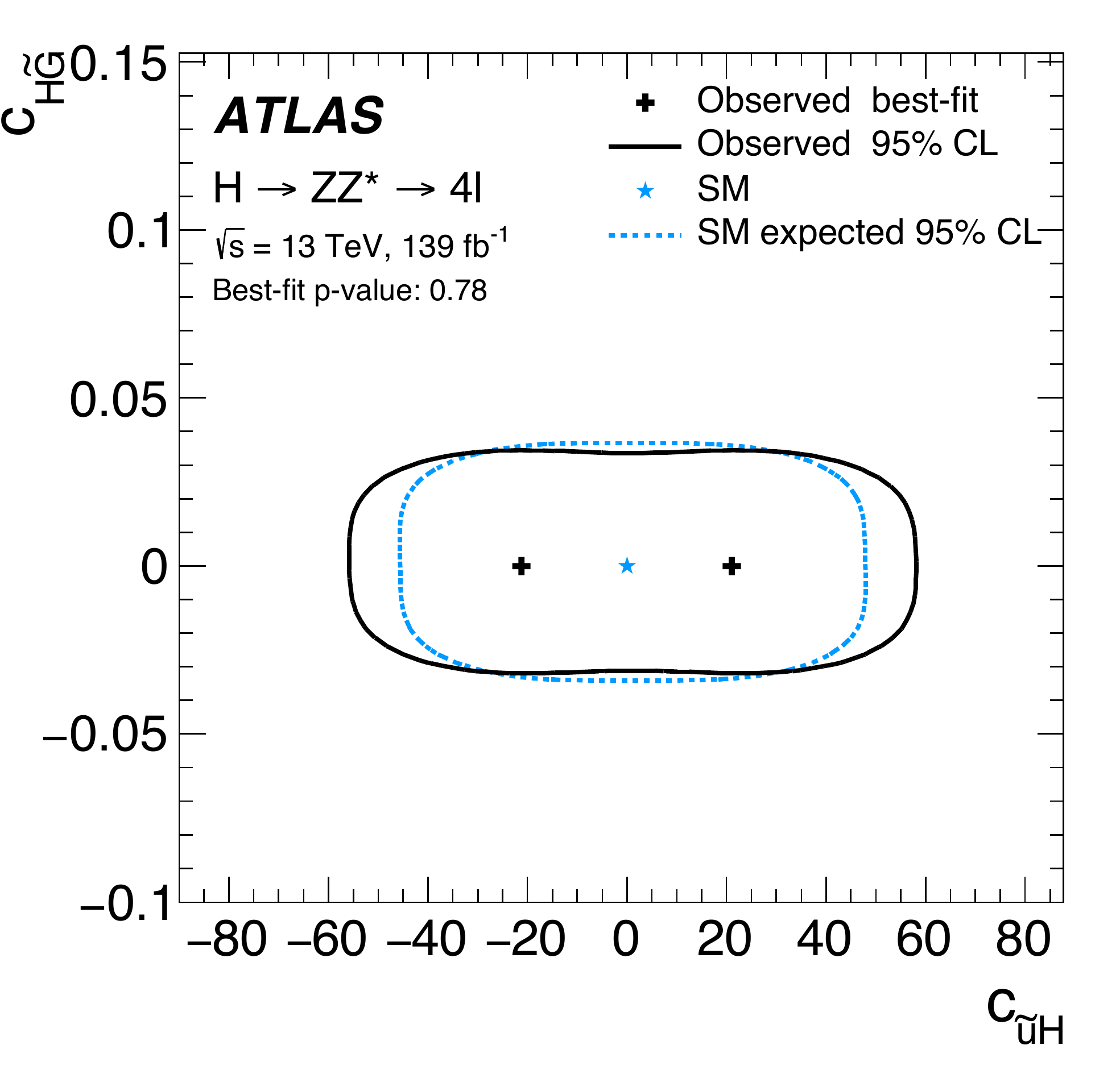}}
\caption{Expected (dashed line) and observed (full line) 2D-fit likelihood curves at the 95\%~CL for the SMEFT Wilson coefficients of CP-odd operators at an integrated luminosity of \LumExact\ and  $\sqrt{s}=13$~\TeV. The best fit to the data (solid cross) and the SM prediction (star) are also indicated. Except for the two fitted Wilson coefficients, all others are set to zero.}
\label{fig:mainEFT_results_2db}
\end{figure}

To explore possible correlations between different Wilson coefficients, the simultaneous fits are also performed on two Wilson coefficients at a time. The corresponding results are shown in \figref{fig:mainEFT_results_2da} for several combinations of two CP-even EFT parameters and in \figref{fig:mainEFT_results_2db} for the corresponding CP-odd operators. The best-fit values as well as the deviation from the SM prediction are shown in Table~\ref{tab:EFTSMEFTresults2D}. Good agreement with the SM predictions is observed for all such possible combinations.

\begin{table}[ht!]
\centering
\caption{The best-fit values and the corresponding deviation from the SM prediction obtained from the two-dimensional likelihood scans of the   CP-odd BSM coupling parameters  performed with $139$~fb$^{-1}$ of data at a centre-of-mass energy of $\sqrt{s}=13$~\TeV.  The limits are computed using the confidence-level interval method.  Except for the two fitted BSM coupling parameters, all others are set to zero.}
\renewcommand{\arraystretch}{1.3}
\begin{tabular}{l p{0.5cm}p{0.1cm}p{2cm} p{0.4cm}p{0.1cm}l c }
\hline\hline
 
\multicolumn{1}{c}{BSM coupling} &
\multicolumn{6}{c}{Observed best fit} &
Best-fit  \\
\multicolumn{1}{c}{parameter} &
&
&
&
&
&
&
$p$-value  \\
\hline

$\cHW, \ \cHB$ & $\cHWhat$&$=$&$\phantom{-}0.57$ & $\cHBhat$&$=$&$\phantom{-}\phantom{0}0.05$  & $0.88$  \\
$\cHG, \ \cHB$ & $\cHGhat$&$=$&$-0.001$ & $\cHBhat$&$=$&\phantom{..}$-0.04$  & $0.78$ \\
$\cHG, \ \cuH$ & $\cHGhat$&$=$&$-0.001$ & $\cuHhat$&$=$&\phantom{..}$-5.7$, $17.7$  & $0.80$  \\
$\cHWtil, \ \cHBtil$ & $~\cHWtilhat$&$=$&$\pm1.12$ & $\cHBtilhat$&$=$&\phantom{..}$\mp0.21$  & $0.91$ \\
$\cHGtil, \ \cHBtil$ & $\cHGtilhat$&$=$&$\phantom{-}0.00$ & $\cHBtilhat$&$=$&$\phantom{-}\phantom{0}0.00$  & $1.00$ \\
$\cHGtil, \ \cuHtil$ & $\cHGtilhat$&$=$&$\phantom{-}0.000$ & $\cuHtilhat$&$=$&$\pm21$  & $0.78$ \\
\noalign{\vspace{0.05cm}}
\hline\hline
\end{tabular}
\label{tab:EFTSMEFTresults2D}
\end{table}


The anti-correlation between the $c_{HW}$ and $c_{HB}$ coefficients, as well as between $c_{H\widetilde{W}}$ and $c_{H\widetilde{B}}$, is driven by their impact on the signal acceptance. The non-ellipsoidal shape is caused by the acceptance correction, which degrades the original branching ratio-driven sensitivity for increasing  parameter values, in particular in the case of the $c_{HW}$ ($c_{H\widetilde{W}}$) coefficient. The sensitivity is, however, partially recouped by the \VBF production vertex.
 
The `V'-shaped correlation between the $c_{HG}$ and $c_{HB}$ parameters is due to the interplay between the EFT parameterisation in the \ggF production vertex and the parameterisation of the branching ratios and acceptances. The \ggF production vertex provides the constraint on the $c_{HG}$ parameter alone, independently of $c_{HB}$. Due to the decay vertex with its acceptance corrections, this constrained range is shifted upward with increasing values of $c_{HB}$. Close to the SM point, the constrained $c_{HG}$ range remains approximately the same as without the decay constraints. An additional constraint on $c_{HB}$ is provided by the \VBF production mode. Around the SM point, the $c_{HB}$ constraints correspond approximately to those from the one-dimensional parameter fit. Additional sensitivity to intermediate values of the $c_{HB}$ parameter is provided by the acceptance corrections, resulting in two additional allowed parameter regions that are disjoint from the region around the SM point. Similar arguments hold also for the CP-odd case with the $c_{H\widetilde{G}}$ and $c_{H\widetilde{B}}$ parameters. As opposed to the CP-even case, however,  the likelihood contours are symmetric around the $c_{H\widetilde{G}}=0$ axis, since there are no linear terms contributing to the \ggF production cross-section.
 
The correlation between the $c_{HG}$ and $c_{uH}$ ($c_{H\widetilde{G}}$ and $c_{\widetilde{u}H}$) parameters is introduced through the interference term in the \ttH vertex. However, the impact of this term on the final result is negligible since the $c_{HG}$ ($c_{H\widetilde{G}}$) parameter is already constrained to very small values compared with $c_{uH}$ ($c_{\widetilde{u}H}$). Therefore, the \ttH production vertex mainly constrains the $c_{uH}$ and $c_{\widetilde{u}H}$ parameters, while the \ggF vertex constrains only the other two. The acceptance correction has no impact on these results. The CP-odd parameter range is less constrained than the CP-even one due to the missing linear $c_{\widetilde{u}H}$ terms in the cross-section parameterisation.
\FloatBarrier

\section{Conclusion}
\label{sec:conclusion}
Higgs boson properties are studied in the four-lepton decay channel using \LumExact\ of LHC proton--proton collision data at $\sqrt{s}=13$~\TeV\ collected by the ATLAS experiment. The Higgs boson candidate events are categorised into several topologies, providing sensitivity to different production modes in various regions of phase space. Additional multivariate discriminants are used to further improve the sensitivity in reconstructed event categories with a sufficiently large number of events. The cross-section times branching ratio for $H\to ZZ^*$ decay measured in dedicated production bins are in good agreement with the SM predictions. The inclusive cross-section times branching ratio for $H\to ZZ^*$ decay in the Higgs boson rapidity range of $|y_H|<2.5$ is measured to be $1.34 \pm 0.12$~pb compared with the SM prediction of $1.33\pm 0.08$~pb. Results are
also interpreted within the $\kappa$-framework with coupling-strength modifiers $\kappa_V$ and $\kappa_F$, showing compatibility with the SM. Based on the product of cross-section, branching ratio and acceptance measured in Reduced Stage-1.1 production bins of simplified template cross-sections, constraints are placed on possible
CP-even and CP-odd BSM interactions of the Higgs boson to vector bosons, gluons and top quarks within an effective field theory framework in the $H\to ZZ^*$ decay. The data are found to be consistent with the SM hypothesis.


\section*{Acknowledgements}
 

We thank CERN for the very successful operation of the LHC, as well as the
support staff from our institutions without whom ATLAS could not be
operated efficiently.
 
We acknowledge the support of ANPCyT, Argentina; YerPhI, Armenia; ARC, Australia; BMWFW and FWF, Austria; ANAS, Azerbaijan; SSTC, Belarus; CNPq and FAPESP, Brazil; NSERC, NRC and CFI, Canada; CERN; CONICYT, Chile; CAS, MOST and NSFC, China; COLCIENCIAS, Colombia; MSMT CR, MPO CR and VSC CR, Czech Republic; DNRF and DNSRC, Denmark; IN2P3-CNRS and CEA-DRF/IRFU, France; SRNSFG, Georgia; BMBF, HGF and MPG, Germany; GSRT, Greece; RGC and Hong Kong SAR, China; ISF and Benoziyo Center, Israel; INFN, Italy; MEXT and JSPS, Japan; CNRST, Morocco; NWO, Netherlands; RCN, Norway; MNiSW and NCN, Poland; FCT, Portugal; MNE/IFA, Romania; MES of Russia and NRC KI, Russia Federation; JINR; MESTD, Serbia; MSSR, Slovakia; ARRS and MIZ\v{S}, Slovenia; DST/NRF, South Africa; MINECO, Spain; SRC and Wallenberg Foundation, Sweden; SERI, SNSF and Cantons of Bern and Geneva, Switzerland; MOST, Taiwan; TAEK, Turkey; STFC, United Kingdom; DOE and NSF, United States of America. In addition, individual groups and members have received support from BCKDF, CANARIE, Compute Canada and CRC, Canada; ERC, ERDF, Horizon 2020, Marie Sk{\l}odowska-Curie Actions and COST, European Union; Investissements d'Avenir Labex, Investissements d'Avenir Idex and ANR, France; DFG and AvH Foundation, Germany; Herakleitos, Thales and Aristeia programmes co-financed by EU-ESF and the Greek NSRF, Greece; BSF-NSF and GIF, Israel; CERCA Programme Generalitat de Catalunya and PROMETEO Programme Generalitat Valenciana, Spain; G\"{o}ran Gustafssons Stiftelse, Sweden; The Royal Society and Leverhulme Trust, United Kingdom.
 
The crucial computing support from all WLCG partners is acknowledged gratefully, in particular from CERN, the ATLAS Tier-1 facilities at TRIUMF (Canada), NDGF (Denmark, Norway, Sweden), CC-IN2P3 (France), KIT/GridKA (Germany), INFN-CNAF (Italy), NL-T1 (Netherlands), PIC (Spain), ASGC (Taiwan), RAL (UK) and BNL (USA), the Tier-2 facilities worldwide and large non-WLCG resource providers. Major contributors of computing resources are listed in Ref.~\cite{ATL-GEN-PUB-2016-002}.
 

\printbibliography

\clearpage 
 
\begin{flushleft}
\hypersetup{urlcolor=black}
{\Large The ATLAS Collaboration}

\bigskip

\AtlasOrcid[0000-0002-6665-4934]{G.~Aad}$^\textrm{\scriptsize 102}$,    
\AtlasOrcid[0000-0002-5888-2734]{B.~Abbott}$^\textrm{\scriptsize 128}$,    
\AtlasOrcid{D.C.~Abbott}$^\textrm{\scriptsize 103}$,    
\AtlasOrcid[0000-0002-2788-3822]{A.~Abed~Abud}$^\textrm{\scriptsize 36}$,    
\AtlasOrcid{K.~Abeling}$^\textrm{\scriptsize 53}$,    
\AtlasOrcid{D.K.~Abhayasinghe}$^\textrm{\scriptsize 94}$,    
\AtlasOrcid[0000-0002-8496-9294]{S.H.~Abidi}$^\textrm{\scriptsize 166}$,    
\AtlasOrcid[0000-0002-8279-9324]{O.S.~AbouZeid}$^\textrm{\scriptsize 40}$,    
\AtlasOrcid{N.L.~Abraham}$^\textrm{\scriptsize 155}$,    
\AtlasOrcid[0000-0001-5329-6640]{H.~Abramowicz}$^\textrm{\scriptsize 160}$
% ...
\bigskip
\\

% ...
$^{36}$CERN, Geneva; Switzerland.\\
% ...
$^{40}$Niels Bohr Institute, University of Copenhagen, Copenhagen; Denmark.\\
% ...
$^{53}$II. Physikalisches Institut, Georg-August-Universit\"{a}t G\"ottingen, G\"ottingen; Germany.\\
% ...
$^{94}$Department of Physics, Royal Holloway University of London, Egham; United Kingdom.\\
% ...
$^{102}$CPPM, Aix-Marseille Universit\'e, CNRS/IN2P3, Marseille; France.\\
$^{103}$Department of Physics, University of Massachusetts, Amherst MA; United States of America.\\
% ...
$^{128}$Homer L. Dodge Department of Physics and Astronomy, University of Oklahoma, Norman OK; United States of America.\\
% ...
$^{155}$Department of Physics and Astronomy, University of Sussex, Brighton; United Kingdom.\\
% ...
$^{160}$Raymond and Beverly Sackler School of Physics and Astronomy, Tel Aviv University, Tel Aviv; Israel.\\
% ...
$^{166}$Department of Physics, University of Toronto, Toronto ON; Canada.\\
% ...

\end{flushleft}


\end{document}